\documentclass[twocolumn]{aastex63}
\def\actaa{Acta Astronomica}

\usepackage{amsmath}
\usepackage{lineno}

\hypersetup{linkcolor=red,citecolor=blue,filecolor=green,urlcolor=magenta}

\begin{document}

\shorttitle{RR Lyrae PLZ \& PWZ Relations}
\shortauthors{Ngeow et al.}

\title{Zwicky Transient Facility and Globular Clusters: The RR Lyrae $gri$-Band Period-Luminosity-Metallicity and Period-Wesenheit-Metallicity Relations}

\correspondingauthor{C.-C. Ngeow}
\email{cngeow@astro.ncu.edu.tw}

\author[0000-0001-8771-7554]{Chow-Choong Ngeow}
\affil{Graduate Institute of Astronomy, National Central University, 300 Jhongda Road, 32001 Jhongli, Taiwan}

\author[0000-0001-6147-3360]{Anupam Bhardwaj}
\affil{Korea Astronomy and Space Science Institute, Daedeokdae-ro 776, Yuseong-gu, Daejeon 34055, Republic of Korea}
\affil{INAF-Osservatorio astronomico di Capodimonte, Via Moiariello 16, 80131 Napoli, Italy}

\author[0000-0002-5884-7867]{Richard Dekany}
\affiliation{Caltech Optical Observatories, California Institute of Technology, Pasadena, CA 91125, USA}

\author[0000-0001-5060-8733]{Dmitry A. Duev} 
\affiliation{Division of Physics, Mathematics, and Astronomy, California Institute of Technology, Pasadena, CA 91125, USA}  

\author[0000-0002-3168-0139]{Matthew J. Graham}
\affiliation{Division of Physics, Mathematics, and Astronomy, California Institute of Technology, Pasadena, CA 91125, USA}

\author[0000-0001-5668-3507]{Steven L. Groom}
\affiliation{IPAC, California Institute of Technology, 1200 E. California Blvd, Pasadena, CA 91125, USA}

\author[0000-0003-2242-0244]{Ashish~A.~Mahabal}
\affiliation{Division of Physics, Mathematics and Astronomy, California Institute of Technology, Pasadena, CA 91125, USA}
\affiliation{Center for Data Driven Discovery, California Institute of Technology, Pasadena, CA 91125, USA}

\author[0000-0002-8532-9395]{Frank J. Masci}
\affiliation{IPAC, California Institute of Technology, 1200 E. California Blvd, Pasadena, CA 91125, USA}

\author[0000-0002-7226-0659]{Michael S. Medford}
\affiliation{University of California, Berkeley, Department of Astronomy, Berkeley, CA 94720}
\affiliation{Lawrence Berkeley National Laboratory, 1 Cyclotron Rd., Berkeley, CA 94720}

\author[0000-0002-0387-370X]{Reed Riddle}
\affiliation{Caltech Optical Observatories, California Institute of Technology, Pasadena, CA 91125, USA}

\begin{abstract}

  Based on time-series observations collected from Zwicky Transient Facility (ZTF), we derived period-luminosity-metallicity (PLZ) and period-Wesenheit-metallicity (PWZ) relations for RR Lyrae located in globular clusters. We have applied various selection criteria to exclude RR Lyrae with problematic or spurious light curves. These selection criteria utilized information on the number of data points per light curve, amplitudes, colors, and residuals on the period-luminosity and/or period-Wesenheit relations. Due to blending, a number of RR Lyrae in globular clusters were found to be anomalously bright and have small amplitudes of their ZTF light curves. We used our final sample of $\sim750$ RR Lyrae in 46 globular clusters covering a wide metallicity range ($-2.36 < \mathrm{[Fe/H]} < -0.54$~dex) to derive PLZ and PWZ relations in $gri$ bands. In addition, we have also derived the period-color-metallicity (PCZ) and for the first time, the PQZ relations where the Q-index is extinction-free by construction. We have compared our various relations to empirical and theoretical relations available in literature, and found a good agreement with most studies. Finally, we applied our derived PLZ relation to a dwarf galaxy, Crater II, and found its true distance modulus should be larger than the most recent determination. 

\end{abstract}


\section{Introduction}

RR Lyrae variables are exclusively old ($\gtrsim 10$~Gyr), low-mass, and short-period pulsating stars that are well-known standard candle with numerous applications in astrophysics. This is because RR Lyrae exhibit the period-luminosity-metallicity (PLZ) relations that can be used to determine distance, where metallicity is approximated by $[\mathrm{Fe/H}]$. In the $V$-band, slope of the PLZ relation almost vanishes due to a nearly flat bolometric correction for a range of temperatures  within the instability strip of RR Lyrae \citep{bono2003,bono2003a}, leaving an absolute $V$-band magnitude and metallicity, the $M_V$-$[\mathrm{Fe/H}]$, relation. Toward the near infrared $K$-band, the bolometric correction is a linear function of temperature, hence exhibiting a PLZ relation \citep{bono2003,bono2001,bono2003a,marconi2003}. Indeed, RR Lyrae exhibit PLZ relations in $R$-band and longer wavelengths, as demonstrated, for examples, in \citet{catelan2004}, \citet{marconi2015}, \citet{neeley2017} and \citet{neeley2019}. A wealth of literature can be found on the calibrations of the $M_V$-$[\mathrm{Fe/H}]$ and/or PLZ relations (especially in the $K$-band), for examples, the reviews presented in \citet{smith2004}, \citet{sandage2006}, \citet{bono2016}, \citet{beaton2018}, \citet{bhardwaj2020,bhardwaj2022}, and references therein.

In optical bands, the PLZ relations, especially the $M_V$-$[\mathrm{Fe/H}]$ relation\footnote{We recall that $M_V$-$[\mathrm{Fe/H}]$ relation is a special case of PLZ relation with vanishing slope.}, are well established in the Johnson-Cousins ($BVRI$) filters, and recently extended to the {\it Gaia's} $G$-band \citep{muraveva2018}. In contrast, there is only a few empirical calibrations of the PLZ relations in the Sloan, or Sloan-like, filters. Based on the Pan-STARRS1 data for $\sim50$ RR Lyrae in five globular clusters, \citet{ses2017} derived the empirical $griz$ PLZ relations. \citet{viv2017} calibrated $ugriz$ period-luminosity (PL) relations for both fundamental mode ab-type and first-overtone c-type RR Lyrae (or RRab and RRc, respectively) in M5 with DECam \citep[Dark Energy Camera,][]{flaugher2015} observations. Recently, \citet{bhardwaj2021} derived the $gi$-band (and $JK_s$-band) PL relations for both types of RR Lyrae in M15 using the archival CFHT (Canada-France-Hawaii Telescope) data. These works were either based on a particular globular cluster or relied on a rather limited sample of globular clusters that host RR Lyrae. For a comparison, \citet{sollima2006} demonstrated that a $K$-band PLZ relation can be derived from a sample of 538 RR Lyrae in 16 globular clusters. Similarly, \citet{dambis2014} derived the PLZ relations in mid-infrared $W1$ and $W2$ band using a sample of 360 RR Lyrae in 15 globular clusters and 275 RR Lyrae in 9 globular clusters, respectively. Finally, \citet{nemec1994} derived the $BVK$-band PLZ relations using more than 1000 RR Lyrae in 22 globular clusters (with $\sim 195$ to $\sim 552$ RR Lyrae, depending on the filters), the Magellanic Clouds, and 5 local dwarf galaxies.

\begin{deluxetable}{lccll}
  \tabletypesize{\scriptsize}
  \tablecaption{Comparison of optical time-domain surveys in the northern sky.\label{tab_survey}}
  \tablewidth{0pt}
  \tablehead{
    \colhead{Survey\tablenotemark{a}} &
    \colhead{Filters\tablenotemark{b}} &
    \colhead{Pixel Scale\tablenotemark{c}} &
    \colhead{Photometry\tablenotemark{d}} &
    \colhead{Depth} 
  }
  \startdata
  ZTF         & $gri$   & 1.01  & PSF \& AP & $r\sim20.6$ \\
  PS1 $3\pi$  & $grizy$ & 0.258 & PSF \& AP & $r\sim21.8$ \\
  ATLAS       & $oc$    & 1.86  &  PSF            & $m\sim19.5$ \\
  ASAS-SN     & $gV$    & 8.0   & AP        & $V\sim17$   \\
  CSS         & ---     & 1.5   & AP        & $V\sim19.5$ \\
  LINEAR      & ---     &  2.25 & AP & $m\sim18$ \\
  SuperWASP   & ---     & 13.7  & AP & $V\sim15$ \\
  \enddata
  \tablenotetext{a}{Abbreviation for each surveys (references are the sources of information entering into this Table): ZTF = Zwicky Transient Facility \citep{bel19,mas19}; PS1 $3\pi$ = Pan-STARRS1 $3\pi$ Survey \citep[][see also \url{https://panstarrs.stsci.edu/}]{chambers2016,magnier2020}; ATLAS = Asteroid Terrestrial-impact Last Alert System \citep{henize2018,tonry2018}; ASAS-SN = All-Sky Automated Survey for Supernovae \citep{kochanek2017}; CSS = Catalina Schmidt Survey \citep[][see also \url{https://catalina.lpl.arizona.edu/}]{drake2013}; LINEAR = Lincoln Near-Earth Asteroid Research \citep{sesar2011}; SuperWASP = Super Wide Angle Search for Planets \citep{pollaco2006}.}
  \tablenotetext{b}{For simplicity, we referred the variants of Sloan-like filters as $ugriz$, for examples the ZTF filters as $gri$ instead of $g_{ZTF}$, $r_{ZTF}$ and $i_{ZTF}$. ``---'' means no filter or clear filter.}
  \tablenotetext{c}{In unit of $\arcsec/$pixel.}
  \tablenotetext{d}{PSF = point-spread function photometry; AP = aperture photometry.}
\end{deluxetable}

Hence, the purpose of this work is to use a large number of globular clusters observed by the Zwicky Transient Facility \citep[ZTF,][]{bellm2017} to improve the derivation of RR Lyrae PLZ relations in Sloan-like filters. In addition to PLZ relations, we also derive the period-Wesenheit-metallicity (PWZ) relation using the same dataset, because the Wesenheit magnitudes are, by construction, extinction-free \citep{madore1982,madore1991}. Table \ref{tab_survey} compares a number of representative time-domain surveys that cover the similar part of the northern sky as observed by ZTF. Given that our target RR Lyrae are located in globular clusters, surveys with large pixel scales and/or catalog products based on aperture photometry are not suitable for our purpose. Time-series data from ATLAS offer competitive quality and quantities similar to ZTF, however main observations of ATLAS were conducted in the customized $o$ and $c$ filters. In case of PS1 $3\pi$ Survey, it has a smaller pixel scale and can reach to a deeper depth than ZTF, but the typical number of observations in each filters is $\lesssim 12$ over its $\sim4.5$~years of operation \citep{ses2017}. In contrast, the average numbers of ZTF observations on our target globular clusters are $\sim218$, $\sim334$ and $\sim95$ in the $gri$-band, respectively (see Section \ref{sec2}). The homogeneous and uniqueness of ZTF data provides several advantages over other time-domain imaging surveys for our purpose, such as observations done in the $gri$ filters and with a fine pixel scale, availability of catalogs based on the PSF (point-spread-function) photometry, a competitive depth to reach most of the globular clusters in northern sky, and numerous observations over a period of $\sim4$ years.

Section \ref{sec2} describes our sample of RR Lyrae in globular clusters, their ZTF light curve data, and the light curve fittings to derive their mean magnitudes. We further filtered out RR Lyrae with problematic light curves using information on amplitudes, colors, and residuals of the period-luminosity (PL) and period-Wesenheit (PW) relations in Section \ref{sec3}. The PLZ and PWZ relations were then derived in Section \ref{sec4} based on the final sample. As byproducts, we also derived the period-color-metallicity (PCZ) relations and investigated the color-color diagram in Section \ref{sec5}. As an example of application, we derived the distance to a dwarf galaxy Crater II using published data together with our derived PLZ relations in Section \ref{sec6}, followed by conclusions given in Section \ref{sec7}.

\section{Sample and Data} \label{sec2}

\subsection{Selections of Globular Clusters and RR Lyrae}\label{sec2.1}

The ``Updated Catalog of Variable Stars in Globular Clusters'' \citep[][hereafter Clement's Catalog]{clement2001,clement2017} was used to select RR Lyrae in globular clusters. We first selected globular clusters that are visible from the Palomar Observatory (i.e. Declination~$>-30^\circ$), and contain at lease one RR Lyrae, either the type RR0 (a.k.a. RRab or fundamental mode) or RR1 (a.k.a RRc or first-overtone mode). The foreground or suspected foreground RR Lyrae labeled in the Clement's Catalog, however, were not included in our selection. We further removed 4 duplicated entries in M80 from the Clement's Catalog, as well as three close ``pairs'' of (blended) RR Lyrae in M3 \citep[4n/s, 250n/s and 270n/s; see][]{guh1994,chi1997,cor2001,ben2006} because they cannot be separated from ZTF observations. Therefore, our initial list contains 961~RR0 and 543~RR1\footnote{Note that the pulsation mode for V4 and V5 in 2MASS-GC02 should be RR1 \citep{bor2007}, however they were mis-labeled as RR0 in the Clement's Catalog. Similarly, V20 in M22 \citep{kunder2013} and V21 in M80 \citep{kopacki2013} should be RR0, they were mis-labeled as RR1 in the Clement's Catalog.}, for a total of 1504 RR Lyrae, in 57 globular clusters. 

\subsection{Adopted Distances, Reddenings and Metallicity}\label{sec2.2}

Homogeneous distances to the selected globular clusters were adopted from the latest compilation given in \citet{baumgardt2021}\footnote{\url{https://people.smp.uq.edu.au/HolgerBaumgardt/globular/orbits.html}}. Based on the adopted globular cluster distances, the reddening $E$ toward each of the RR Lyrae was obtained using the {\tt Bayerstar2019} 3D reddening map\footnote{\url{http://argonaut.skymaps.info/}} \citep{green2019}, via the {\tt dustmaps}\footnote{\url{https://dustmaps.readthedocs.io/en/latest/} \citep{green2018} python package}. Finally, homogeneous values of $[\mathrm{Fe/H}]$ for these globular cluster (G.C.) were adopted from the GOTHAM (GlObular clusTer Homogeneous Abundances Measurements) survey\footnote{\url{http://www.sc.eso.org/~bdias/files/dias+16\_MWGC.txt}} \citep{dias2015, dias2016a, dias2016b, vasquez2018}.

\subsection{ZTF Light Curves Extraction}\label{sec2.3}

ZTF\footnote{Technically, ZTF includes two phases of operation due to difference in funding profiles: before and after 01 December 2020 are known as ZTF-I and ZTF-II, respectively. For simplicity, we collectively referred them as ZTF in this work.} is a dedicated time domain wide-field synoptic sky survey aimed to explore the transient universe. ZTF utilizes the Palomar 48-inch Samuel Oschin Schmidt telescope, together with a new mosaic CCD camera, that provides a field-of-view of 47 squared degree to observe the northern sky in customized $gri$ filters. Observing time and the data right of ZTF was divided into three high-level surveys -- partner surveys, public surveys, and Caltech (California Institute of Technology) surveys. Further details regarding ZTF can be found in \citet{bel19}, \citet{gra19} and \citet{dec20}, and will not be repeated here. Imaging data taken from ZTF were processed with a dedicated reduction pipeline, which is described in detail in \citet{mas19}. The final ZTF data products included reduced images, and catalogs based on both aperture and PSF photometry.

\begin{figure}
  \epsscale{1.1}
  \plotone{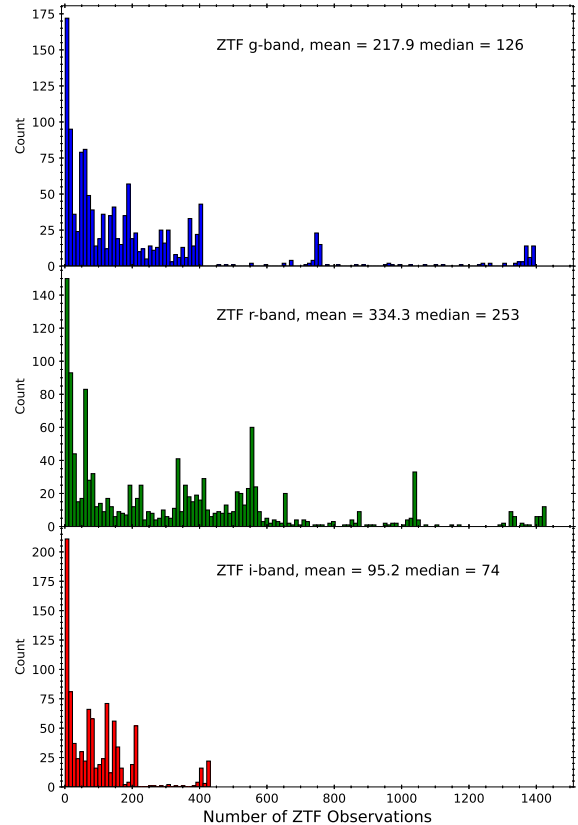}
  \caption{Histograms of the number of ZTF observation for our sample of RR Lyrae in the globular clusters. The mean and median numbers of observations were also given in the upper-right corner in each panels.}\label{fig_hist}
\end{figure}

ZTF $gri$-band light curves were extracted for our sample of RR Lyrae from the PSF catalogs available from both of the ZTF Public Data Release 7\footnote{\url{https://www.ztf.caltech.edu/page/dr7}} (including both public and Caltech data) and the partner surveys until 30 September 2021. Using a matching radius of $1\arcsec$, $gri$-band light curves of 1329 RR Lyrae in our initial list, whenever available, were extracted. Distributions of the number of data points per light curves $N$ (in the $gri$-band separately) are presented in Figure \ref{fig_hist}, where $N$ can be as large as $\sim1390$ in the $gr$-band. The number of data points in the $i$-band light curves is much smaller due to combination of several reasons, such as late arrival of the $i$-band filter, the ZTF public surveys were conducted in the $gr$-band only, and the observing strategies for ZTF partnership surveys were focused on $gr$-band in the early time. Out of the 1329 RR Lyrae in our sample, 1328 of them have ZTF data in either $g$- or $r$-band (or both), but only 907 RR Lyrae have ZTF observations in the $i$-band.

\subsection{Light Curve Fittings}\label{sec2.4}

Template light curve fitting approach was adopted to derive the mean magnitudes in $gri$-band for our sample of RR Lyrae. Since we do not have well-constrained amplitudes for RR Lyrae in our sample, we solved for the amplitude, phase shift, and mean magnitude as free parameters. Given that the minimum number of data points per phased light curve to fit with a template light curve is $3$ to account for the scaling (i.e. amplitude), the $x$-shift (i.e. the phase shift) and the $y$-shift (i.e. the mean magnitude) of the light curves, we further restricted the fitting to be performed only on the ZTF light curve in any bands that contains at least $3$ data points. The {\tt RRLyraeTemplateModelerMultiband} module available in the {\tt astroML/gatspy}\footnote{\url{https://github.com/astroML/gatspy}, also see \citet{vdp2016}.} package \citep[hereafter {\tt gatspy};][]{vdp2015} was employed to fit the ZTF light curves using the $gri$-band template light curves derived in \citet{ses2010}. During the light curves fitting, periods of the RR Lyrae were also revised from the initial values given in the Clement's Catalog, as improvements in the fitted light curves can be seen on these RR Lyrae.

By default, {\tt gatspy} will fit a given light curve with both RR0 and RR1 template light curves. Since the pulsation types of our sample of RR Lyrae are known from the Clement's Catalog, we modified the {\tt gatspy} code to only use either the RR0 or RR1 template light curves for a given RR Lyrae. To remove dubious outliers in the light curves, we adopted a two-iterations process: we first fit the ZTF light curves with the template light curves and data points that are more than $3s$ from the fitted light curves were removed ($s$ is the dispersion of the fitted light curve\footnote{We use $s$ to represent the light curve dispersion and $\sigma$ for the dispersion of the PWZ and PLZ relation.}), and refit the template light curves on the remaining data points. The outputs of our light curve fitting procedures include the improved pulsation period $P$, the intensity-averaged mean magnitudes $\langle m \rangle$ and amplitudes $AMP_m$, where $m=\{g,\ r,\ i\}$ (whenever available), for each RR Lyrae based on the best-fit template light curves. 

\subsection{Errors Estimation on Mean Magnitudes}\label{sec2.5}

\begin{figure}
  \epsscale{1.1}
  \plotone{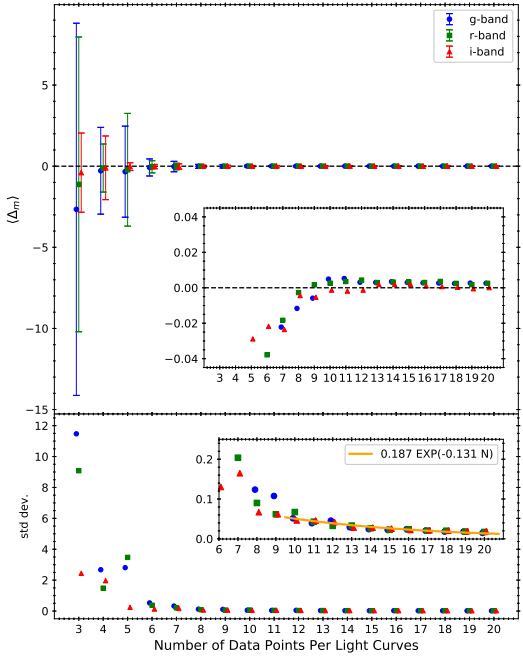}
  \caption{Results on errors estimation, based on simulated light curves (see text for details), for the mean magnitudes obtained with template light curve fittings. For a better visualization, the $g$- and $i$-band data points were offset slightly on the $x$-axis. Upper panel presents the averaged $\Delta_m$, obtained from 1000 simulated light curves, as a function of $N$, the number of data points per light curves. The error bars represent $1\sigma$ standard deviation on the averaged values. The horizontal dashed lines represent $\langle \Delta_m \rangle =0$. The inset figure in the upper panel is an enlarged version with error bars being omitted for clarity. In the lower panel, the corresponding standard deviations were plotted as a function of $N$. The orange curve in the inset figure is the best-fit exponential decay function to describe the trend of the standard deviations, at which the fittings of the exponential decay function were only done on data points with $N\geq10$.}\label{fig_merr}
\end{figure}

We ran Monte Carlo simulations to evaluate the errors on the mean magnitudes based on our procedures described above. Since template light curves fitting in {\tt gatspy} is computationally intensive, we selected nine RR Lyrae that have more than 100 data points per light curves, as well as the dispersions on the final fitted light curves were smaller than $0.05$, in all three bands as our sample to run the Monte Carlo simulations. The fitted mean magnitudes of these nine RR Lyrae were considered as the ``true values'' in our simulations. To create a ``simulated'' light curve, we randomly selected one RR Lyrae out of the nine RR Lyrae, and for this selected RR Lyrae, we randomly picked $N$ data points without replacement, using a uniform random number generator, from the ZTF light curves. The simulated light curve was fitted with the two-steps template light curves fitting procedure as described previously to determine its mean magnitude which was compared with the ``true value'' to determine the difference $\Delta_m=FIT_m-TRUE_m$. We generated 1000 such simulated light curves, separately in the $gri$-band, to built up the distributions of $\Delta_m$, for $N$ spanning from $3$ to $20$. Averaged $\Delta_m$ and the associated standard deviations, were then calculated for each $N$ separately in the $gri$-band and presented in Figure \ref{fig_merr}.

It is clear from Figure \ref{fig_merr} that the largest averaged $\Delta_m$ and standard deviation occurred when $N=3$ and hence the template light curves fitting should not be applied to light curves with only three data points. In the case of a small number of clustered data points, the templates fits can give very large amplitudes resulting in unrealistically large scatter in the mean magnitudes. Furthermore, the averaged $\Delta_m$ values exhibit a trend for $N<10$, and stabilized when $N\geq 10$ with values within $\pm 0.005$~mag. Hence, the template light curves fitting procedures should only be applied to light curves with $N\geq 10$. Large standard deviation can be seen when $N$ is less than 9, which decreases substantially for large $N$. An exponential decay function in the form of $f(N) = 0.187\mathrm{e}^{-0.131N}$ can be used to fit the standard deviations when $N\geq10$. Therefore, the calculated values of standard deviation based on $f(N)$ is $0.051$~mag at $N=10$, reduced to $0.014$~mag at $N=20$, and smaller than $0.001$~mag when $N > 40$. Standard deviations calculated from $f(N)$ would be adopted as an error estimate for the mean magnitudes.

\section{Further Filtering of RR Lyrae Sample}\label{sec3}

\begin{deluxetable*}{lcrrrrrrrrr}
  \tabletypesize{\scriptsize}
  \tablecaption{List of globular clusters and the number of RR Lyrae in the sample. \label{tab_gc}}
  \tablewidth{0pt}
  \tablehead{
    \colhead{G.C.} &
    \colhead{$\mathrm{[Fe/H]}$ (dex)\tablenotemark{a}} &
    \colhead{$n^{RR0}_g$} &
    \colhead{$n^{RR0}_r$} &
    \colhead{$n^{RR0}_i$} &
    \colhead{$n^{RR1}_g$} &
    \colhead{$n^{RR1}_r$} &
    \colhead{$n^{RR1}_i$} &
    \colhead{$n_{\mathrm{tot}}$\tablenotemark{b}} &
    \colhead{$D$ (kpc)\tablenotemark{a}}&
    \colhead{$\langle E \rangle$\tablenotemark{c}} 
  }
  \startdata
NGC6426		& $ -2.36 \pm0.03$ & 7	& 7	& 0	& 5	& 5	& 0	& 12	& $20.71\pm0.35$ & 0.366 \\
M30		& $ -2.33 \pm0.03$ & 3	& 4	& 0	& 2	& 2	& 0	& 6	& $8.46\pm0.09$ & 0.000 \\
M15		& $ -2.27 \pm0.01$ & 53	& 50	& 47	& 53	& 48	& 39	& 109	& $10.71\pm0.10$ & 0.141 \\
M92		& $ -2.27 \pm0.06$ & 11	& 10	& 10	& 6	& 6	& 6	& 17	& $8.50\pm0.07$ & 0.000 \\
M68		& $ -2.25 \pm0.02$ & 14	& 14	& 0	& 16	& 16	& 0	& 30	& $10.40\pm0.10$ & 0.084 \\
NGC5053		& $ -2.24 \pm0.16$ & 6	& 6	& 5	& 4	& 4	& 4	& 10	& $17.54\pm0.23$ & 0.001 \\
NGC2419		& $ -2.12 \pm0.16$ & 26	& 27	& 18	& 30	& 30	& 15	& 57	& $88.47\pm2.40$ & 0.134 \\
NGC5897		& $ -1.99 \pm0.03$ & 3	& 3	& 0	& 3	& 3	& 0	& 6	& $12.55\pm0.24$ & 0.089 \\
NGC4147		& $ -1.95 \pm0.09$ & 5	& 5	& 5	& 10	& 10	& 10	& 15	& $18.54\pm0.21$ & 0.000 \\
M22		& $ -1.92 \pm0.02$ & 10	& 10	& 0	& 13	& 14	& 0	& 24	& $3.30\pm0.04$ & 0.356 \\
NGC6293		& $ -1.92 \pm0.10$ & 2	& 2	& 0	& 2	& 2	& 0	& 4	& $9.19\pm0.28$ & 0.410 \\
M53		& $ -1.90 \pm0.05$ & 29	& 29	& 28	& 29	& 30	& 28	& 59	& $18.50\pm0.18$ & 0.004 \\
Palomar13	& $ -1.85 \pm0.16$ & 4	& 4	& 4	& 0	& 0	& 0	& 4	& $23.48\pm0.40$ & 0.109 \\
NGC5466		& $ -1.82 \pm0.08$ & 12	& 13	& 13	& 7	& 7	& 7	& 20	& $16.12\pm0.16$ & 0.005 \\
M56		& $ -1.80 \pm0.15$ & 1	& 1	& 1	& 2	& 2	& 2	& 3	& $10.43\pm0.14$ & 0.191 \\
NGC5634		& $ -1.77 \pm0.04$ & 8	& 9	& 7	& 6	& 6	& 6	& 15	& $25.96\pm0.62$ & 0.073 \\
M80		& $ -1.72 \pm0.04$ & 6	& 5	& 0	& 6	& 6	& 0	& 12	& $10.34\pm0.12$ & 0.210 \\
M19		& $ -1.70 \pm0.05$ & 1	& 1	& 0	& 0	& 0	& 0	& 1	& $8.34\pm0.16$ & 0.330 \\
M9		& $ -1.70 \pm0.08$ & 7	& 7	& 0	& 8	& 8	& 0	& 15	& $8.30\pm0.14$ & 0.373 \\
NGC7492		& $ -1.67 \pm0.12$ & 1	& 1	& 1	& 2	& 2	& 2	& 3	& $24.39\pm0.57$ & 0.029 \\
Palomar3	& $ -1.60 \pm0.16$ & 6	& 6	& 3	& 0	& 0	& 0	& 6	& $94.84\pm3.23$ & 0.004 \\
M79		& $ -1.57 \pm0.02$ & 5	& 5	& 0	& 3	& 3	& 0	& 8	& $13.08\pm0.18$ & 0.009 \\
NGC7006		& $ -1.57 \pm0.05$ & 48	& 41	& 39	& 4	& 3	& 2	& 53	& $39.32\pm0.56$ & 0.136 \\
M13		& $ -1.55 \pm0.05$ & 1	& 1	& 1	& 8	& 7	& 6	& 9	& $7.42\pm0.08$ & 0.000 \\
M2		& $ -1.51 \pm0.02$ & 21	& 21	& 15	& 10	& 11	& 8	& 32	& $11.69\pm0.11$ & 0.016 \\
M3		& $ -1.48 \pm0.05$ & 101& 132	& 88	& 24	& 35	& 18	& 167	& $10.18\pm0.08$ & 0.035 \\
NGC6934		& $ -1.48 \pm0.11$ & 61	& 62	& 55	& 8	& 8	& 7	& 72	& $15.72\pm0.17$ & 0.098 \\
NGC6355		& $ -1.46 \pm0.06$ & 4	& 4	& 0	& 1	& 1	& 0	& 5	& $8.65\pm0.22$ & 0.865 \\
Palomar5	& $ -1.38 \pm0.16$ & 0	& 0	& 0	& 5	& 5	& 0	& 5	& $21.94\pm0.51$ & 0.086 \\
NGC6235		& $ -1.37 \pm0.08$ & 1	& 2	& 0	& 1	& 1	& 0	& 3	& $11.94\pm0.38$ & 0.379 \\
M72		& $ -1.35 \pm0.08$ & 34	& 35	& 0	& 6	& 6	& 0	& 41	& $16.66\pm0.18$ & 0.041 \\
NGC6229		& $ -1.35 \pm0.16$ & 40	& 36	& 33	& 15	& 15	& 15	& 55	& $30.11\pm0.47$ & 0.096 \\
NGC288		& $ -1.33 \pm0.03$ & 1	& 1	& 1	& 1	& 1	& 0	& 2	& $8.99\pm0.09$ & 0.015 \\
M14		& $ -1.28 \pm0.07$ & 4	& 5	& 0	& 6	& 7	& 0	& 12	& $9.14\pm0.25$ & 0.552 \\
M28		& $ -1.18 \pm0.05$ & 10	& 10	& 9	& 7	& 7	& 6	& 17	& $5.37\pm0.10$ & 0.457 \\
NGC6717		& $ -1.17 \pm0.09$ & 1	& 1	& 0	& 0	& 0	& 0	& 1	& $7.52\pm0.13$ & 0.258 \\
M4		& $ -1.12 \pm0.02$ & 31	& 31	& 0	& 14	& 14	& 0	& 45	& $1.85\pm0.02$ & 0.385 \\
M5		& $ -1.12 \pm0.01$ & 85	& 81	& 75	& 36	& 36	& 35	& 121	& $7.48\pm0.06$ & 0.094 \\
NGC6401		& $ -1.08 \pm0.06$ & 22	& 23	& 15	& 11	& 9	& 7	& 34	& $8.06\pm0.24$ & 1.038 \\
NGC6284		& $ -1.07 \pm0.05$ & 4	& 4	& 0	& 0	& 0	& 0	& 4	& $14.21\pm0.42$ & 0.318 \\
NGC6642		& $ -1.03 \pm0.17$ & 7	& 7	& 0	& 3	& 3	& 0	& 10	& $8.05\pm0.20$ & 0.469 \\
M75		& $ -1.01 \pm0.04$ & 15	& 13	& 0	& 5	& 6	& 0	& 21	& $20.52\pm0.45$ & 0.225 \\
M107		& $ -1.00 \pm0.02$ & 15	& 15	& 0	& 6	& 6	& 0	& 21	& $5.63\pm0.08$ & 0.424 \\
NGC6712		& $ -0.97 \pm0.05$ & 7	& 6	& 6	& 6	& 6	& 3	& 13	& $7.38\pm0.24$ & 0.414 \\
NGC6638		& $ -0.89 \pm0.05$ & 3	& 5	& 0	& 9	& 10	& 0	& 15	& $9.78\pm0.34$ & 0.424 \\
NGC6366		& $ -0.59 \pm0.04$ & 1	& 1	& 0	& 0	& 0	& 0	& 1	& $3.44\pm0.05$ & 0.730 \\
\hline
NGC6544		& $ -1.37 \pm0.07$ & 1	& 1	& 1	& 0	& 0	& 0	& 1	& $2.58\pm0.06$ & 0.767 \\
2MASS-GC02	& $ -1.05 \pm0.16$ & 0	& 0	& 0	& 0	& 1	& 1	& 1	& $5.50\pm0.44$ & 1.650 \\
Terzan10	& $ -0.97 \pm0.16$ & 0	& 2	& 2	& 0	& 0	& 0	& 2	& $10.21\pm0.40$ & 2.029 \\
Djorg2		& $ -0.91 \pm0.05$ & 1	& 1	& 0	& 2	& 2	& 0	& 3	& $8.76\pm0.18$ & 1.059 \\
NGC6540		& $ -0.89 \pm0.73$ & 2	& 2	& 2	& 1	& 1	& 1	& 3	& $5.91\pm0.27$ & 0.619 \\
IC1276		& $ -0.53 \pm0.05$ & 1	& 1	& 1	& 0	& 0	& 0	& 1	& $4.55\pm0.25$ & 1.156 \\
NGC6316		& $ -0.46 \pm0.04$ & 1	& 1	& 0	& 1	& 1	& 0	& 2	& $11.15\pm0.39$ & 0.707 \\
NGC6304		& $ -0.43 \pm0.05$ & 0	& 0	& 0	& 1	& 1	& 0	& 1	& $6.15\pm0.15$ & 0.658 \\
  \enddata
  \tablenotetext{a}{Homogeneous metallicities and distances for each globular cluster (G.C.) were adopted from the GOTHAM survey and \citet{baumgardt2021}, respectively, see Section \ref{sec2.2} for more details.}
  \tablenotetext{b}{Number of RR Lyrae in a given globular cluster after removing RR Lyrae with $N<10$ in all $gri$-band ZTF light curves in our two-iteration template light curve fitting procedures (see Section \ref{sec2.4}).}
  \tablenotetext{c}{$\langle E \rangle$ is the averaged reddening value returned from the {\tt Bayerstar2019} 3D reddening map \citep{green2019} for the RR Lyrae in a given globular cluster, where the number of RR Lyrae used in calculating the averaged reddening value was listed in the column marked with $n_{\mathrm{tot}}$.}
\end{deluxetable*}

\begin{figure}
  \epsscale{1.1}
  \plotone{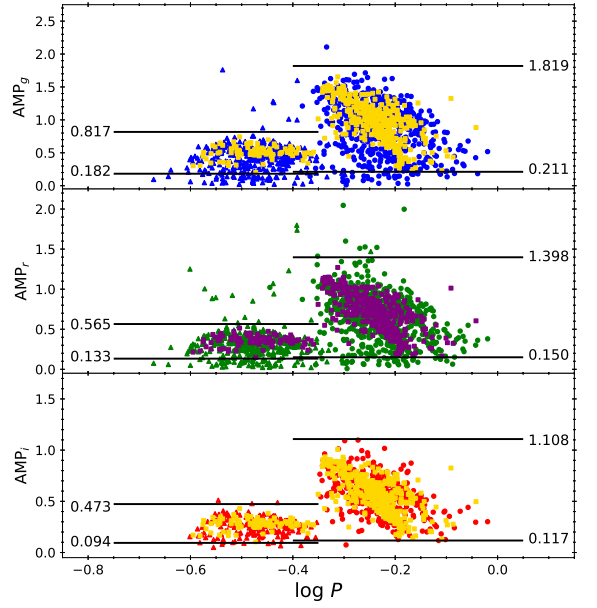}
  \caption{Comparison of the amplitudes for the RR Lyrae in our sample (represented in blue, green and red colors for the $gri$-band, respectively) and the RR Lyre located in the SDSS stripe 82 region \citep[adopted from][shown as square symbols with either gold or purple colors]{ses2010}. RR Lyrae in our sample with type RR0 (or RRab) and RR1 (or RRc) are marked with circles and triangles, respectively. The horizontal lines represent the adopted amplitude cuts in each filters, separated for RR0 and RR1, with values given next to these lines (see text for more details).}\label{fig_amp}
\end{figure}

\begin{figure}
  \epsscale{1.1}
  \plotone{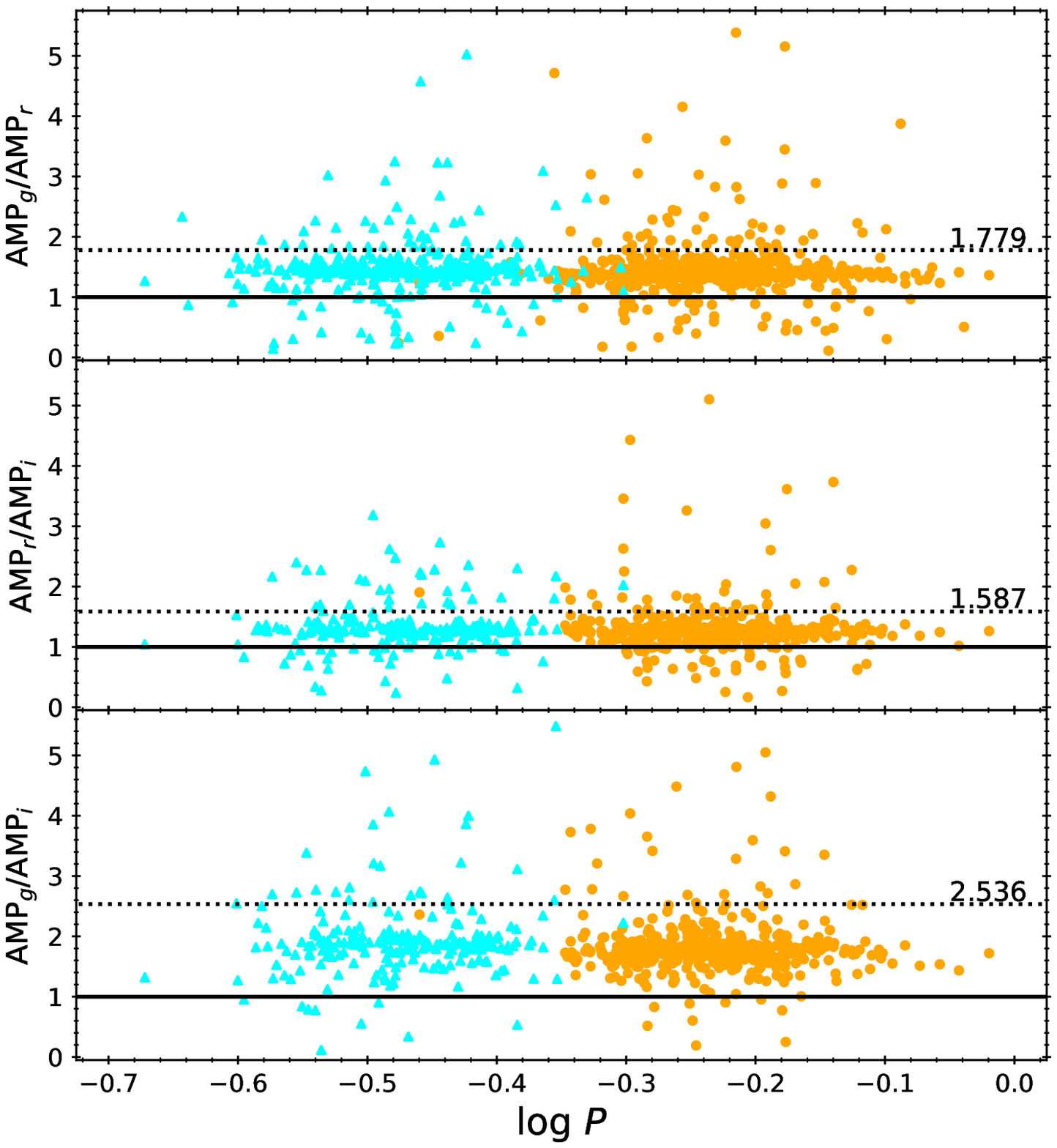}
  \caption{The $gri$-band amplitude ratios as a function of pulsation periods for the RR Lyrae in our sample, marked as filled orange circles and cyan triangles for the RR0 and RR1, respectively. The horizontal solid lines represent the amplitude ratios of unity. The dotted lines are the adopted upper limits for the amplitude ratio cuts (see text for more details).}\label{fig_ampratio}
\end{figure}

Based on the results obtained in Section \ref{sec2.5}, we removed 120 RR Lyrae that have $N<10$ in all three filters after the second iteration in our template light curves fitting procedure, which leaves 1209 RR Lyrae in our sample. These remaining RR Lyrae are located in 54 globular clusters, as summarized in Table \ref{tab_gc} along with their number, and the metallicity, extinction and the distance to each cluster. We further visually inspected a subset of the light curves for this sample of RR Lyrae and found that a fraction of the fitted light curves were problematic. For examples, some of these problematic light curves contain small number of data points that clustered at certain pulsational phases, or the light curves having mis-matched data points from very close-by neighboring sources (due to crowded nature of globular clusters). In many cases, these problematic light curves were found to be caused by blending. Due to highly crowded nature of globular clusters, blending is an unavoidable issue for RR Lyrae located near the center, or core, of the globular clusters. The inclusion of the fluxes from very close and yet unresolved constant stars will have two effects on the RR Lyrae light curves: making the RR Lyrae to appear brighter and the amplitudes of the RR Lyrae light curves will be greatly ``damped'' or reduced. A similar discussion on Cepheid's light curve due to blending can be found in \citet{riess2020}. Therefore, those RR Lyrae affected by blending, as well as other problematic light curves, should be excluded. 

\subsection{Filtering Based on Amplitudes}\label{sec3.1}

We first examined amplitudes of the RR Lyrae remained in our sample. Since the light curve amplitudes for pulsating stars cannot be arbitrary large, fitted light curves with large amplitudes indicated there were some problems related to the sampling of the light curves. Similarly, if the light curve amplitudes are too small this would hint the presence of blending. The $gri$-band amplitudes for our sample of RR Lyrae are shown in Figure \ref{fig_amp}, overlaid with a sample of 483 RR Lyrae located in the Sloan Digital Sky Survey (SDSS) Stripe 82 region \citep{ses2010}. The SDSS Stripe 82 RR Lyrae samples are located in the Galactic halo, hence they are expected to be unaffected by blending. Clearly, amplitudes for some of our sample of RR Lyrae were either larger or smaller than those laid out by the SDSS Stripe 82 RR Lyrae, which can be used to define the amplitude cuts. To account for the possibility that the RR Lyrae in the globular clusters may have larger amplitudes, the maximum amplitudes found in the SDSS Stripe 82 RR Lyrae were then increased by $10\%$ and adopted as the maximum amplitude cuts. Similarly, the minimum amplitude cuts were based on the minimum amplitudes found in the SDSS Stripe 82 RR Lyrae but decreased by $10\%$. These adopted amplitude cuts were shown as horizontal lines in Figure \ref{fig_amp}. 

Amplitude ratios between different filters can also be used to filter out the problematic light curves. Figure \ref{fig_ampratio} presents the derived amplitude ratios for our sample of RR Lyrae, at which majority of the RR Lyrae fall in a tight sequence. Since the amplitudes for RR Lyrae are larger at shorter wavelength filters \citep[for examples, see][]{braga2016,bhardwaj2020a}, amplitude ratios as defined in Figure \ref{fig_ampratio} should have values larger than unity (indicated with solid lines in Figure \ref{fig_ampratio}). However, there are a number of outliers with unusually large amplitude ratios. Therefore, we first exclude those RR Lyrae in our sample with amplitude ratios smaller than unity, the remaining RR Lyrae were used to estimate the averaged amplitude ratios via an iterative $3\sigma$-clipping algorithm, which were found to be:

\begin{equation*}
  \langle \frac{AMP_g}{AMP_r} \rangle = 1.419,  \ \langle \frac{AMP_r}{AMP_i} \rangle = 1.266, \ \langle \frac{AMP_g}{AMP_i} \rangle = 1.782,
\end{equation*}

\noindent with corresponding standard deviations of $0.120$, $0.107$, and $0.251$, respectively. The lower limit of the amplitude ratio cut is 1. For the upper limit, we adopted a value that is three times of the standard deviation larger than the averaged amplitude ratio, indicated by the dotted lines in Figure \ref{fig_ampratio}. While deriving these averaged values (and their standard deviations), we did not separate out the RR0 and RR1 because the derived averaged values do not show a significant deviation between them (for example, the averaged $gr$-band amplitude ratios are $1.405$ and $1.452$ for RR0 and RR1, respectively). 

\begin{figure}
  \epsscale{1.1}
  \plotone{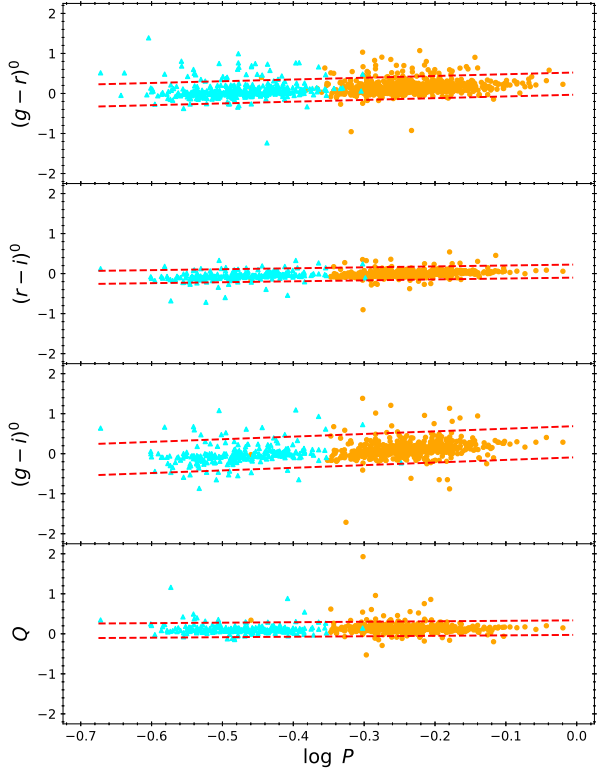}
  \caption{Extinction corrected period-color (PC, top three panels) relations and the extinction-free $Q$-indices as a function of pulsation periods (i.e. the PQ relation, bottom panel) for the RR Lyrae in our sample, separated for RR0 (filled orange circles) and RR1 (filled cyan triangles). The red dashed lines indicate the selection boundaries based on the provisional fitted PC/PQ relations to the RR0 and RR1 combined samples (see text for more details).}\label{fig_iniPC}
\end{figure}

Combining the amplitude cuts as given in Figure \ref{fig_amp} and the amplitude ratio cuts as shown in Figure \ref{fig_ampratio} (the dotted and the solid lines), there are 331 RR Lyrae in our sample (out of 1209) which did not satisfy the selection criteria based on either the amplitudes or amplitude ratios, or both. These RR Lyrae were flagged as ``A'' in our sample.

\subsection{Filtering Based on Colors}\label{sec3.2}

\begin{figure*}
  \centering
  \begin{tabular}{ccc}
    \includegraphics[width=0.71\columnwidth]{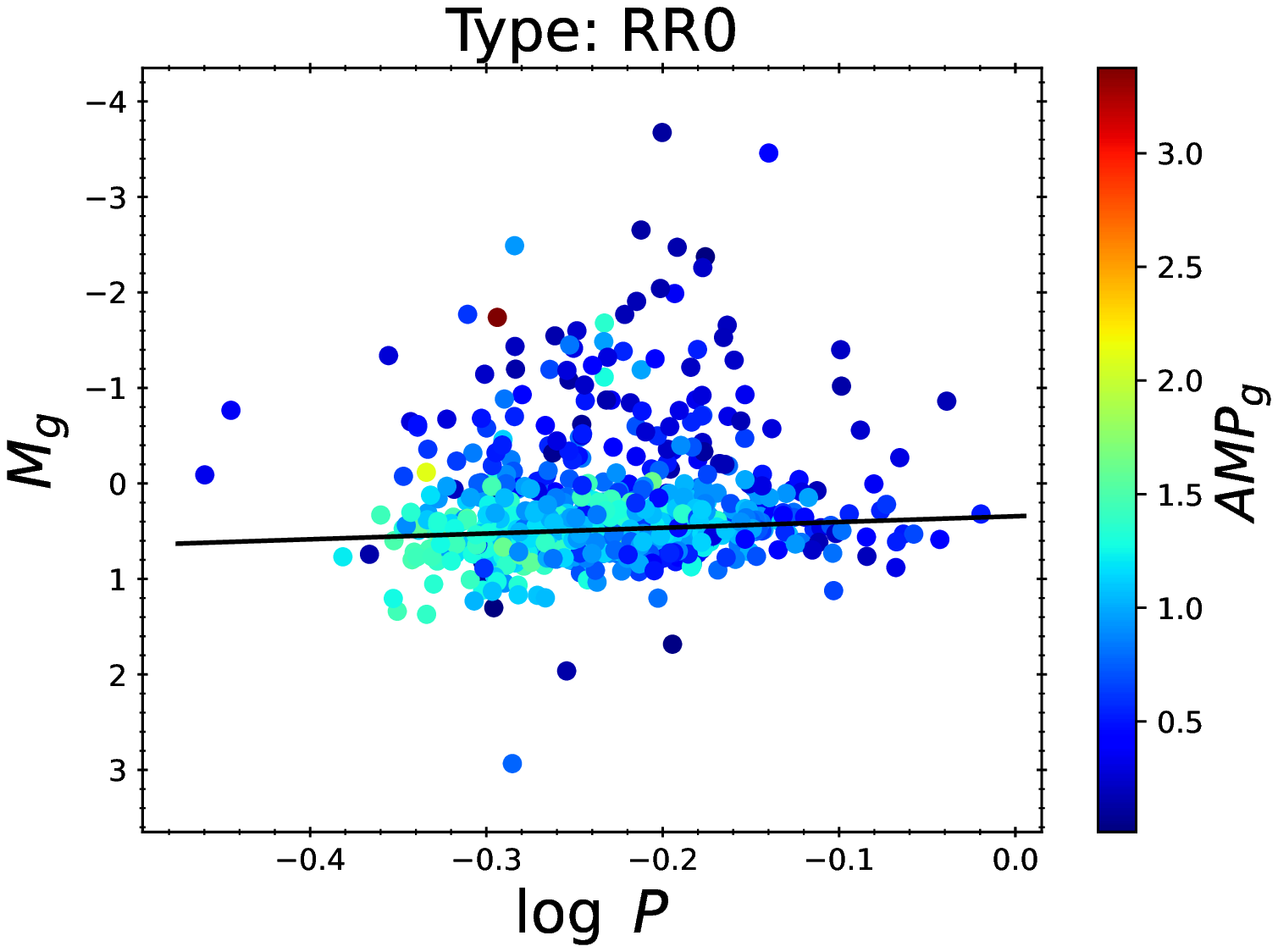} &   \includegraphics[width=0.71\columnwidth]{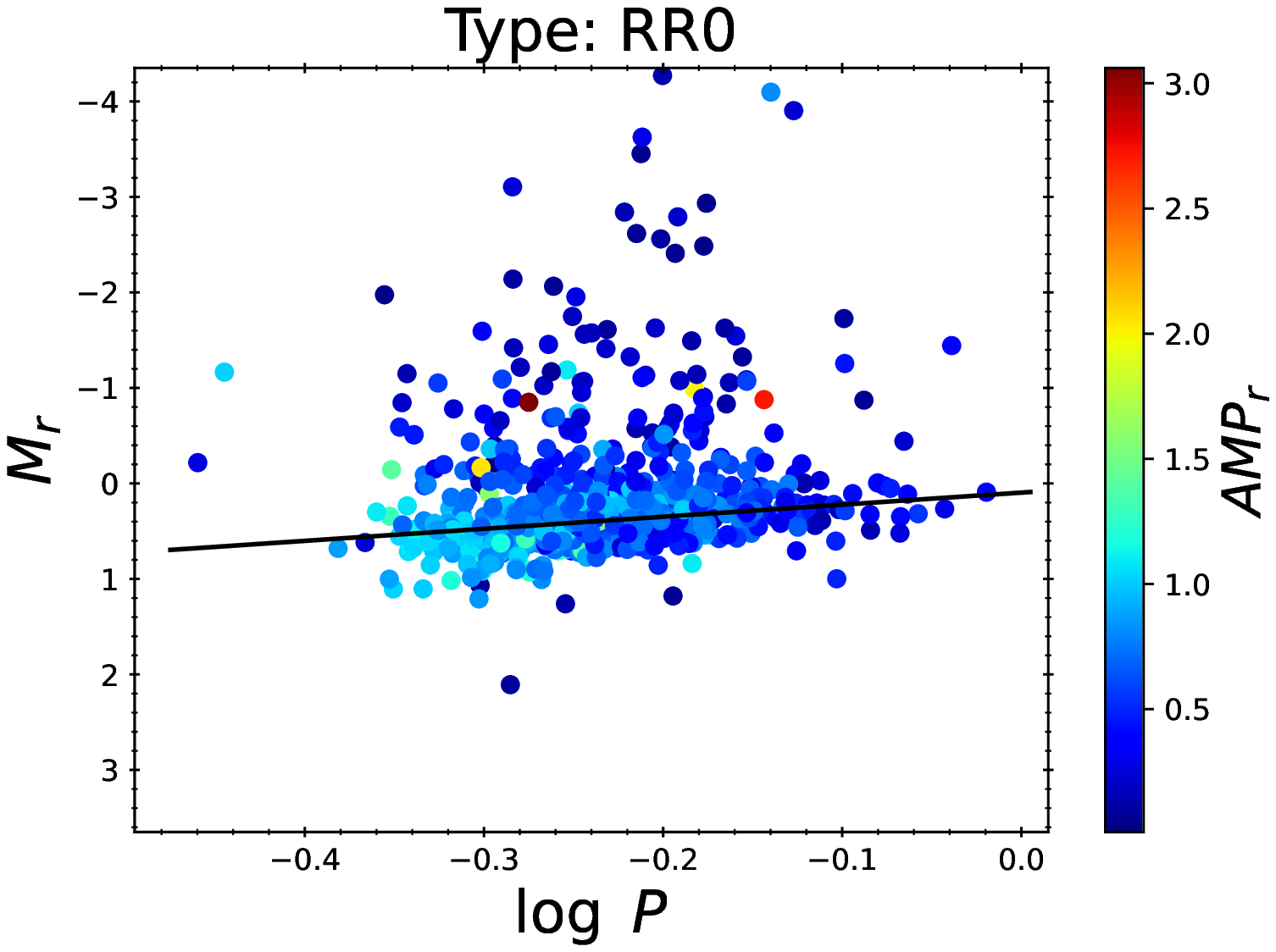} &  \includegraphics[width=0.71\columnwidth]{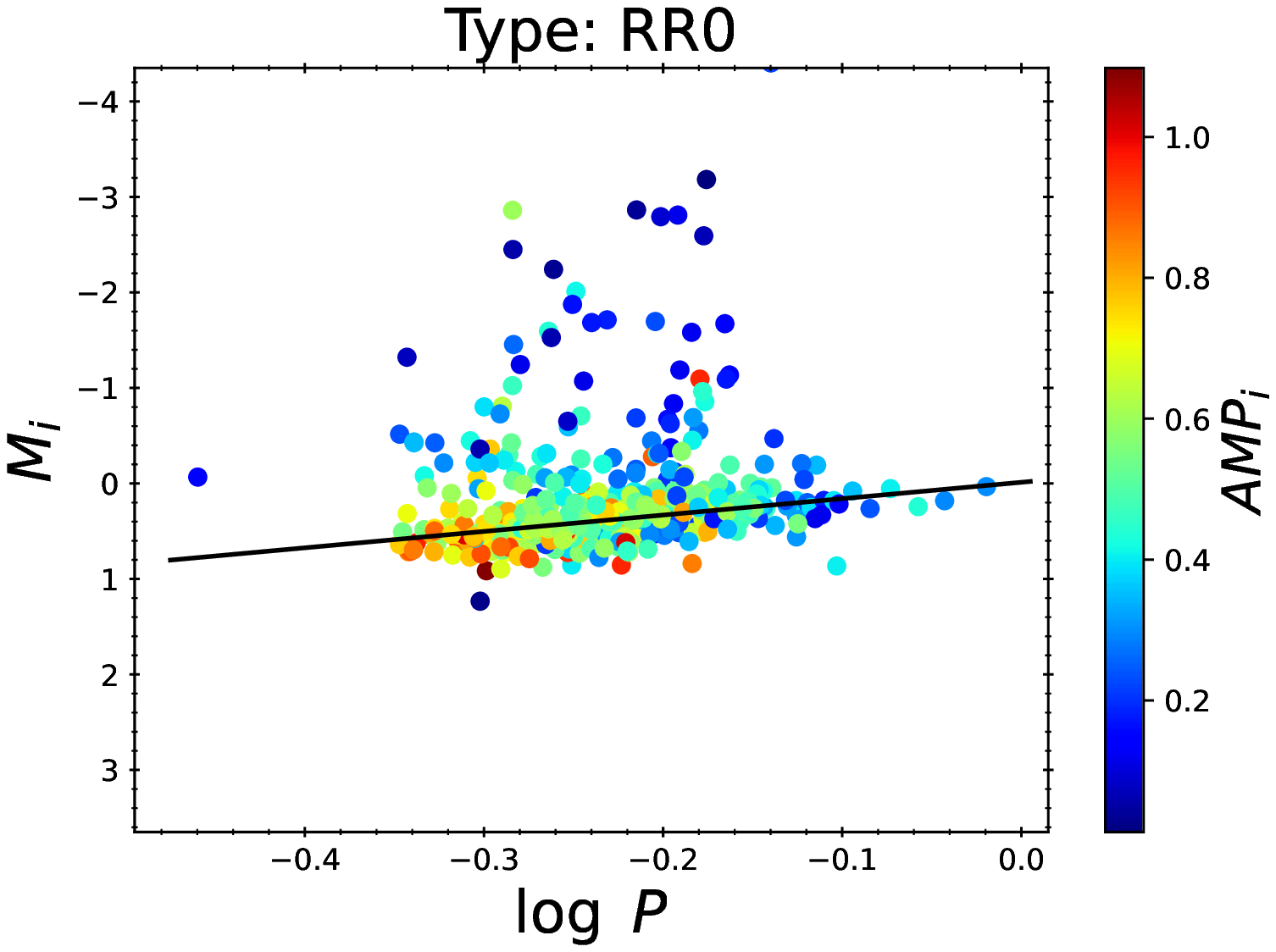} \\
    \includegraphics[width=0.71\columnwidth]{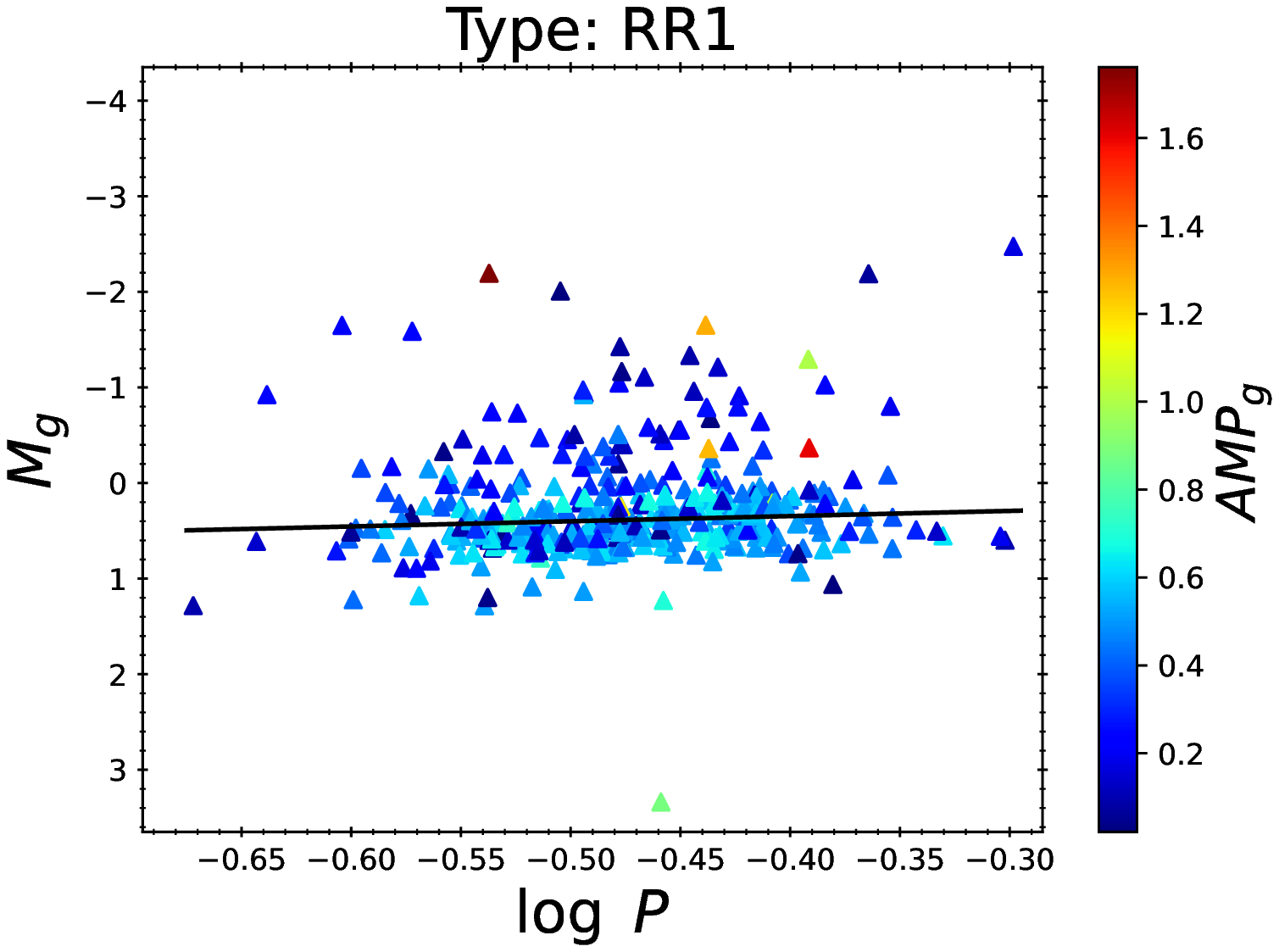} &   \includegraphics[width=0.71\columnwidth]{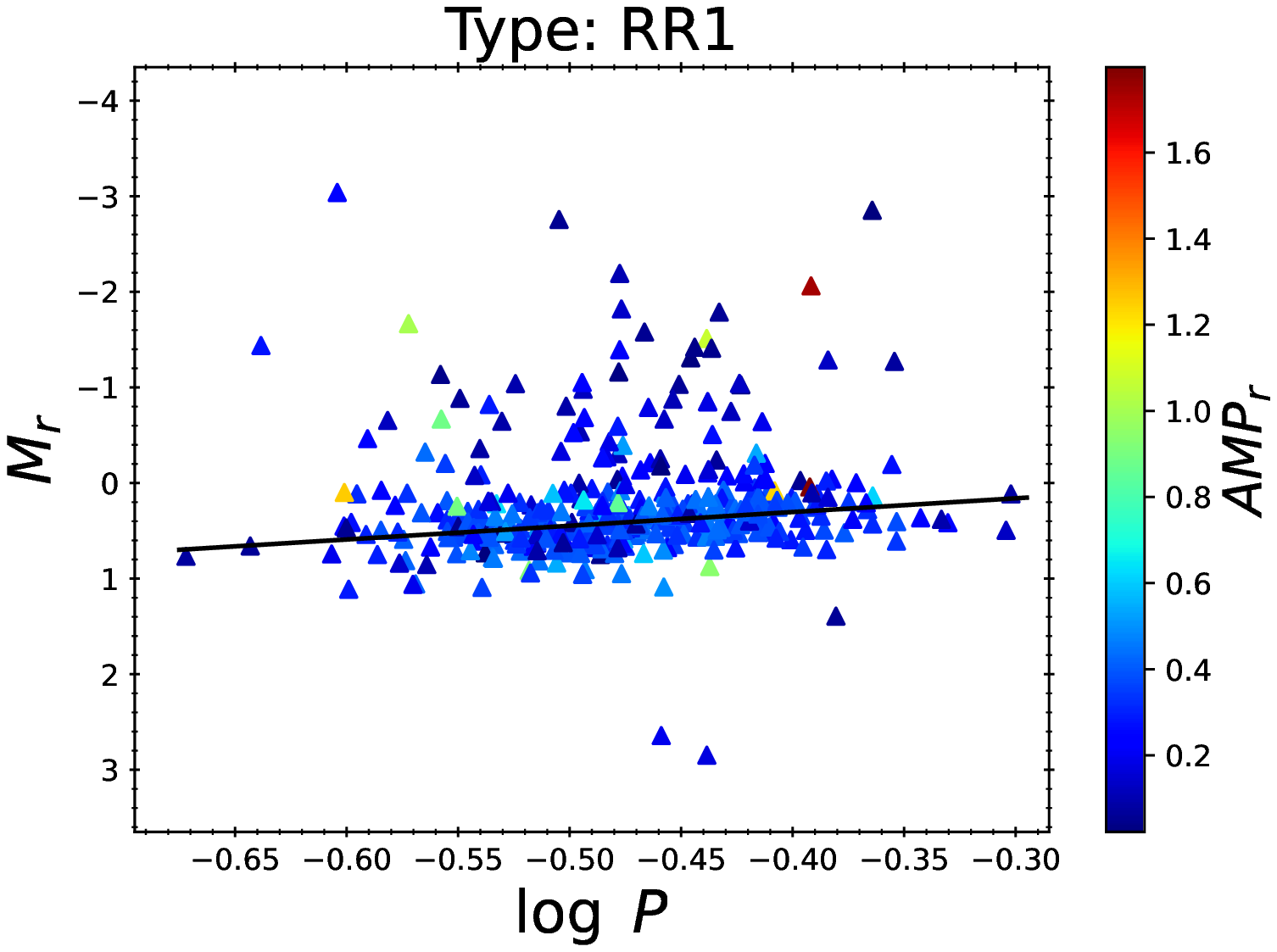} &  \includegraphics[width=0.71\columnwidth]{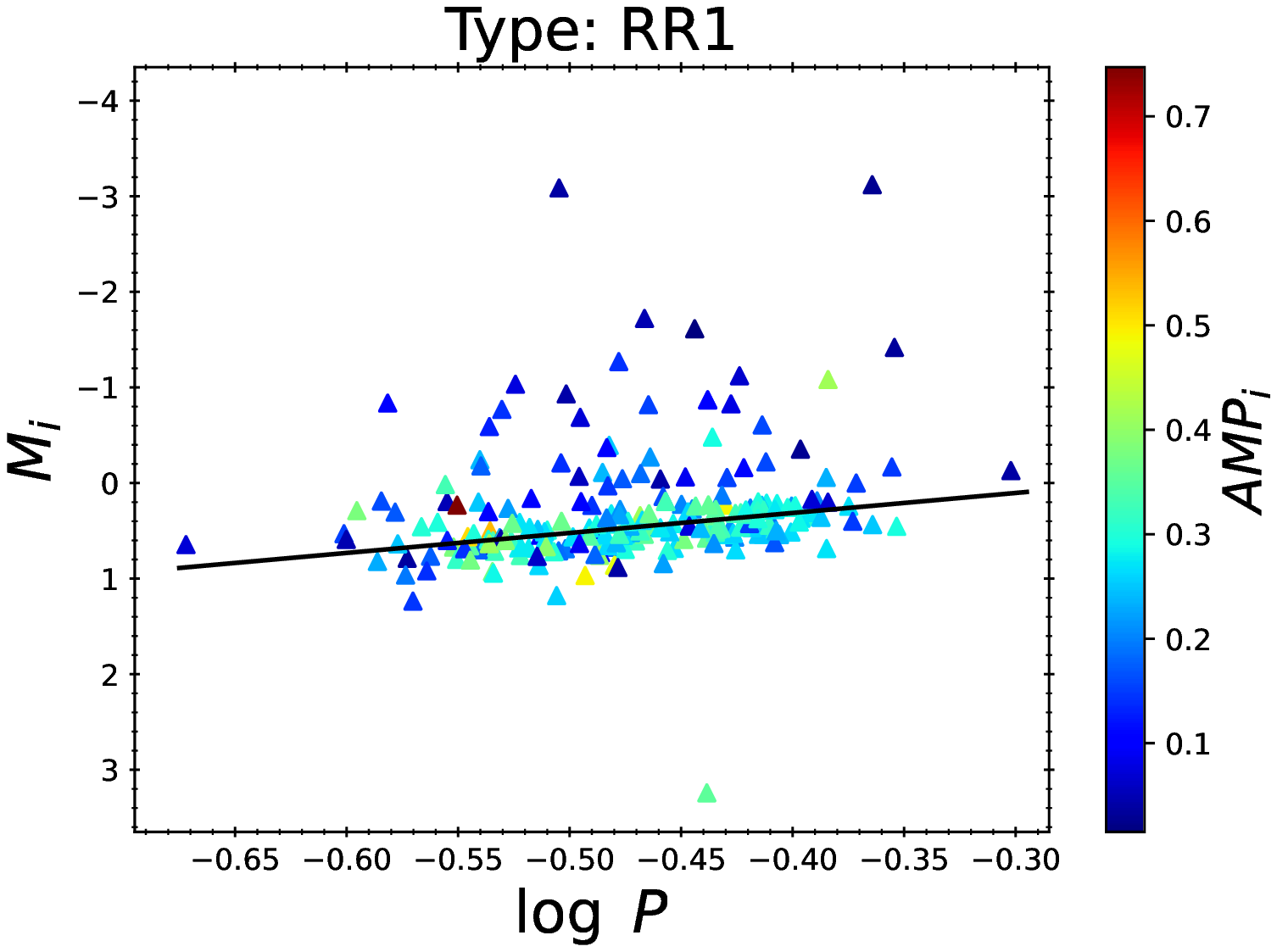} \\
    \end{tabular}
  \caption{The composite $gri$-band period-luminosity (PL) relations based on the 1209 RR Lyrae in different globular clusters, separated for RR0 (upper panels) and RR1 (lower panels). The solid lines are the provisional fitted PL relations (see text for details). For clarity, error bars are omitted. Colors on the data points represent the amplitudes derived from the best-fit template light curves.}\label{fig_plamp}
\end{figure*}

\begin{figure*}
  \epsscale{1.15}
  \plotone{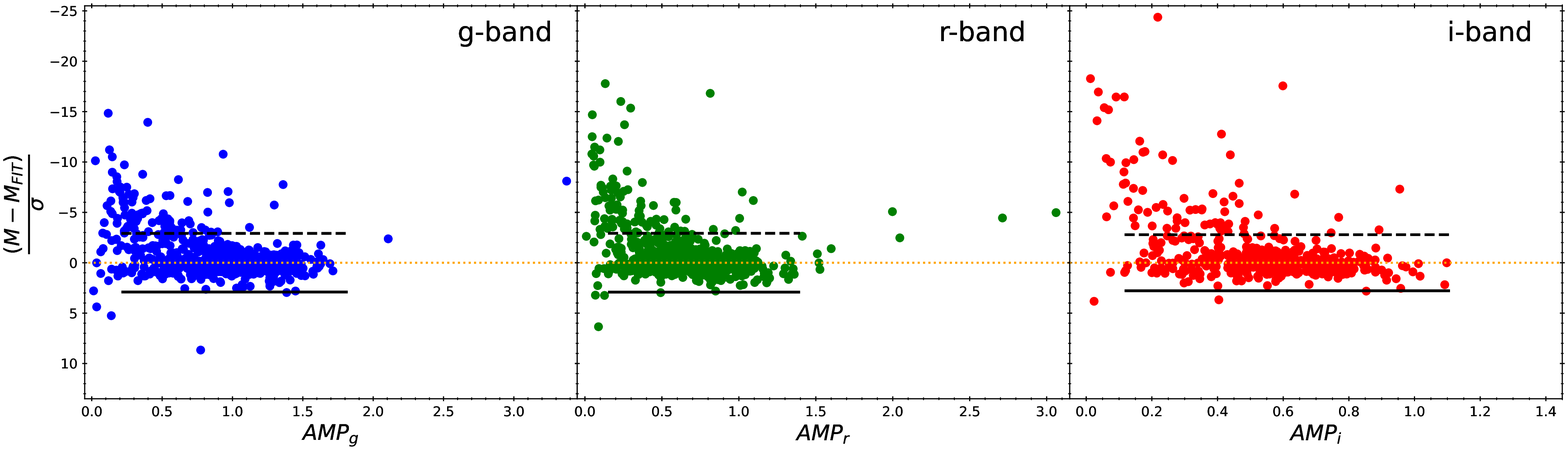}
  \plotone{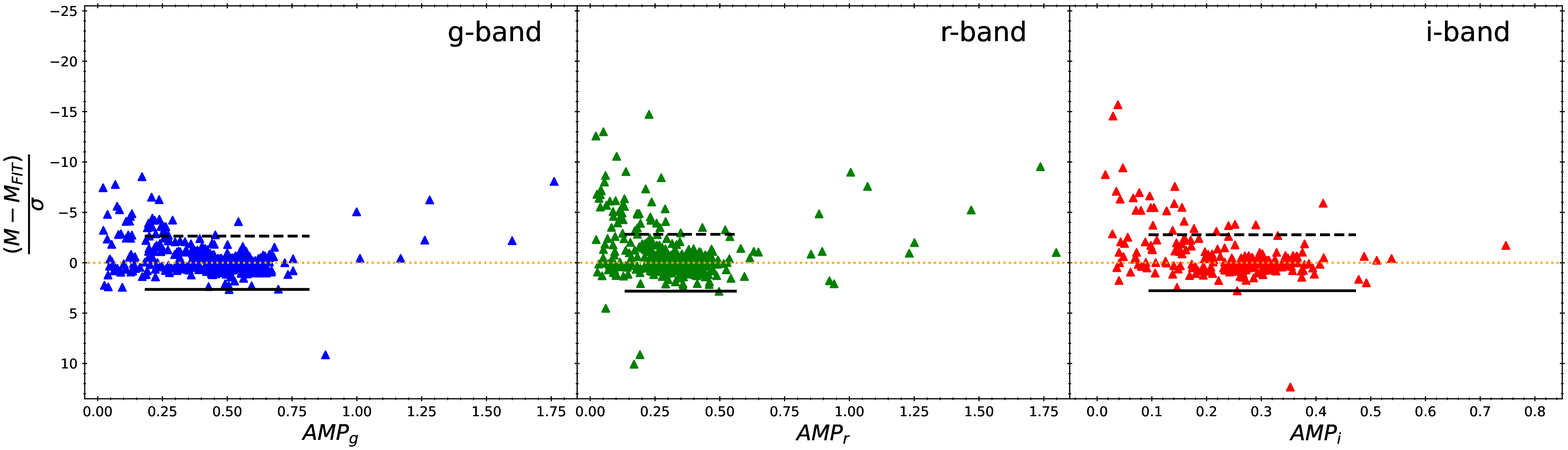}
  \caption{Residuals of the fitted PL relations given in Figure \ref{fig_plamp}, in the unit of the PL dispersion $\sigma$, as a function of the light curve amplitudes. The dotted (orange) horizontal line represents the zero residuals to guide the eyes. The solid horizontal lines are the boundaries on the positive residuals determined using the {\tt boundfit} code \citep[][see text for more details]{cardiel2009}. The dashed horizontal lines are the boundaries on the negative residuals mirrored from the solid horizontal lines. Size of these horizontal lines were based on the adopted amplitude cuts (see Figure \ref{fig_amp}). Upper and lower panels are for RR0 (in filled circles) and RR1 (in filled triangles), respectively.}\label{fig_plres}
\end{figure*}

\begin{figure*}
  \centering
  \begin{tabular}{ccc}
    \includegraphics[width=0.71\columnwidth]{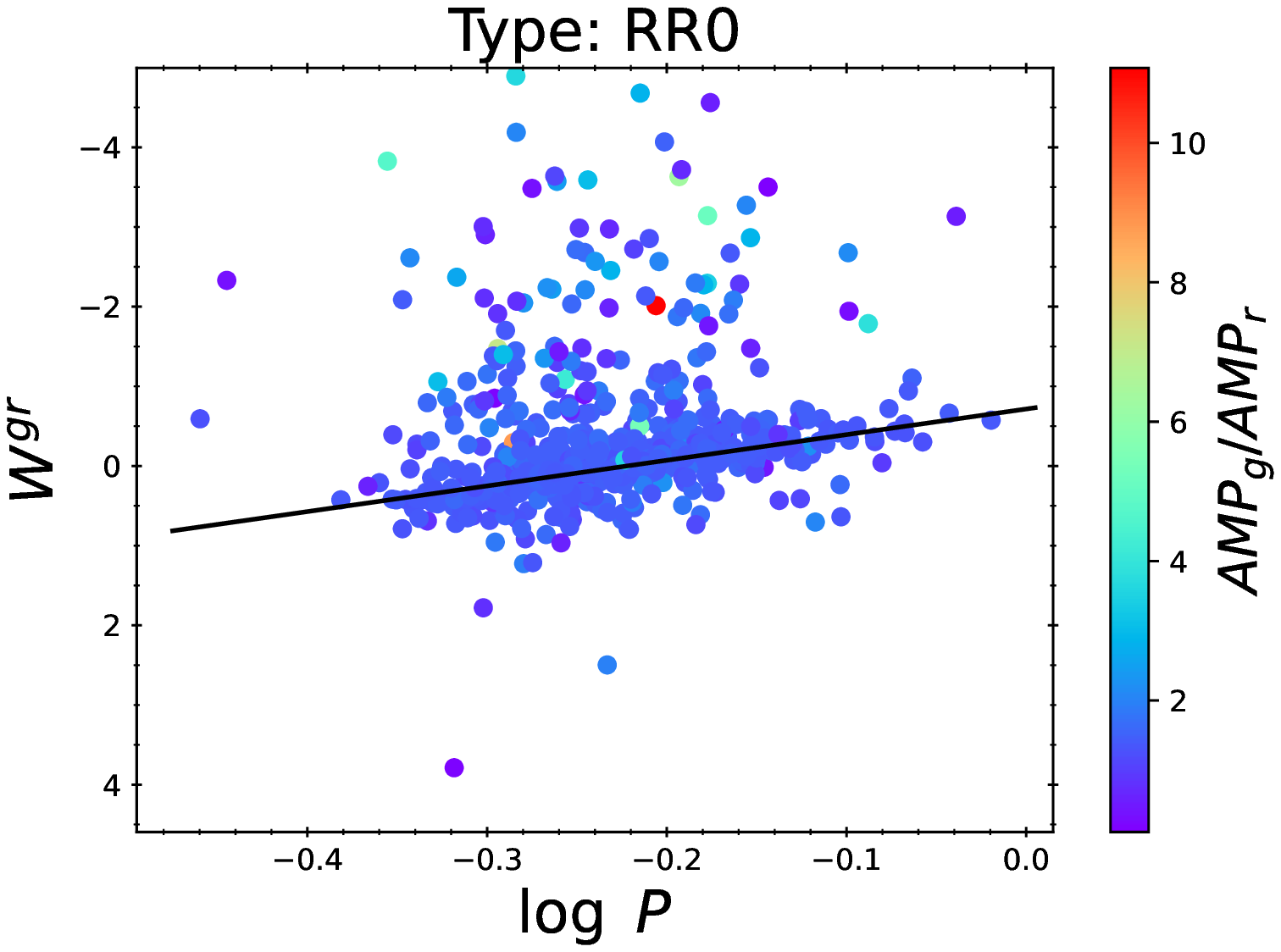} &   \includegraphics[width=0.71\columnwidth]{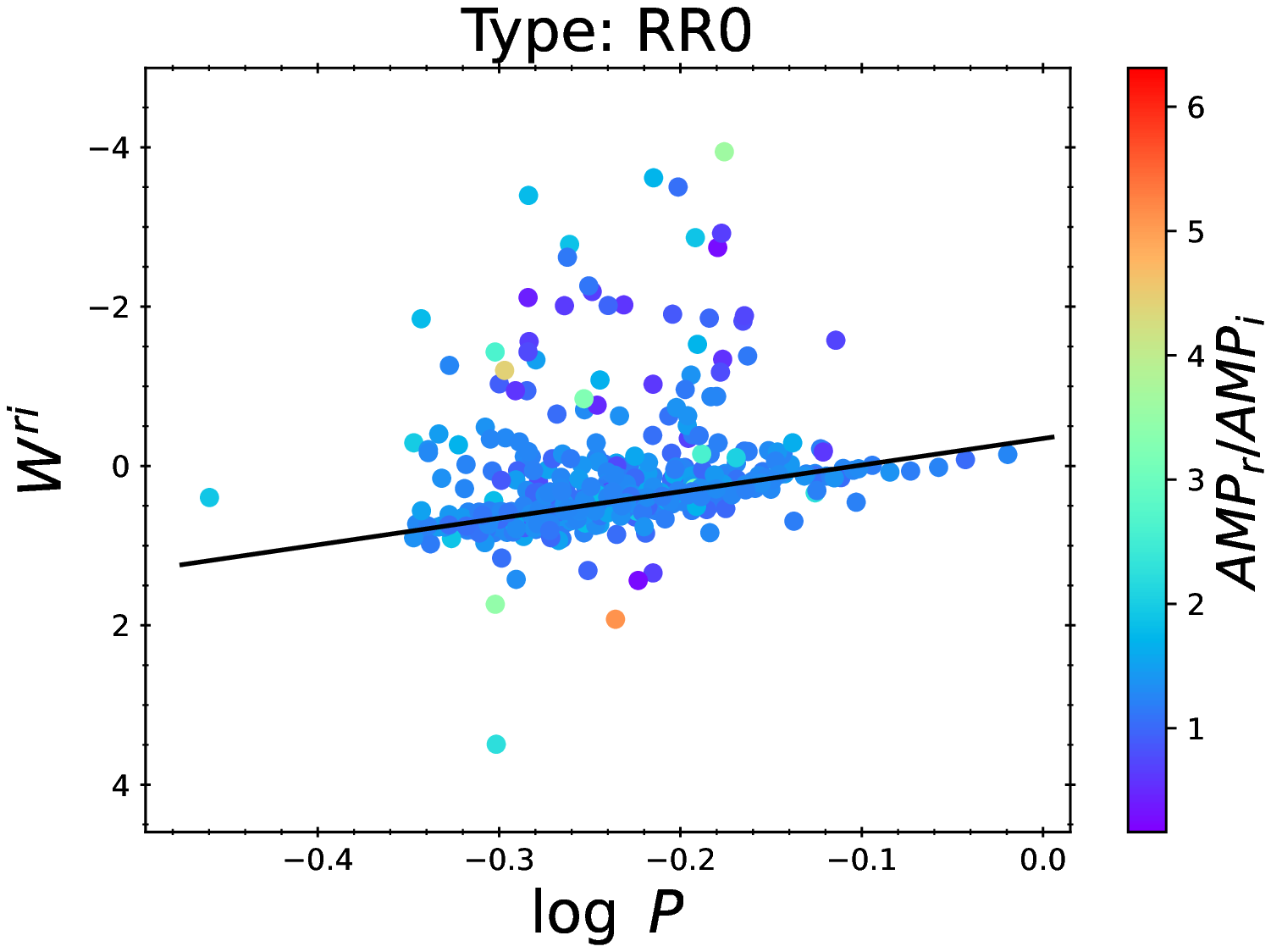} &  \includegraphics[width=0.71\columnwidth]{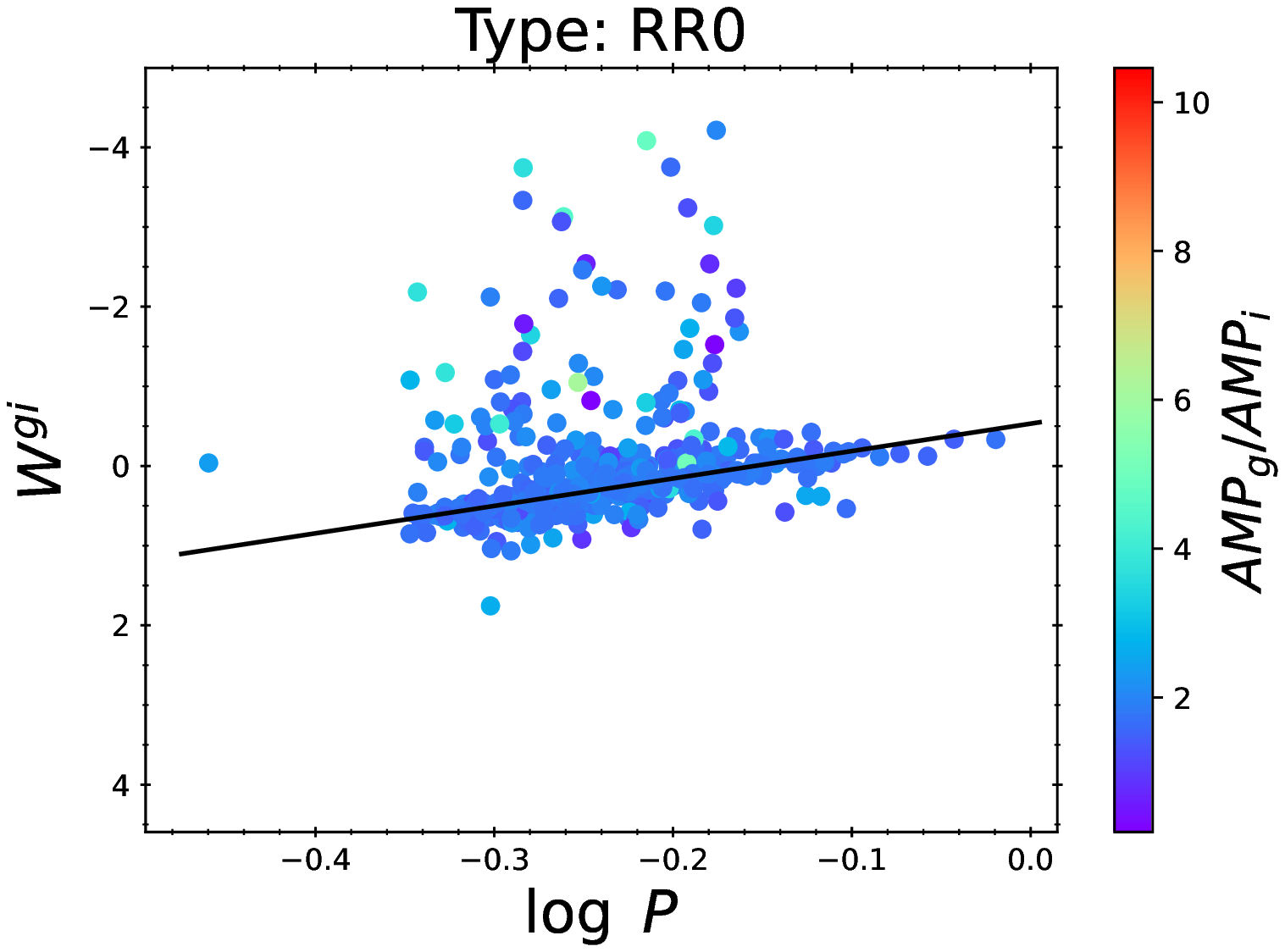} \\
    \includegraphics[width=0.71\columnwidth]{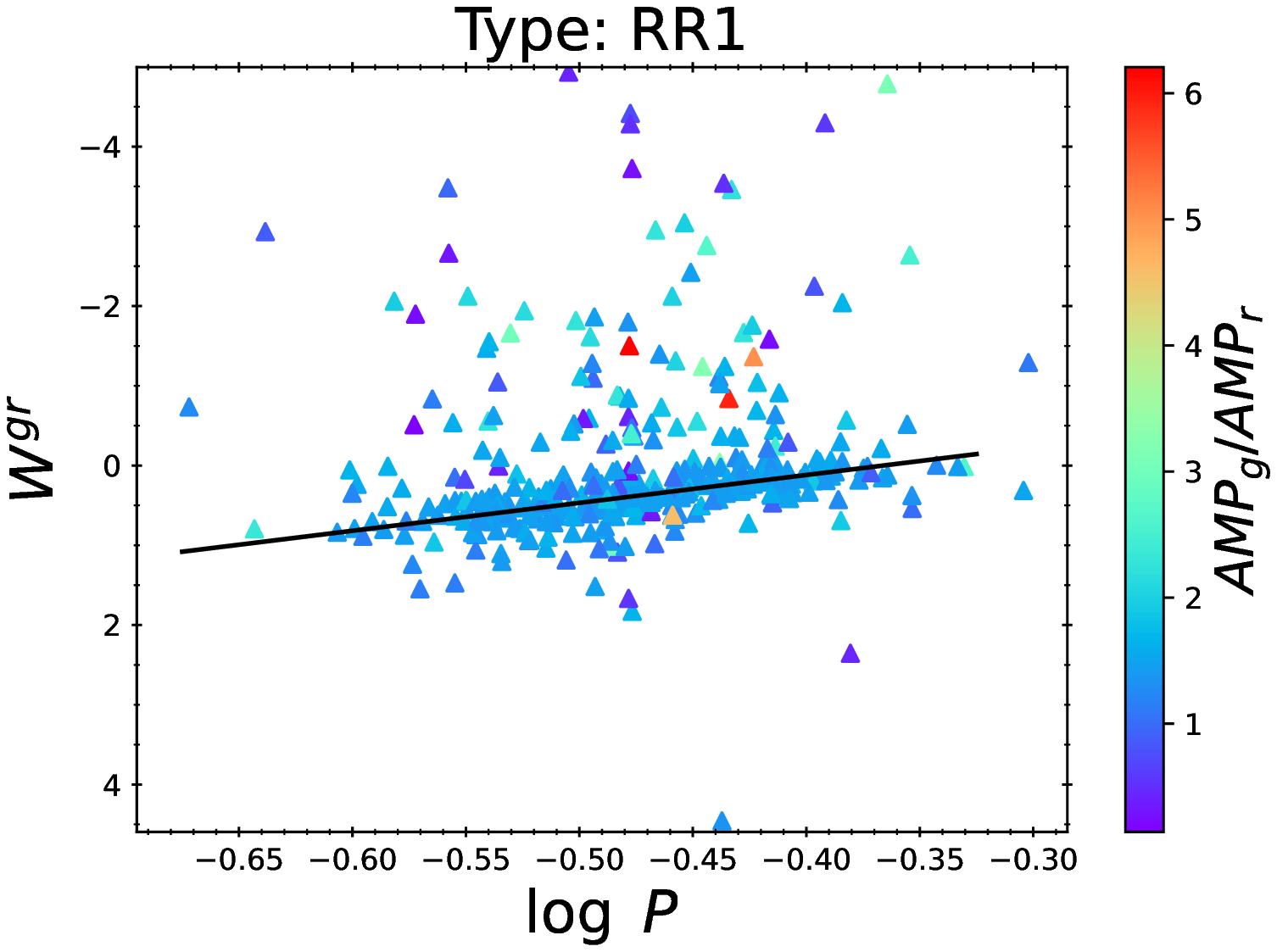} &   \includegraphics[width=0.71\columnwidth]{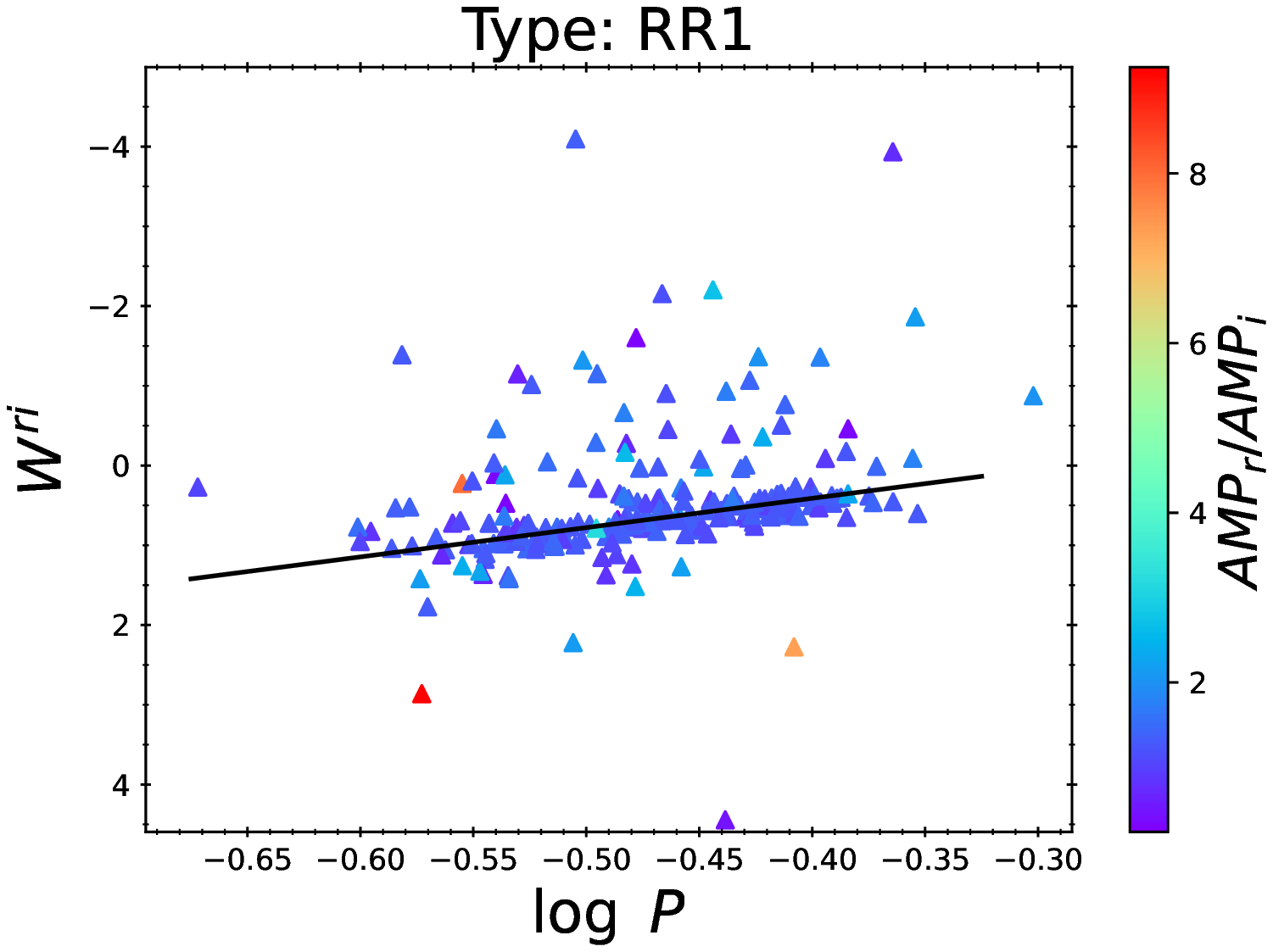} &  \includegraphics[width=0.71\columnwidth]{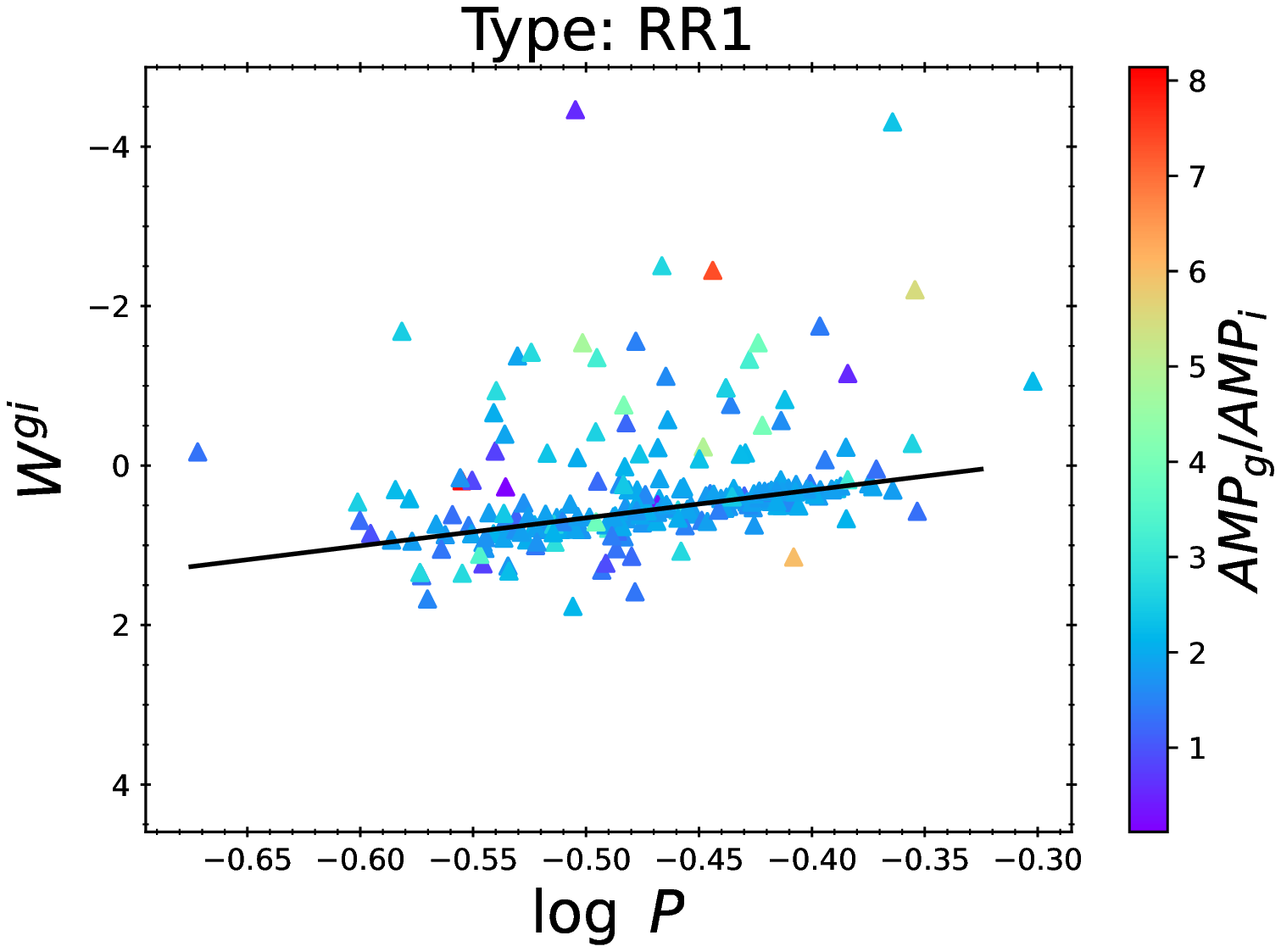} \\
    \end{tabular}
  \caption{The composite period-Wesenheit (PW) relations based on the 1209 RR Lyrae in different globular clusters, separated for RR0 (upper panels) and RR1 (lower panels). The solid lines are the provisional fitted PW relations (see text for details). For clarity, error bars are omitted. Colors on the data points represent the amplitude ratios derived from the best-fit template light curves.}\label{fig_pwaratio}
\end{figure*}

\begin{figure*}
  \epsscale{1.15}
  \plotone{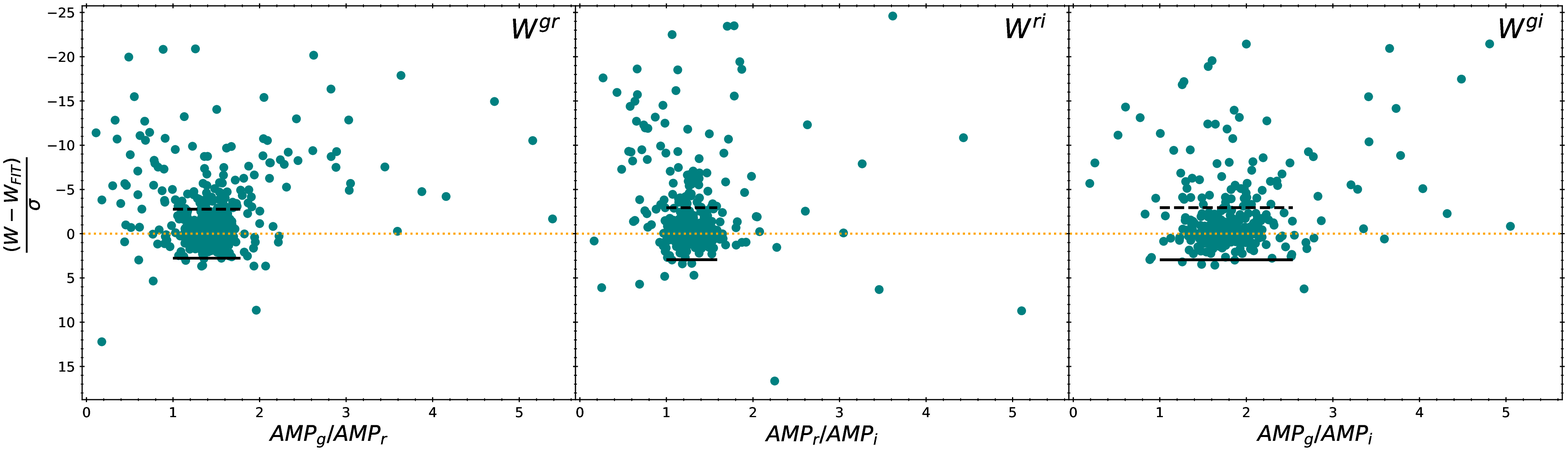}
  \plotone{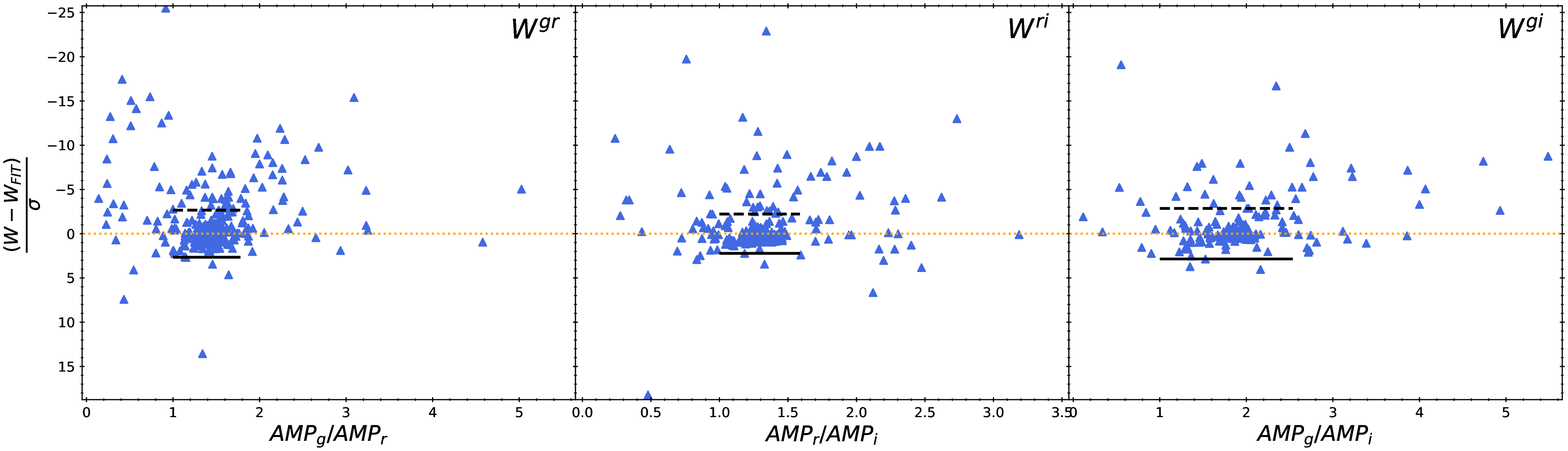}
  \caption{Residuals of the fitted PW relations given in Figure \ref{fig_pwaratio}, in the unit of the PW dispersion $\sigma$, as a function of the amplitude ratios. The dotted (orange) horizontal line represents the zero residuals to guide the eyes. The solid horizontal lines are the boundaries on the positive residuals determined using the {\tt boundfit} code \citep[][see text for more details]{cardiel2009}. The dashed horizontal lines are the boundaries on the negative residuals mirrored from the solid horizontal lines. Size of these horizontal lines were based on the adopted amplitude ratio cuts (see Figure \ref{fig_ampratio}). Upper and lower panels are for RR0 (in filled circles) and RR1 (in filled triangles), respectively.}\label{fig_pwres}
\end{figure*}

As pulsating stars, RR Lyrae are expected to obey a period-color (PC) relation. Therefore, outliers presented on the plot of the PC relation indicate their corresponding light curves could be problematic. Top panels of Figure \ref{fig_iniPC} present the extinction-corrected PC relations for the RR Lyrae in our sample, at which the mean magnitudes in each filters were corrected using $R_{\{g,r,i\}}E$. The values of extinction coefficient are $R_{\{g,r,i\}} = \{3.518,\ 2.617,\ 1.971\}$ \citep{green2019} because ZTF photometry is calibrated to the Pan-STARRS1 system \citep{mas19}, and the reddening $E$ toward each of the RR Lyrae was obtained using the {\tt Bayerstar2019} 3D reddening map \citep[][see Section \ref{sec2.2}]{green2019}.

In Figure \ref{fig_iniPC}, majority of the RR Lyrae show a tight and continuous PC relation for the RR0 and RR1 combined sample, but outliers were also presented in these PC relations. Most of these outliers were either having the amplitude or the amplitude ratios beyond the respected cuts as defined in previous subsection. To remove these outliers, an iterative linear regression with $3\sigma$ outliers rejection algorithm was applied to fit both RR0 and RR1. The $3\sigma$ boundaries from the provisional fitted PC relations, displayed as dashed lines in Figure \ref{fig_iniPC}, were adopted as the period-dependent color cuts. Note that fitting of the PC relations, separately for RR0 and RR1, on our final sample will be done in Section \ref{sec5}.

To circumvent the possibility that the outliers shown in Figure \ref{fig_iniPC} were (partly) due to the extinction, we plotted the extinction-free $Q$-indices as a function of pulsation periods in the bottom panel of Figure \ref{fig_iniPC} for RR Lyrae that have mean magnitudes in all $gri$ bands. Given the adopted values of $R_{\{g,r,i\}}$, the extinction-free $Q$-index is defined as:

\begin{eqnarray}
  Q & = & (g-r) - 1.395(r-i). \nonumber
\end{eqnarray}

\noindent Similar to the PC relations, we derived a provisional PQ relation, and the $3\sigma$ boundaries were adopted as the period-dependent $Q$-index cuts, shown as the dashed lines in the bottom panel of Figure \ref{fig_iniPC}.

In total, there are 147 RR Lyrae located outside the $3\sigma$ boundaries of either the three PC relations or the PQ relation. These RR Lyrae were flagged as ``C'' in our sample.

\subsection{Filtering Based on PL/PW Residuals}\label{sec3.3}

Figure \ref{fig_plamp} presents the extinction-corrected $gri$-band PL relations for our sample of RR Lyrae. The mean magnitudes $\langle m \rangle$ for these RR Lyrae were converted to absolute magnitudes using $M_{\{g,r,i\}} = \langle m_{\{g,r,i\}}\rangle - R_{\{g,r,i\}}E - 5\log D + 5$, where the extinction term ($R_{\{g,r,i\}}E$) was described in Section \ref{sec3.2}, and the distance $D$ was adopted from \citet[][see Section \ref{sec2.2}]{baumgardt2021}. Errors from $\langle m \rangle$, $E$, and $D$ were propagated to the total errors on $M$.

As mentioned, blending would cause a RR Lyrae to appear brighter, at the same time the amplitude of such RR Lyrae would be reduced. As can be seen from Figure \ref{fig_plamp}, there are a number of RR Lyrae that seem to be brighter than the rest of the RR Lyrae at a given period. These RR Lyrae also tend to have smaller amplitudes obtained from the best-fit template light curves, suggesting these RR Lyrae were most likely affected by blending and should be excluded. A provisional period-luminosity (PL) relation in the form of $M_{FIT} = a\log P + b$, separately for RR0 and RR1, was fitted to our sample of RR Lyrae, where the fitting was performed using an iterative $3\sigma$-rejection procedure to remove the outliers until the fitted parameters $a$ and $b$ converged. The fitted provisional PL relations were marked as solid lines in Figure \ref{fig_plamp}.

To better visualize the effects of blending, residuals of the fitted PL relations, in the unit of PL dispersion $\sigma$, i.e. $(M-M_{FIT})/\sigma$, were plotted against the amplitudes determined from the best-fit template light curves in Figure \ref{fig_plres}, showing that more negative residuals (i.e. brighter than expected) tend to have a smaller amplitude. After excluding RR Lyrae with $(M-M_{FIT})/\sigma > 3$ (based on the adopted iterative $3\sigma$-rejection linear regression fitting) and those with amplitudes outside the amplitude cuts defined in Figure \ref{fig_amp}, we determined the one-site boundaries on the positive residuals by using the {\tt boundfit} code\footnote{\url{https://github.com/nicocardiel/boundfit}} \citep{cardiel2009}. These boundaries are indicated by solid horizontal lines in Figure \ref{fig_plres}. Since the residuals of the PL relation should be symmetric, these boundaries were then mirrored to the negative residuals and presented as dashed horizontal lines in Figure \ref{fig_plres}. RR Lyrae with negative residuals that were more negative than the dashed horizontal lines are brighter than expected, implying these RR Lyrae could be affected by blending (which also tend to have smaller amplitudes).

In addition to the PL relations, we have also examined the PW relation for our sample of RR Lyrae. The extinction-free Wesenheit indexes were defined as \citep{ngeow2021}:

\begin{eqnarray}
  W^{gr} & = & r - 2.905 (g-r), \nonumber \\
  W^{ri} & = & r - 4.051 (r-i), \nonumber \\
  W^{gi} & = & g - 2.274 (g-i). \nonumber
\end{eqnarray}

The corresponding PW relations are presented in Figure \ref{fig_pwaratio}, together with the provisional PW relations (fitted with the same iterative $3\sigma$-rejection algorithm as in the case of provisional PL relations) shown in the solid lines. Data points in Figure \ref{fig_pwaratio} were also color-coded with amplitude ratios. There are two features revealed from Figure \ref{fig_pwaratio}. First of all, majority of the outliers are brighter than those RR Lyrae located along the ridge lines at a given period, again indicating the presence of blending. Secondly, RR Lyrae along the ridge lines have similar, or consistent amplitude ratios, while the outliers either have a large or a small (i.e., smaller than unity) amplitude ratio. 

Residuals of the provisional PW relations, in unit of PW dispersion $\sigma$, as a function of amplitude ratios were presented in Figure \ref{fig_pwres}. It is clear that RR Lyrae with small PW residuals (e.g., $< |3\sigma |$) also tend to concentrate on a small range of amplitude ratio, iterating the similar concentration in Figure \ref{fig_ampratio}. Similar to the PL residuals, we used {\tt boundfit} to determine the one-site boundaries on the positive residuals after excluding the outlying residuals with $(W-W_{FIT})/\sigma > 3$ and outside the amplitude ratio cuts given in Figure \ref{fig_ampratio}. The determined one-site boundaries were then mirrored to the negative residuals. These boundaries were shown as solid and dashed horizontal lines in Figure \ref{fig_pwres}.

Based on the residuals from the PL or PW relations, there were 306 RR Lyrae with residuals beyond the boundaries determined in either Figure \ref{fig_plres} or \ref{fig_pwres} (both solid and dashed lines). These RR Lyrae were flagged as ``R'' in our sample.

\subsection{The Final Sample of RR Lyrae}\label{sec3.4}

\begin{figure}
  \epsscale{1.15}
  \plotone{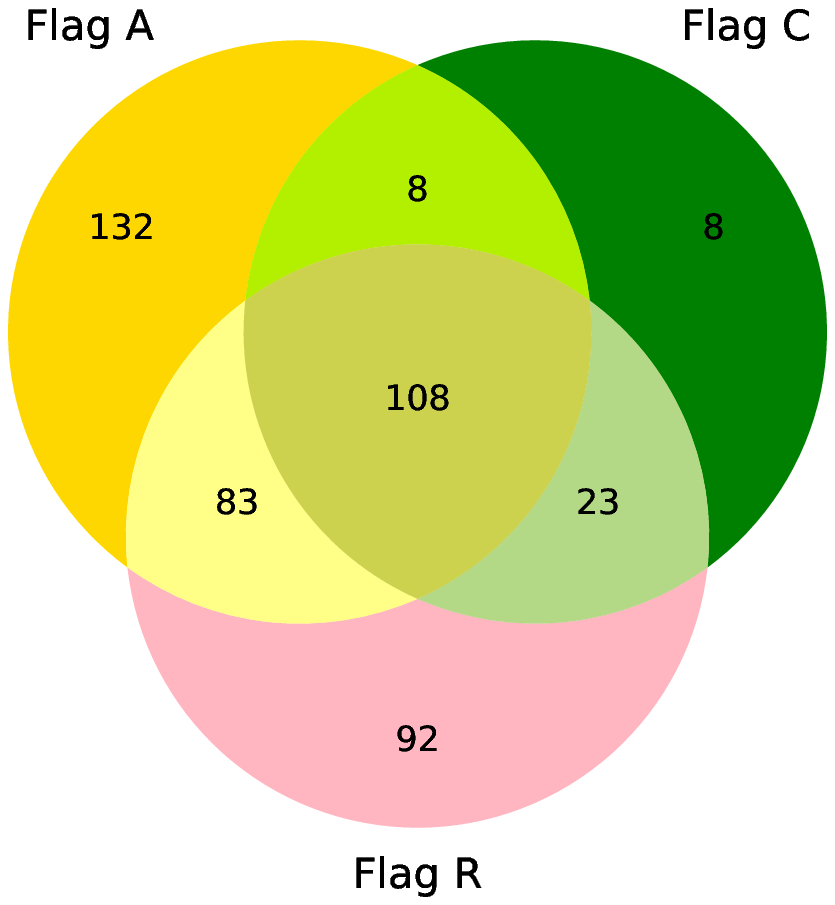}
  \caption{Venn diagram for the {\it excluded} RR Lyrae in our sample (see text for details on flag ``A'', ``C'', and ``R'').}\label{fig_venn}
\end{figure}

\begin{figure*}
  \gridline{\fig{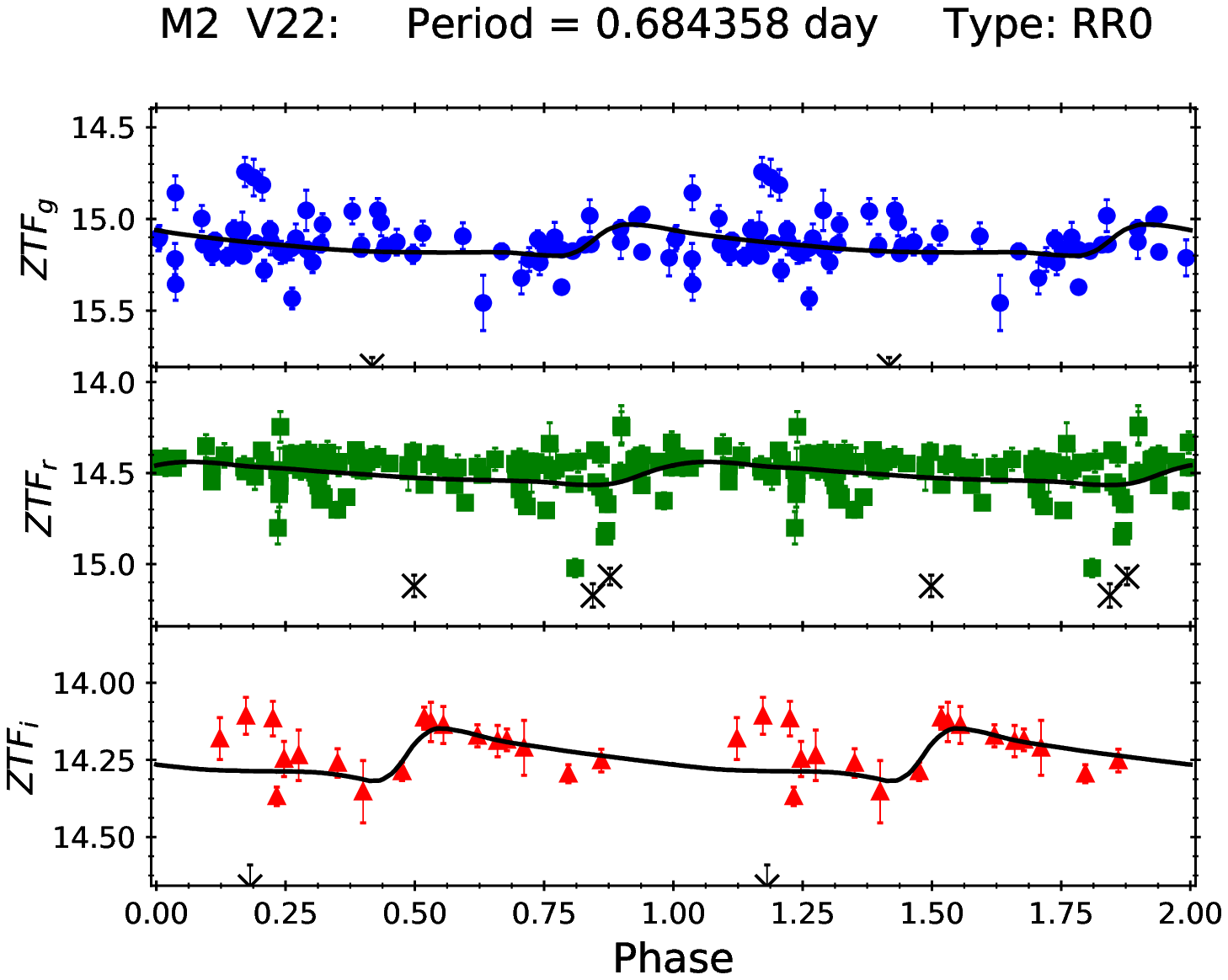}{0.32\textwidth}{With ``ACR'' flags.}
    \fig{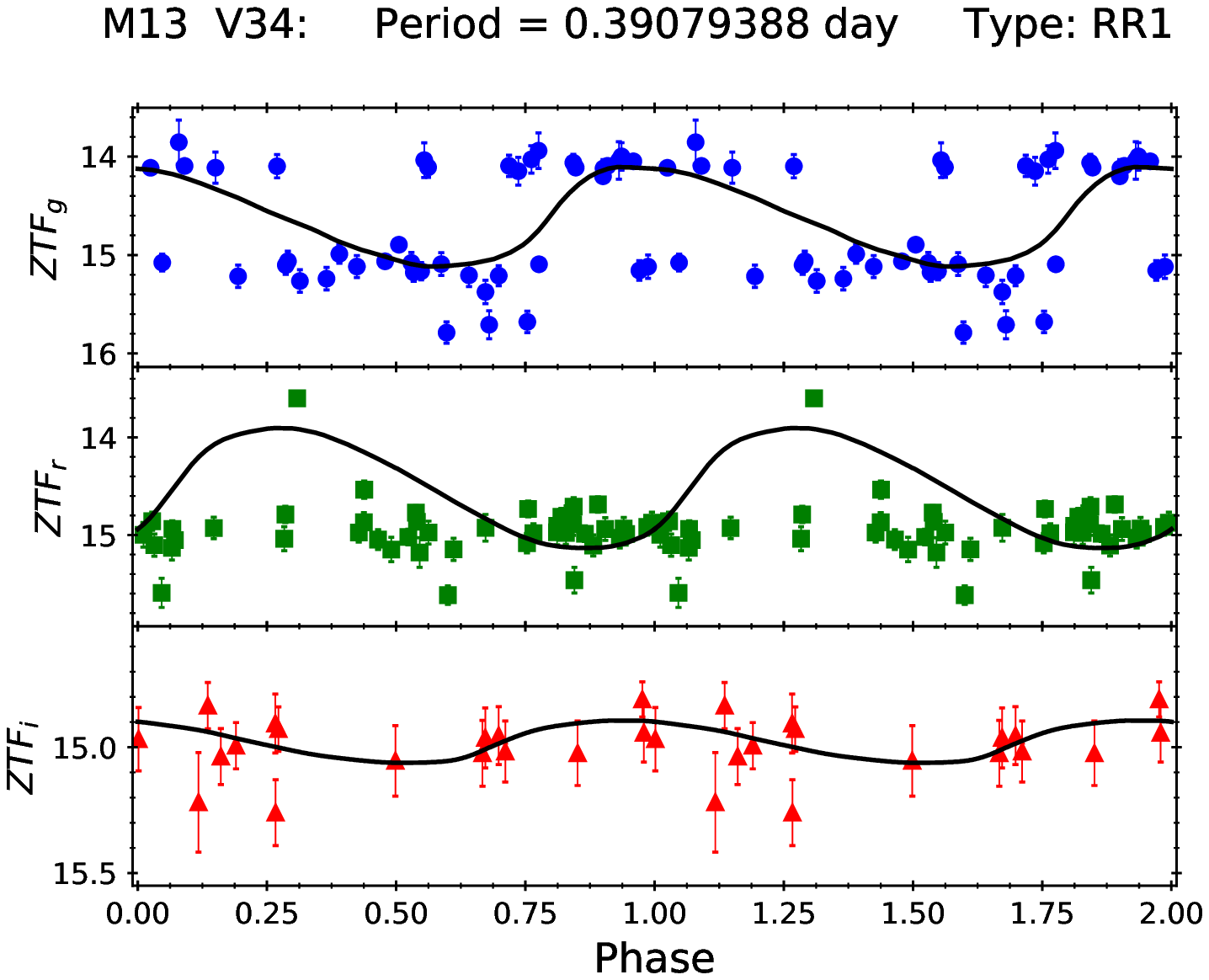}{0.32\textwidth}{With ``ACR'' flags.}
    \fig{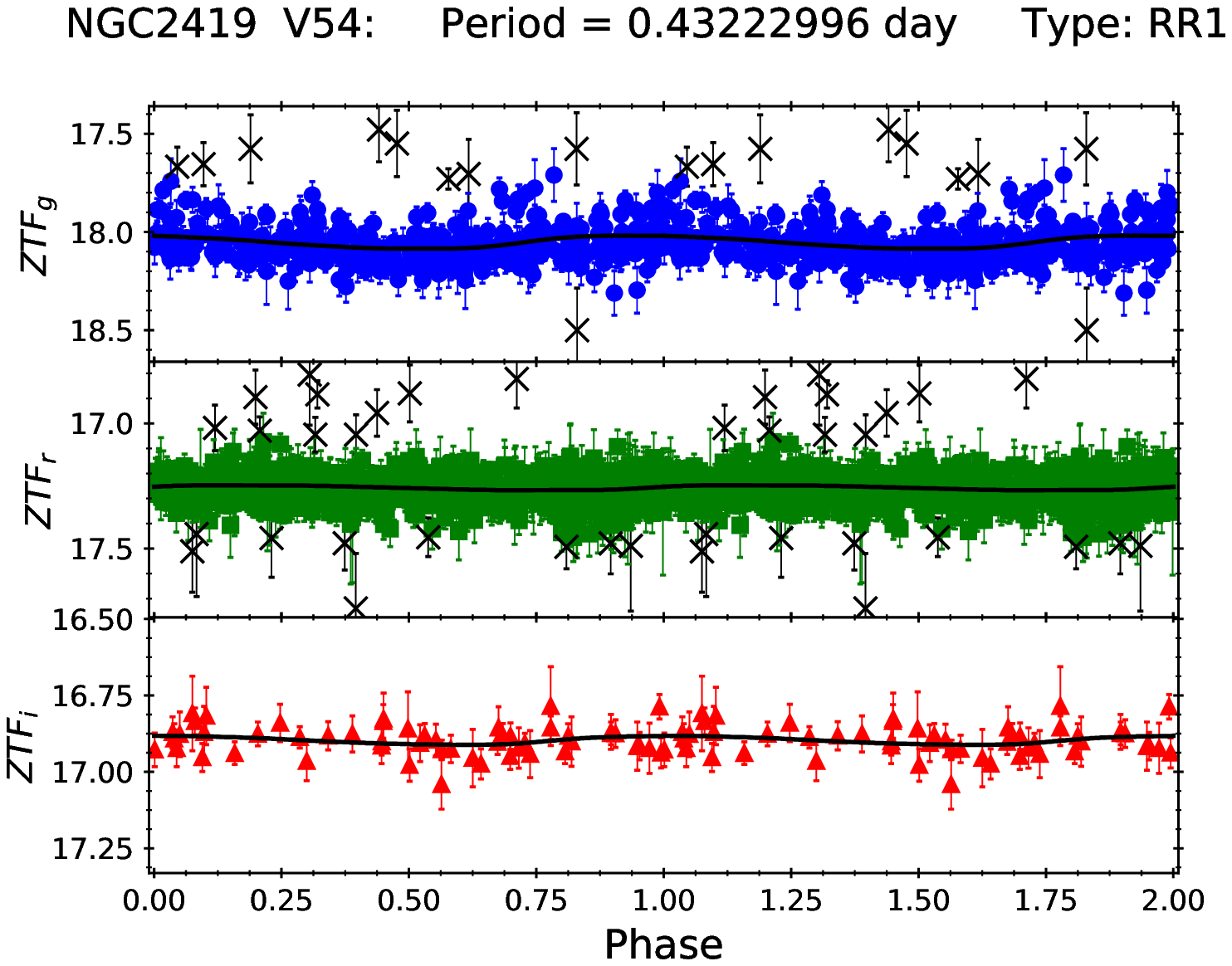}{0.32\textwidth}{With ``ACR'' flags.}}
  \gridline{\fig{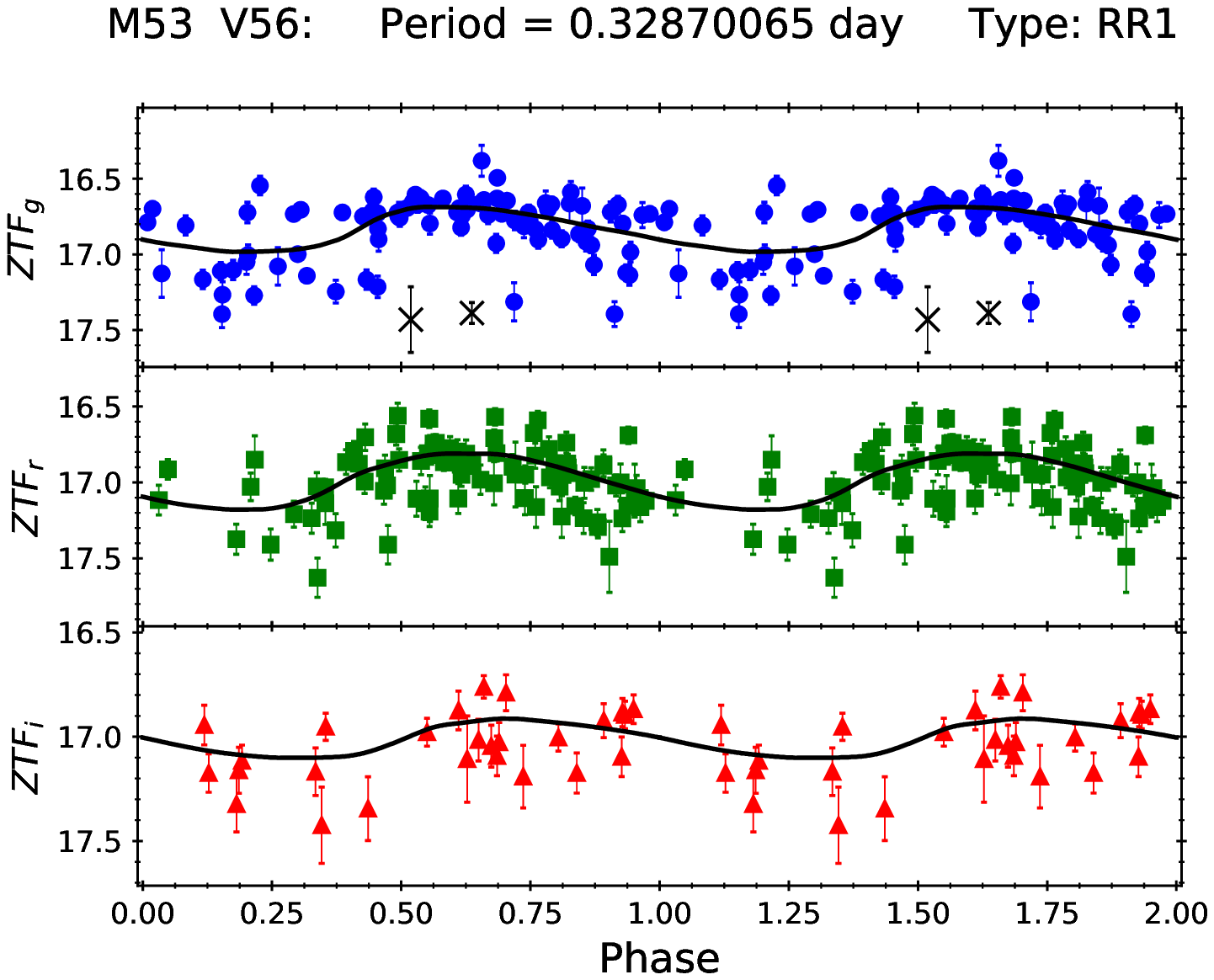}{0.32\textwidth}{With ``AC'' flags.}
    \fig{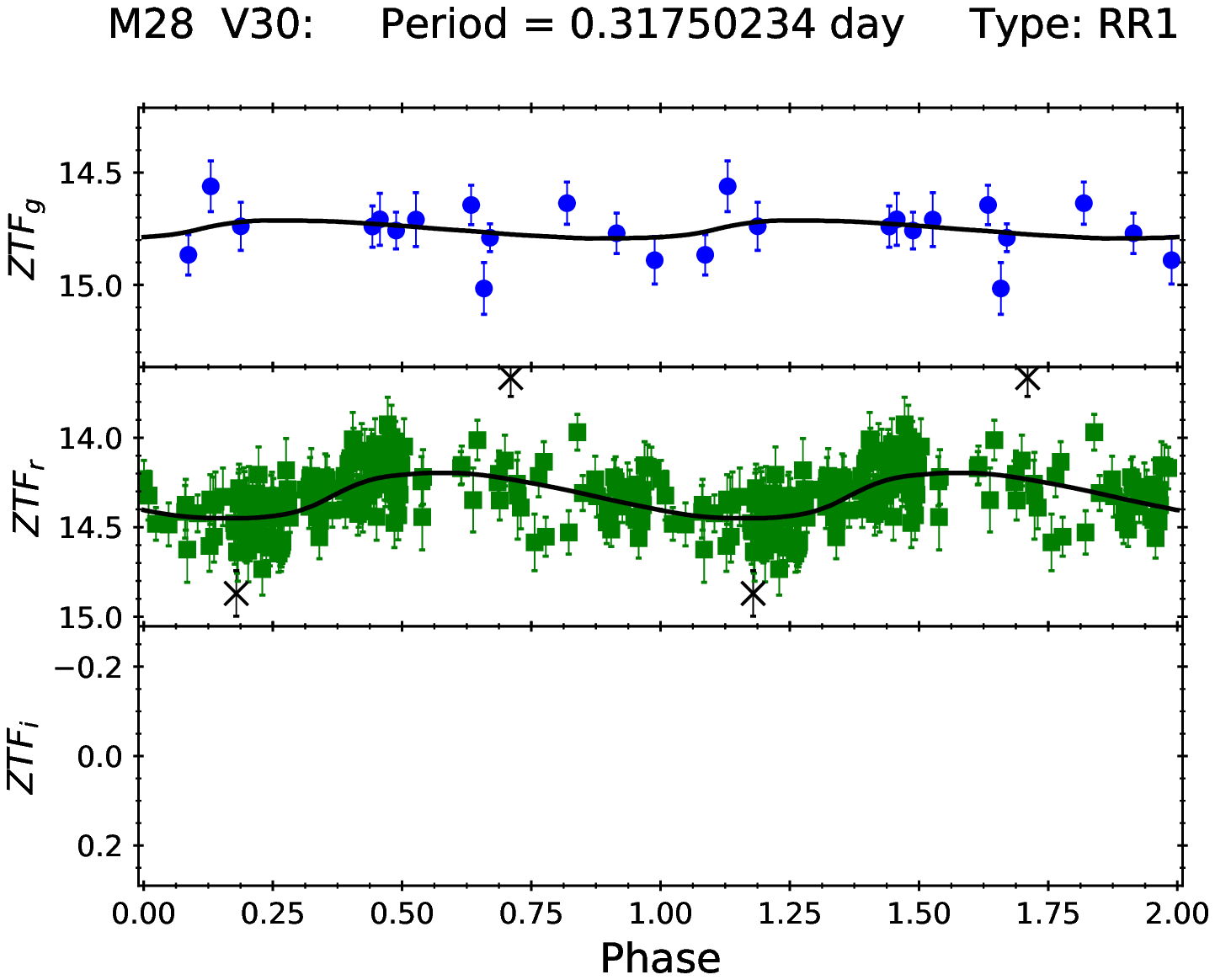}{0.32\textwidth}{With ``AR'' flags.}
    \fig{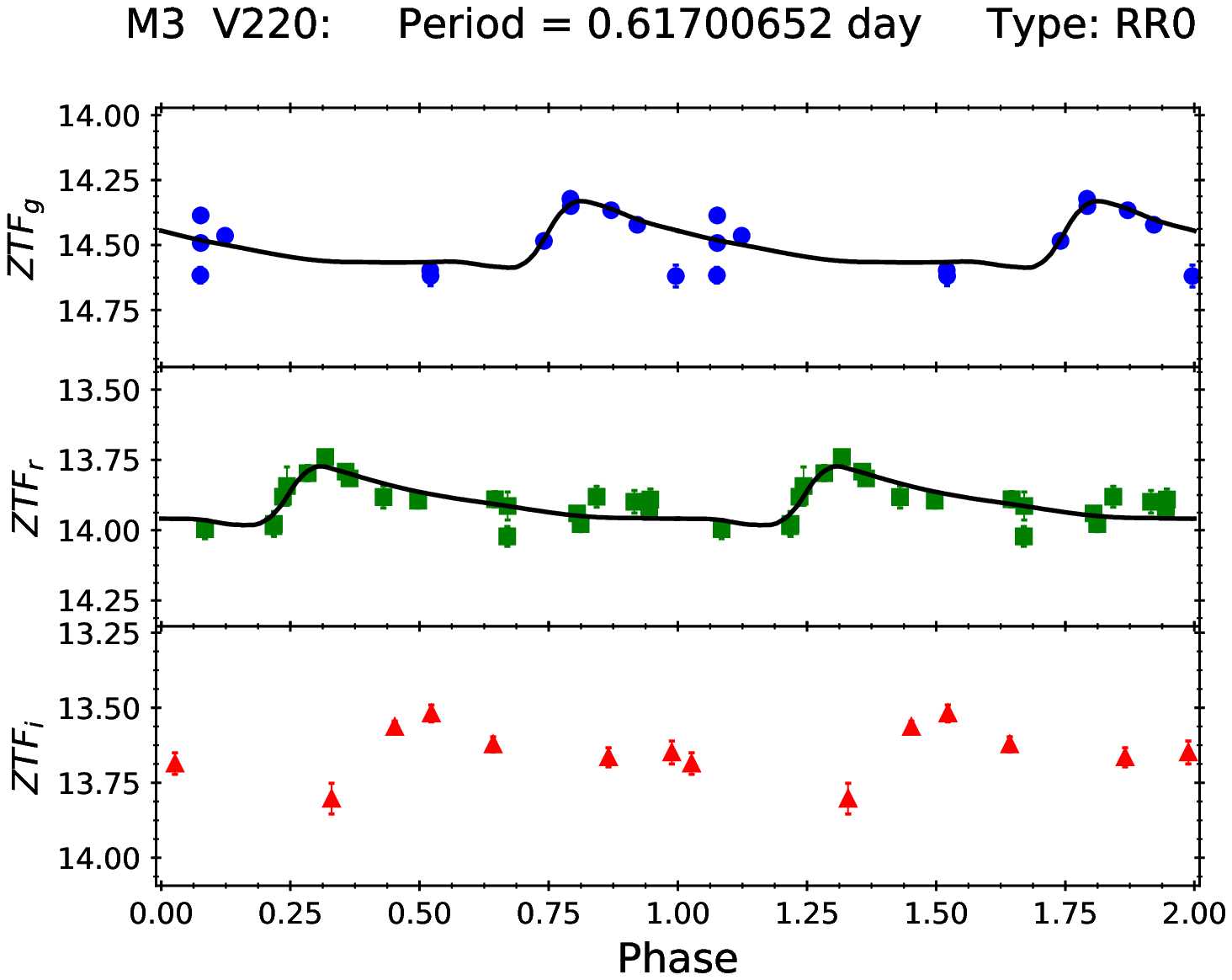}{0.32\textwidth}{With ``CR'' flags.}}
  \gridline{\fig{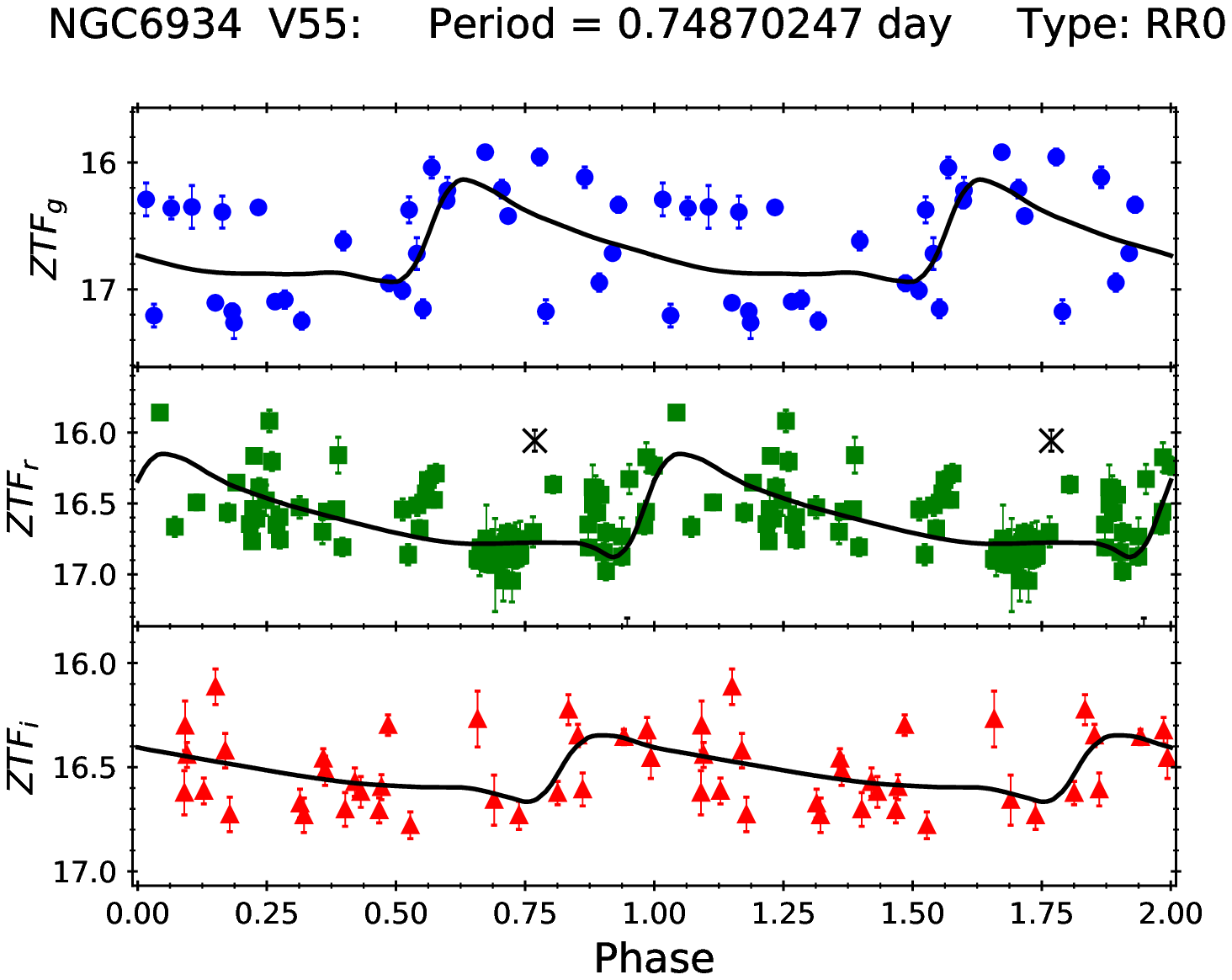}{0.32\textwidth}{With ``A'' flag.}
    \fig{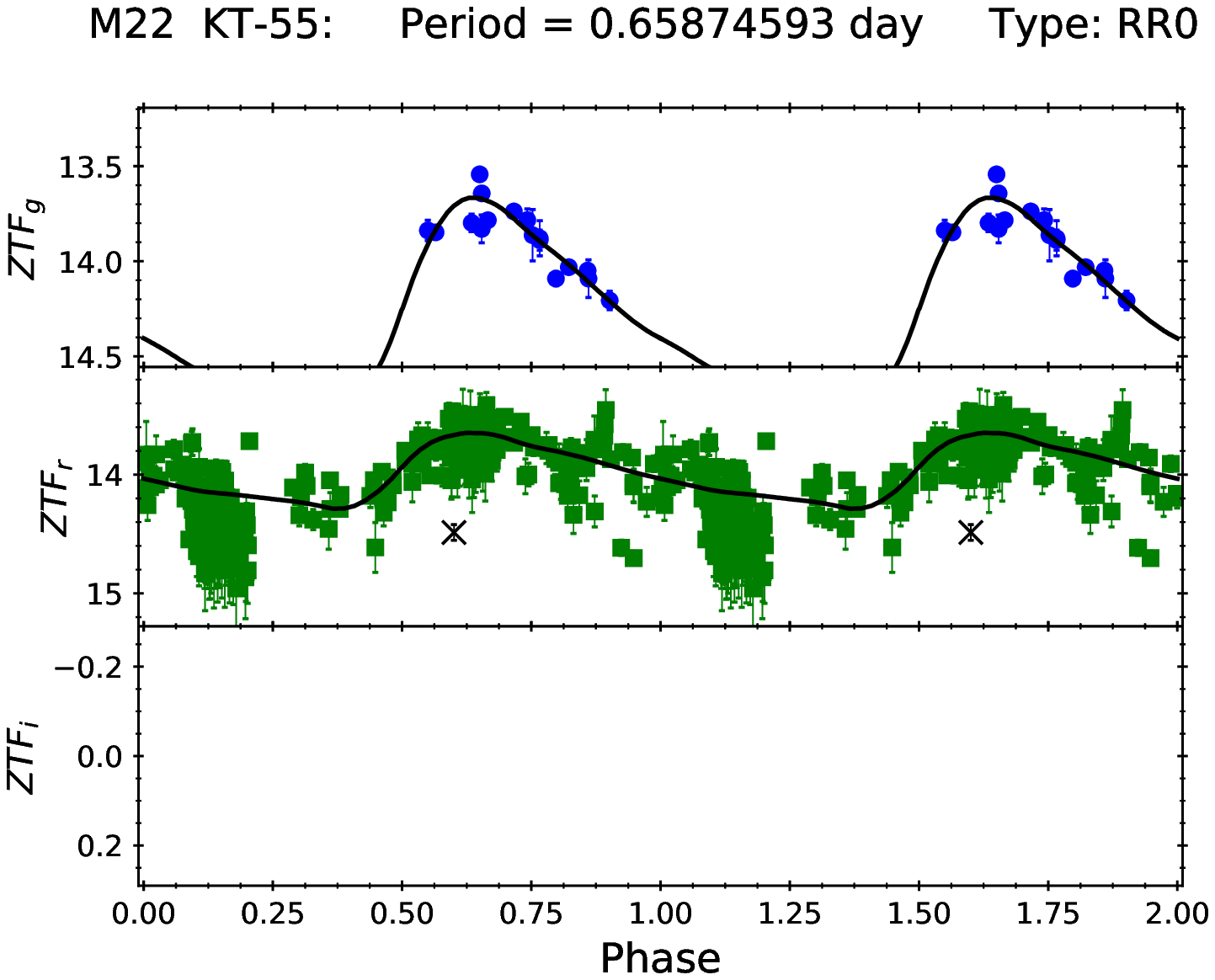}{0.32\textwidth}{With ``C'' flag.}
    \fig{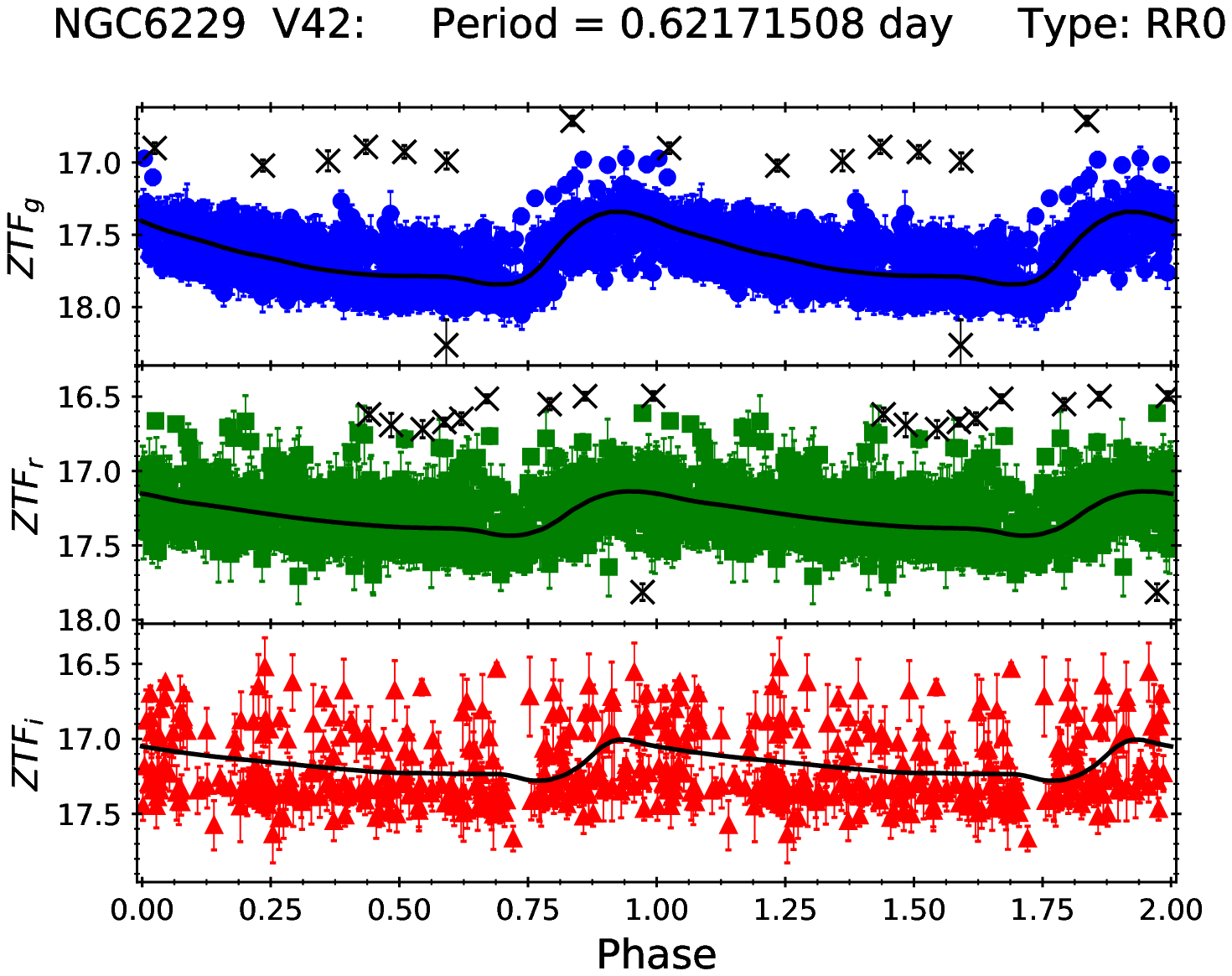}{0.32\textwidth}{With ``R'' flag.}}
\caption{Examples of the light curves for RR Lyrae with 3 (top panels), 2 (middle panels), and 1 (bottom panels) of the ``ACR'' flags. The black crosses are rejected data points during the first pass of the two-iterations template light curve fitting procedure, and the black curves are the best-fit template light curves after the second pass of the procedure (see Section \ref{sec2.4} for more details).}\label{fig_flags}
\end{figure*}

\begin{figure*}
  \centering
  \begin{tabular}{ccc}
    \includegraphics[width=0.68\columnwidth]{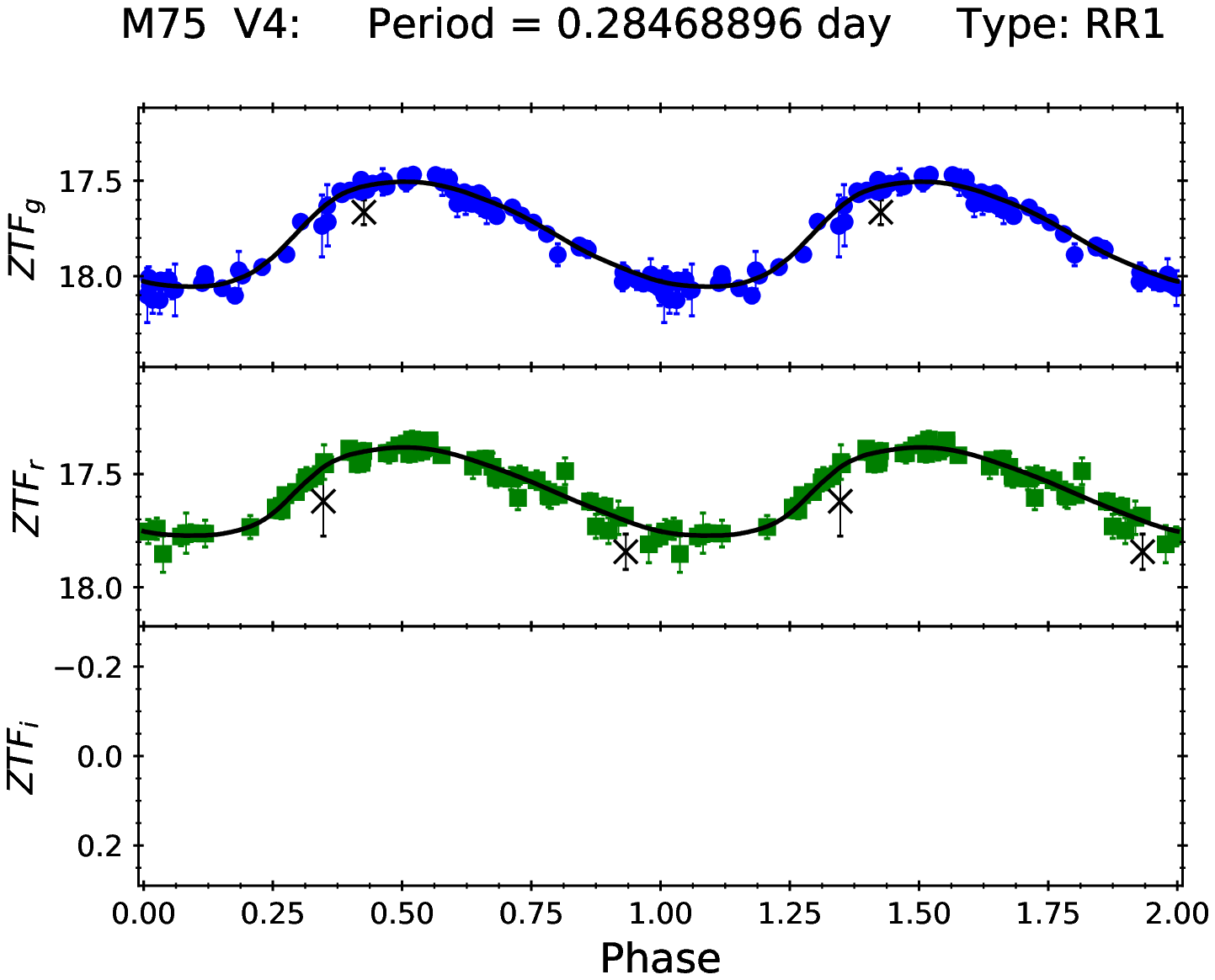} &   \includegraphics[width=0.68\columnwidth]{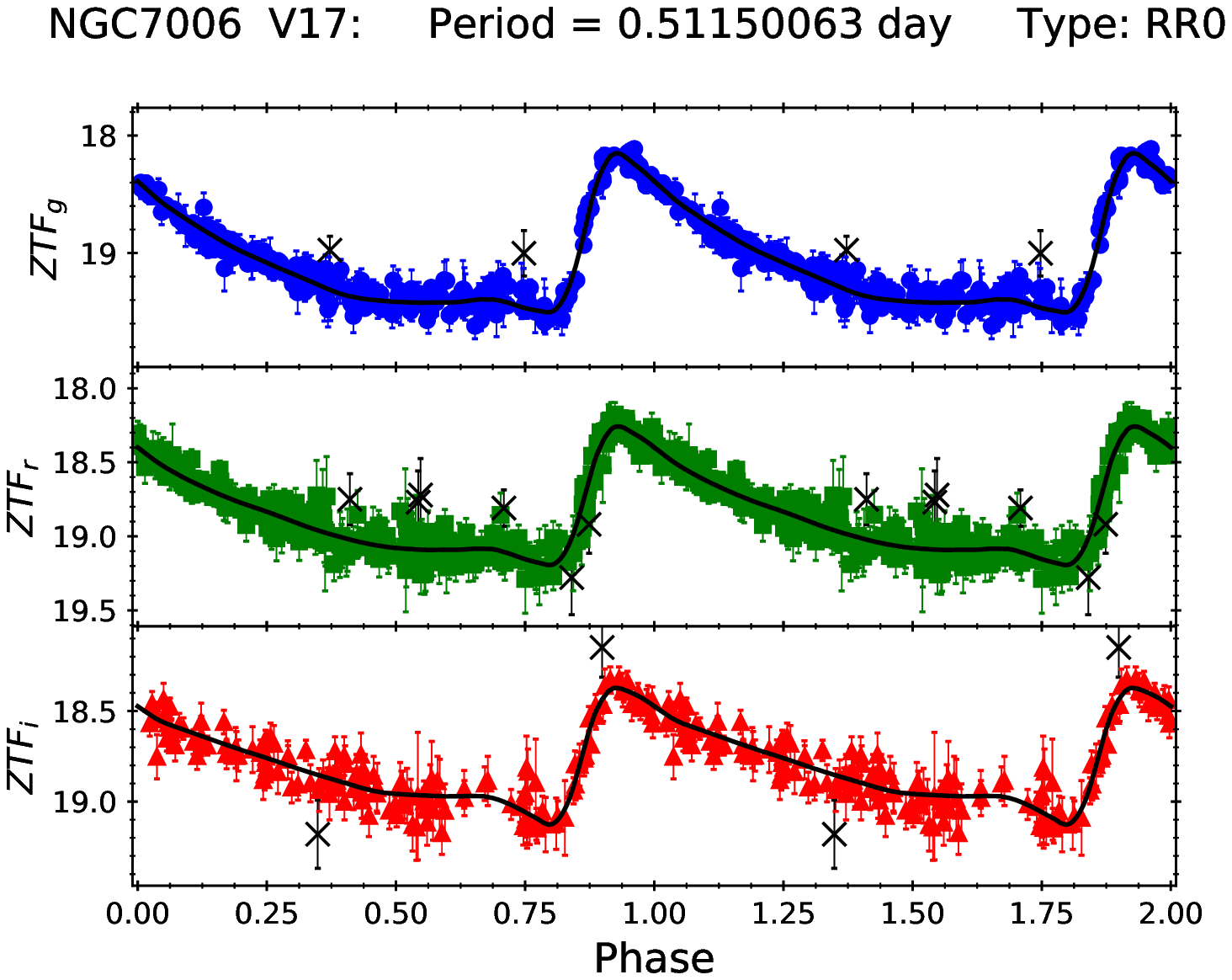} &  \includegraphics[width=0.68\columnwidth]{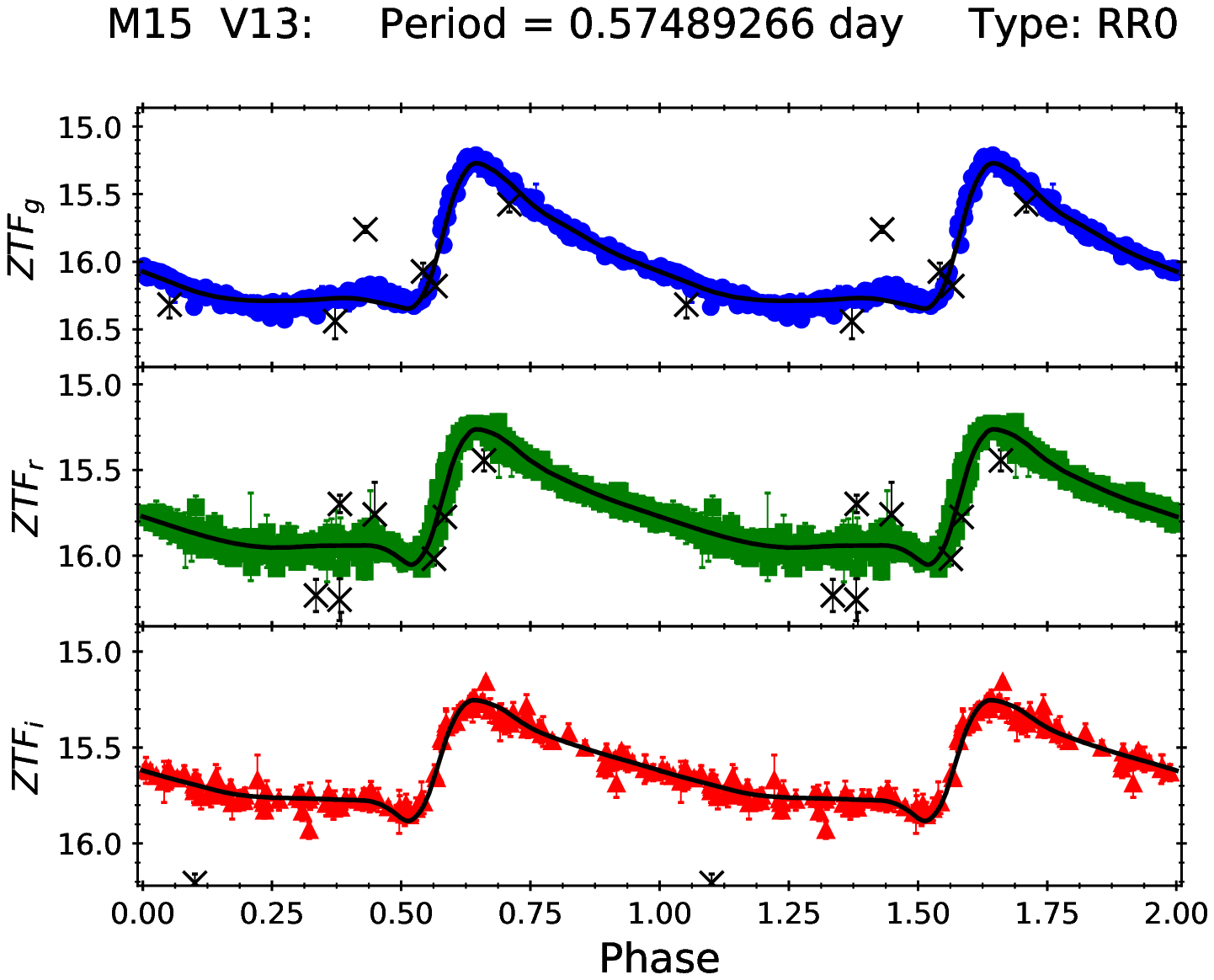} \\
    \includegraphics[width=0.68\columnwidth]{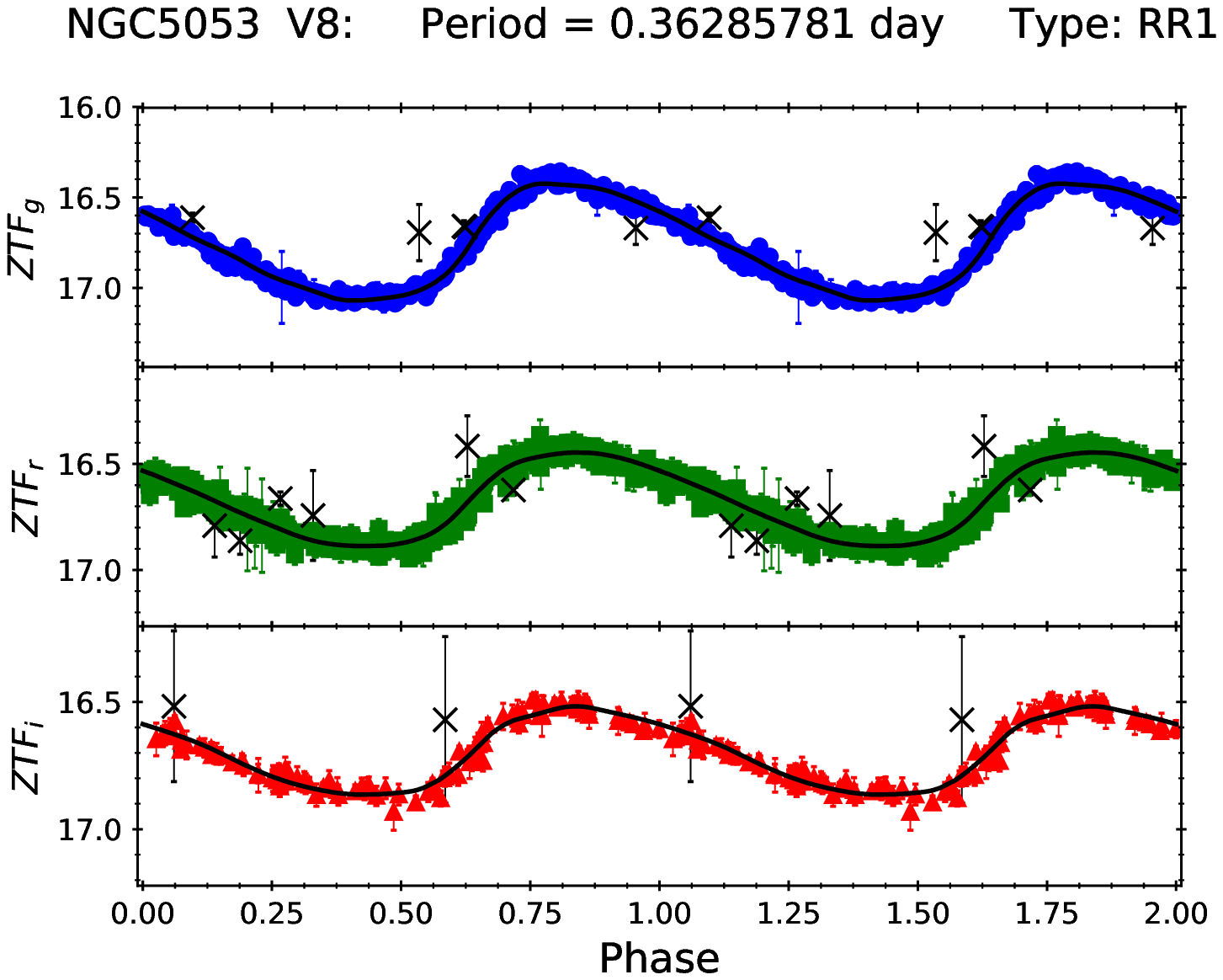} &   \includegraphics[width=0.68\columnwidth]{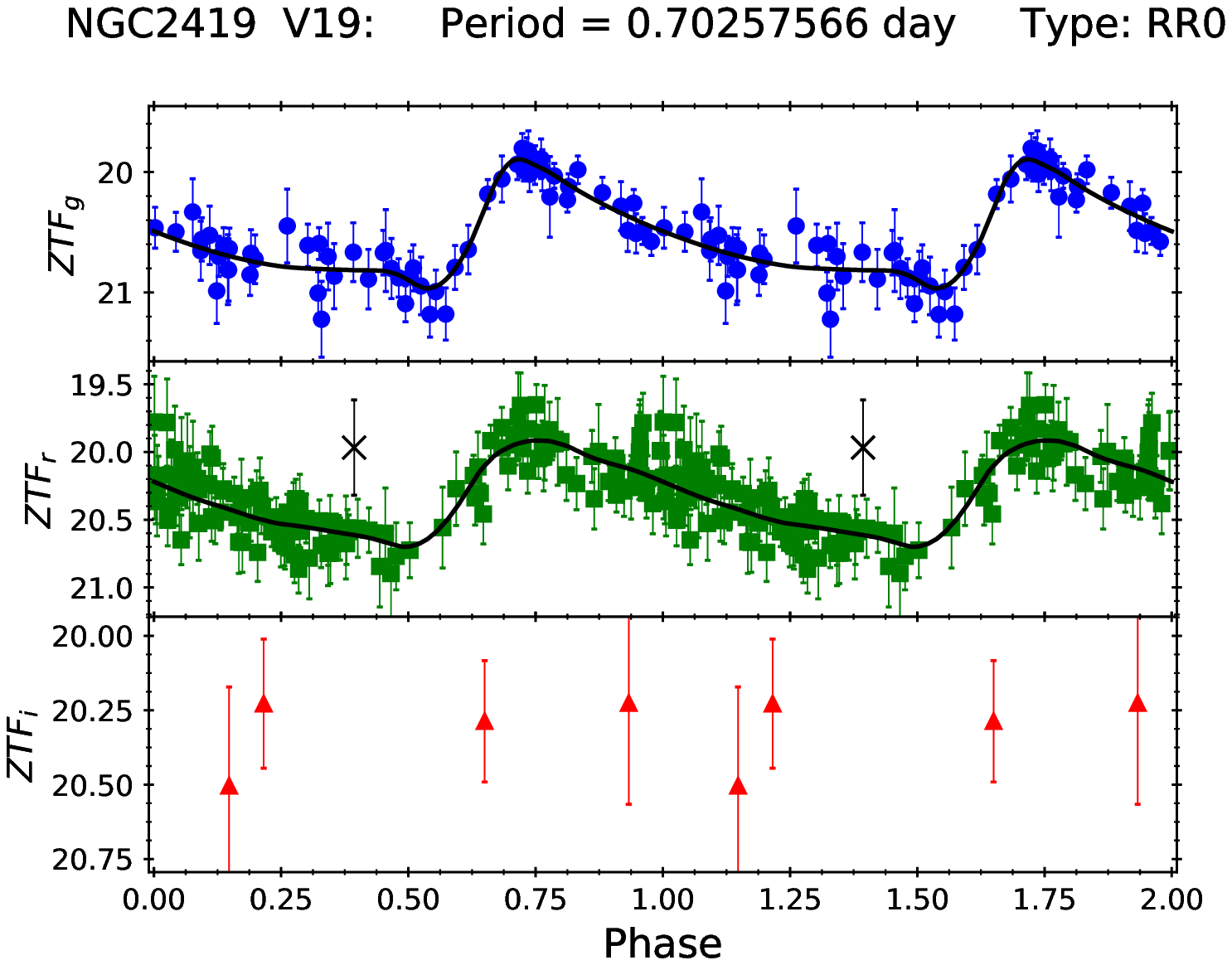} &  \includegraphics[width=0.68\columnwidth]{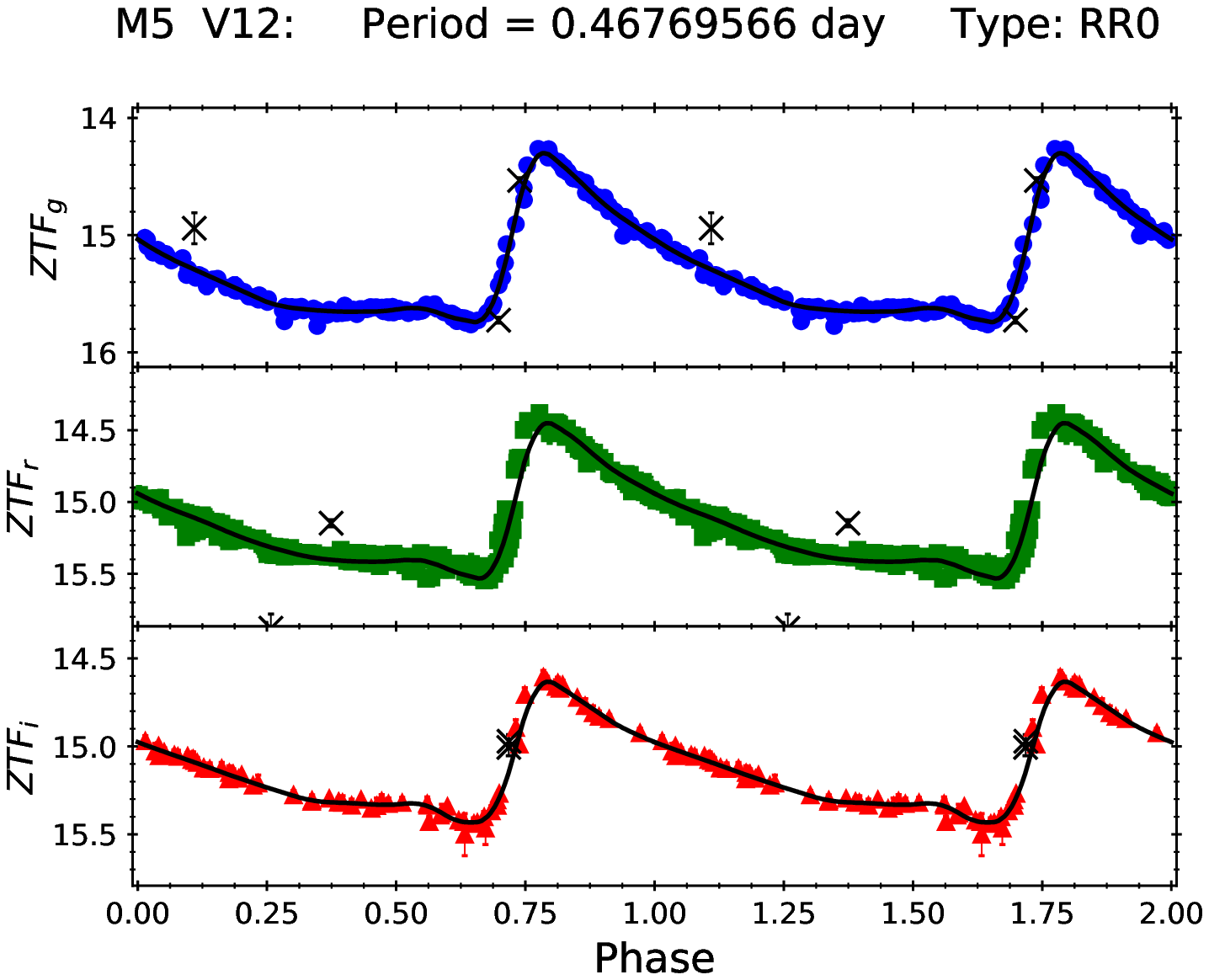} 
    \end{tabular}
  \caption{Examples of the randomly selected light curves for RR Lyrae without any flags. The black crosses are rejected data points during the first pass of the two-iterations template light curve fitting procedure, and the black curves are the best-fit template light curves after the second pass of the procedure (see Section \ref{sec2.4} for more details).}\label{fig_goodlc}
\end{figure*}

The refined periods and reddenings, as well as the mean magnitudes, amplitudes and number of data points left in the $gri$-band light curves for the 1209 RR Lyrae in our sample are presented in Table \ref{tab_rrl}. In the last column of Table \ref{tab_rrl} we also included the flags for each RR Lyrae based on the analysis given in previous subsections. Out of the 1209 RR Lyrae in our sample, 108 have all three ``ACR'' flags. The number of RR Lyrae with two flags is 114 (8, 23 and 83 with ``AC'', ``CR'' and ``AR'', respectively), while the number of RR Lyrae with only one flag is 232 (132, 8 and 92 with ``A'', ``C'' and ``R'', respectively). A Venn diagram showing the distribution of these flagged RR Lyrae is presented in Figure \ref{fig_venn}. Example light curves for these flagged RR Lyrae are shown in Figure \ref{fig_flags} for the cases with 3, 2, and 1 flag, showing that the light curves for RR Lyrae with any of the ``ACR'' flags, or a combination of them, were problematic due to a variety of reasons and should be excluded. 

\begin{figure}
  \epsscale{1.15}
  \plotone{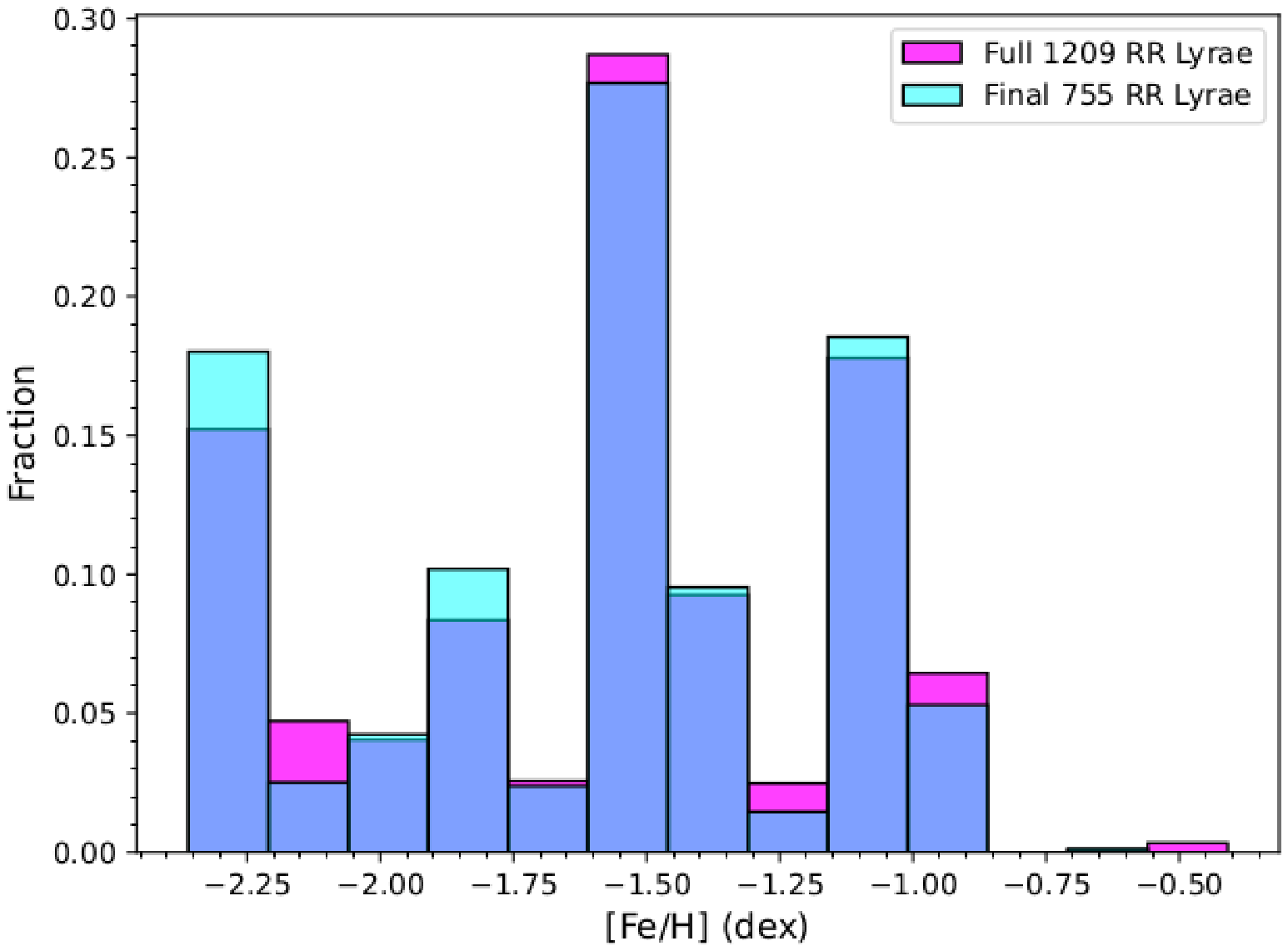}
  \caption{Histograms of $[\mathrm{Fe/H}]$ distribution for the full sample of 1209 RR Lyrae and the remaining sample of 755 RR Lyrae in our sample. The $[\mathrm{Fe/H}]$ values for the RR Lyrae are based on their host globular clusters (see Table \ref{tab_gc}).}\label{fig_hfeh}
\end{figure}

The remaining 755 RR Lyrae in our sample do not carry any of the ``ACR'' flags. Examples of their light curves are presented in Figure \ref{fig_goodlc}, demonstrating the good quality of the ZTF light curves for them. These sample of RR Lyrae will be used to derive the PLZ and PWZ relation in the next section, and they are located in 46 (out of 54) globular clusters. The excluded globular clusters are (as listed in the bottom of Table \ref{tab_gc}): 2MASS-GC02, Djorg2, IC1276, NGC6304, NGC6316, NGC6540, NGC6544 and Terzan10, as all of their RR Lyrae have one or more flags. These also removed the three globular clusters with highest metallicities (IC1276, NGC6316, and NGC6304), hence the range of metallicity covered by the 755 RR Lyrae is from $-2.36$~dex to $-0.54$~dex, with a median of $-1.48$~dex. Figure \ref{fig_hfeh} compares the metallicity distribution between the ``full'' sample of 1209 RR Lyrae and the remaining sample of 755 RR Lyrae, no substantial difference can be seen between their distributions. This ensures there is no bias introduced when fitting the PLZ and PWZ relations to the remaining 755 RR Lyrae sample.

\begin{deluxetable*}{lccccccccccccc}
  \tabletypesize{\scriptsize}
  \tablecaption{Periods, mean magnitudes, amplitudes and reddenings for our sample of RR Lyrae.\label{tab_rrl}}
  \tablewidth{0pt}
  \tablehead{
    \colhead{Var. Name\tablenotemark{a}} &
    \colhead{Type\tablenotemark{b}} &
    \colhead{Period (days)} &
    \colhead{$\langle m_g \rangle$} &
    \colhead{$AMP_g$} &
    \colhead{$N_g$} &
    \colhead{$\langle m_r \rangle$} &
    \colhead{$AMP_r$} &
    \colhead{$N_r$} &
    \colhead{$\langle m_i \rangle$} &
    \colhead{$AMP_i$} &
    \colhead{$N_i$} &
    \colhead{$E$} &
    \colhead{Flag}
  }
  \startdata
  Palomar3\_V2 & RR0 & 0.59867205 & 20.512 & 0.742 & 92 & 20.301 & 0.597 & 121 & 20.215 & 0.293 & 10 & $0.004\pm0.002$ & A\_\_ \\
  Palomar3\_V3 & RR0 & 0.56690511 & 20.469 & 0.872 & 107 & 20.269 & 0.661 & 120 & 20.150 & 0.431 & 12 & $0.004\pm0.002$ & \_\_R \\
  Palomar3\_V5 & RR0 & 0.58168014 & 20.518 & 0.846 & 97 & 20.327 & 0.585 & 128 & 20.209 & 0.379 & 10 & $0.004\pm0.002$ & \_\_\_ \\
  Palomar3\_V6 & RR0 & 0.59336755 & 20.493 & 0.916 & 83 & 20.311 & 0.583 & 60 & -99.999 & -9.999 & 7 & $0.004\pm0.002$ & \_\_\_ \\
  $\cdots$  &  $\cdots$ & $\cdots$ &  $\cdots$  &  $\cdots$ & $\cdots$ &   $\cdots$  &  $\cdots$ & $\cdots$ &  $\cdots$  &  $\cdots$ & $\cdots$ &  $\cdots$ &  $\cdots$ \\
  \enddata
  \tablenotetext{a}{Format for the names of the RR Lyrae (RRL) in globular clusters (G.C.) is [G. C.]\_[RRL Name], where [RRL Name] is adopted from the Clement's Catalog. Note that we renamed 23 P1 in NGC6712 as 23$\_$P1 in our Table.}
  \tablenotetext{b}{Pulsation types for RR Lyrae, either as RR0 (for fundamental mode) or RR1 (for first-overtone mode).}
  \tablecomments{Table \ref{tab_rrl} is published in its entirety in the machine-readable format. A portion is shown here for guidance regarding its form and content. $AMP_{\{g,r,i\}}$ is the amplitude based on the best-fit template light curve. Values of $-99.990$ and $-9.990$ denote no data or rejected light curves. $N_{\{g,r,i\}}$ is the number of data points in a given filter after the two-steps template fitting processes as described in Section \ref{sec2.3}. $E$ is the reddening value returned from the {\tt Bayerstar2019} 3D reddening map \citep{green2019} using the globular cluster distance listed in Table \ref{tab_gc}.}
\end{deluxetable*}

\section{The PLZ and PWZ Relations}  \label{sec4}

\begin{figure*}
  \epsscale{1.1}
  \plotone{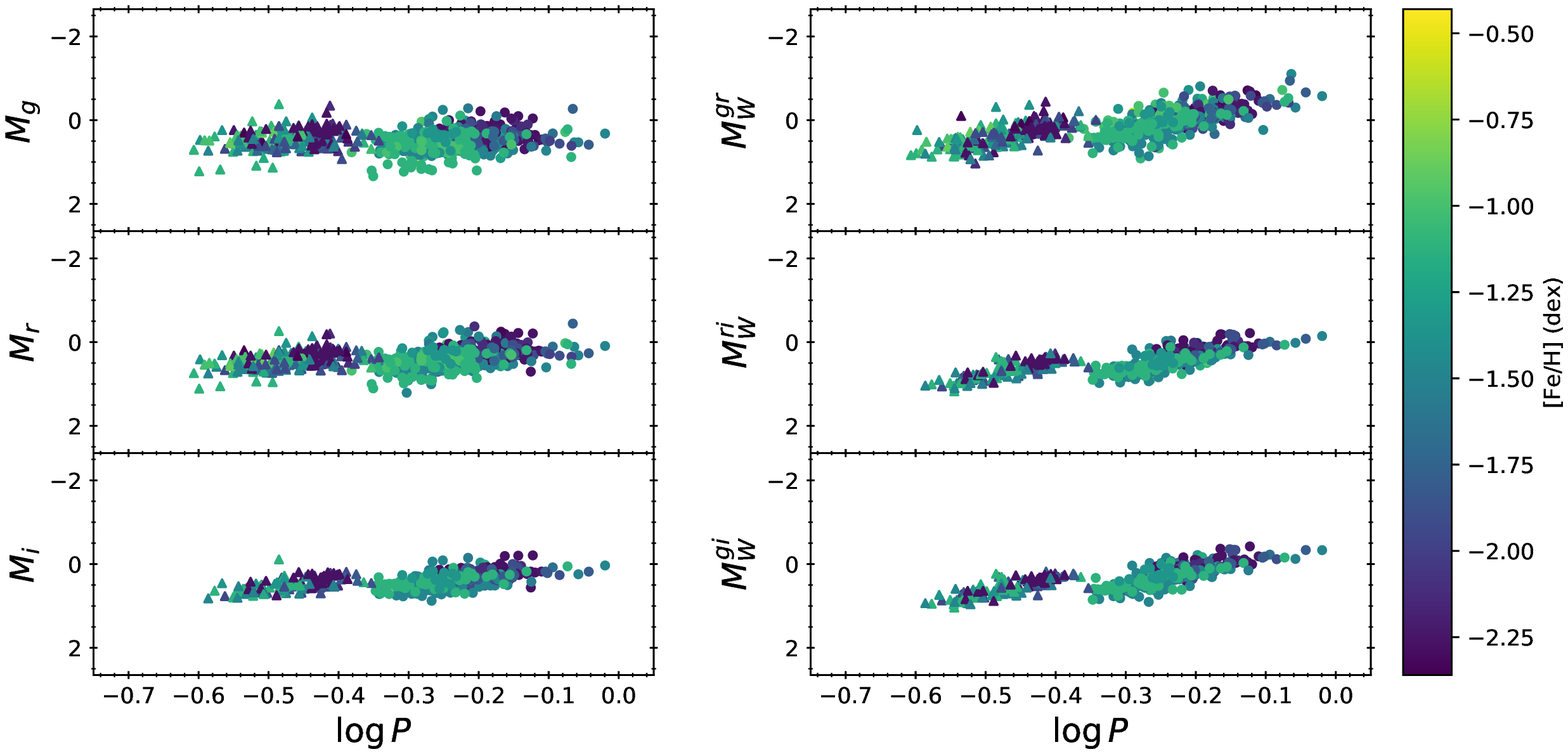}
  \caption{The PL relations (left panels, after corrected for extinction) and PW relations for all of the RR Lyrae listed in Table \ref{tab_rrl}, by adopting the distance of the globular clusters as listed in Table \ref{tab_gc}. The color-bar represents the metallicity for the host globular clusters. Circular and triangle symbols are for the RR Lyrae with pulsation types of RR0 and RR1, respectively. Error bars are omitted for clarity.}
  \label{fig_plpw}
\end{figure*}

\begin{deluxetable}{lrrrcc}
  \tabletypesize{\scriptsize}
  \tablecaption{Fitted coefficients for various relations.\label{tab_plpw}}
  \tablewidth{0pt}
  \tablehead{
    \colhead{} &
    \colhead{$a$} &
    \colhead{$b$} &
    \colhead{$c$} &
    \colhead{$\sigma$} &
    \colhead{$N_{\mathrm{ini}}/N_{\mathrm{fit}}$} 
  }
  \startdata
  \multicolumn{6}{c}{RR0 only} \\
  $M_g$   & $0.649\pm0.073$ &  $-0.302\pm0.193$  & $0.159\pm0.024$ &  0.227  &  493/490 \\
  $M_r$   & $0.337\pm0.059$ &  $-1.090\pm0.154$  & $0.139\pm0.020$ &  0.187  &  516/508 \\
  $M_i$   & $0.243\pm0.055$ &  $-1.432\pm0.144$  & $0.144\pm0.018$ &  0.148  &  326/321 \\
  $W^{gr}$ & $-0.644\pm0.058$&  $-3.324\pm0.155$  & $0.052\pm0.020$ &  0.205  &  483/478 \\
  $W^{ri}$ & $0.093\pm0.051$ &  $-2.600\pm0.133$  & $0.193\pm0.017$ &  0.145  &  326/325 \\
  $W^{gi}$ & $-0.198\pm0.054$&  $-2.908\pm0.142$  & $0.142\pm0.018$ &  0.146  &  325/323 \\
  $(g-r)$ & $0.333\pm0.023$ &  $0.764\pm0.060$   & $0.035\pm0.007$ &  0.076  &  483/483 \\
  $(r-i)$ & $0.040\pm0.014$ &  $0.361\pm0.038$   & $-0.017\pm0.004$&  0.035  &  326/323 \\
  $(g-i)$ & $0.323\pm0.033$ &  $1.110\pm0.088$   & $-0.001\pm0.010$&  0.093  &  325/325 \\
  $Q$     & $0.222\pm0.016$ &  $0.220\pm0.044$   & $0.042\pm0.005$ &  0.046  &  325/320 \\
  \multicolumn{6}{c}{RR1 only} \\
  $M_g$   & $0.411\pm0.166$ &  $-0.342\pm0.289$  & $0.092\pm0.028$ &  0.189  &  218/212  \\
  $M_r$   & $-0.082\pm0.138$&  $-1.393\pm0.238$  & $0.091\pm0.024$ &  0.164  &  227/224  \\
  $M_i$   & $-0.205\pm0.139$&  $-1.706\pm0.246$  & $0.077\pm0.023$ &  0.128  &  123/122  \\
  $W^{gr}$ & $-1.327\pm0.143$&  $-3.680\pm0.248$  & $0.013\pm0.024$ &  0.172  &  216/210 \\
  $W^{ri}$ & $-0.884\pm0.141$&  $-3.503\pm0.249$  & $0.067\pm0.023$ &  0.132  &  123/123 \\
  $W^{gi}$ & $-1.086\pm0.126$&  $-3.696\pm0.223$  & $0.055\pm0.020$ &  0.111  &  123/120 \\
  $(g-r)$ & $0.398\pm0.059$ &  $0.781\pm0.103$   & $0.017\pm0.009$ &  0.069  &  216/215 \\
  $(r-i)$ & $0.194\pm0.039$ &  $0.569\pm0.070$   & $-0.005\pm0.006$&  0.030  &  123/121 \\
  $(g-i)$ & $0.670\pm0.101$ &  $1.545\pm0.179$   & $0.014\pm0.016$ &  0.080  &  123/122 \\
  $Q$     & $0.192\pm0.029$ &  $0.162\pm0.052$   & $0.024\pm0.004$ &  0.039  &  123/121 \\
  \multicolumn{6}{c}{RR0 + RR1} \\
  $M_g$   & $0.801\pm0.048$ & $-0.032\pm0.110$  & $0.190\pm0.019$  &  0.216  &  711/704 \\
  $M_r$   & $0.432\pm0.039$ & $-0.874\pm0.089$  & $0.154\pm0.016$  &  0.180  &  743/731 \\
  $M_i$   & $0.249\pm0.041$ & $-1.362\pm0.093$  & $0.115\pm0.016$  &  0.138  &  449/444 \\
  $W^{gr}$ & $-0.727\pm0.041$& $-3.286\pm0.093$  & $0.010\pm0.016$  &  0.191  &  699/687 \\
  $W^{ri}$ & $0.010\pm0.037$ & $-2.756\pm0.086$  & $0.149\pm0.015$  &  0.130  &  448/444 \\
  $W^{gi}$ & $-0.288\pm0.039$& $-3.066\pm0.090$  & $0.101\pm0.015$  &  0.137  &  449/444 \\
  $(g-r)$ & $0.333\pm0.014$ & $0.803\pm0.034$   & $0.030\pm0.005$  &  0.073  &  699/697 \\
  $(r-i)$ & $0.063\pm0.009$ & $0.428\pm0.023$   & $-0.012\pm0.003$ &  0.033  &  449/444 \\
  $(g-i)$ & $0.370\pm0.022$ & $1.263\pm0.055$   & $0.007\pm0.007$  &  0.089  &  448/447 \\
  $Q$     & $0.218\pm0.009$ & $0.222\pm0.023$   & $0.038\pm0.003$  &  0.043  &  448/440 \\
   \enddata
  \tablecomments{The fitting parameters $a$, $b$, and $c$ are defined in equation (1), and $\sigma$ is the dispersion of the fitted relation. $N_{\mathrm{ini}}$ and $N_{\mathrm{fit}}$ are the number of RR Lyrae  before and after applying the $3\sigma$-rejection algorithm (the number of rejected RR Lyrae varies from 0 to 12). Note that $M_{gri}$ and various colors have been corrected for extinction.}
\end{deluxetable}

The extinction-corrected $gri$-band PL and the PW relations for the final sample of 755 RR Lyrae were presented in the left and right panels of Figure \ref{fig_plpw}, respectively, where the coded-colors represent the metallicity of the host globular clusters. These RR Lyrae were fitted with PLZ and PWZ relations in the following form:

\begin{eqnarray}
 y  & = & a + b \log P  + c [Fe/H], 
\end{eqnarray}

\noindent where $y$ is either $M_{gri}$ or $W$, and all RR Lyrae in a given clusters are assumed to have the same distance as given in Table \ref{tab_gc}. We solved equation (1) using a $\chi^2$ minimization within a matrix formalism of dependent and independent parameters including their associated uncertainties in the covariance matrix \citep[similar to equation (5) in][]{bhardwaj2016}. As in \citet{bhardwaj2021}, we also applied an iterative $3\sigma$-rejection algorithm to remove the single largest outlier in each iteration until convergence. Figure \ref{fig_plpw} shows that there are still a few outliers presented in the PL and PW relations that were not flagged based on the selection criteria presented in Section \ref{sec3}. Their light curves look normal implying other physical reasons (such as location on the near-site of the host globular clusters, incorrect estimation of extinction, etc) causing them to become outliers on the PL and/or PW relations. Nevertheless, they should be excluded to obtain a more robust relation. The fitted coefficients and the corresponding dispersion are summarized in Table \ref{tab_plpw} for RR0 and RR1 separately, as well as the RR0 and RR1 combined sample. When combining RR0 and RR1 samples, periods for RR1 were fundametalized using $\log P_0 = \log P_1 + 0.127$ \citep[for examples, see][]{iben1974,coppola2015}, where $P_1$ and $P_0$ represent the first-overtone period and the corresponding fundametalized period.

\begin{figure}
  \epsscale{1.1}
  \plotone{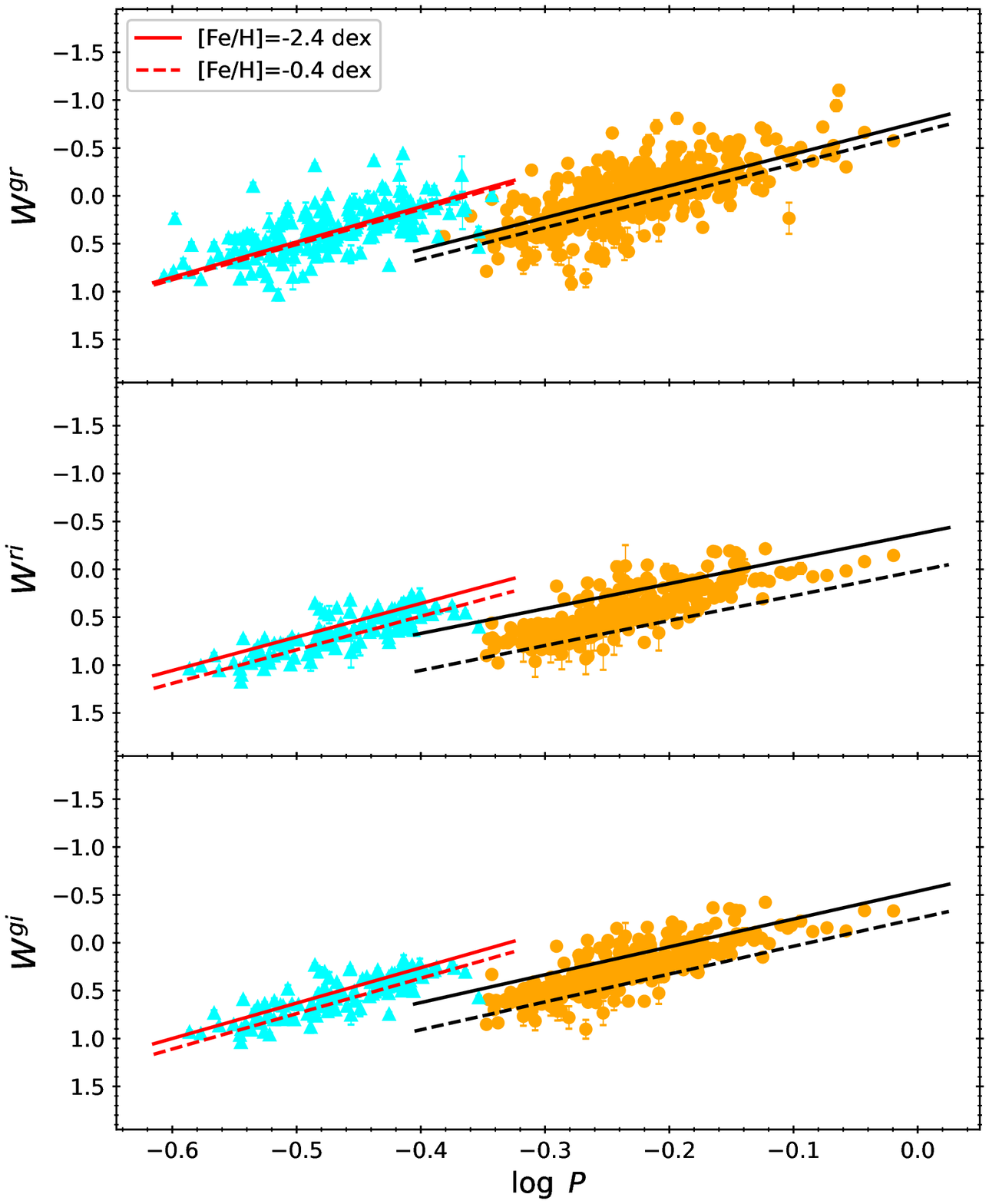}
  \caption{The metallicity-dependent period-Wesenheit (PWZ) relations for our sample of RR Lyrae. The solid and dashed lines (in both red and black colors) represent the PWZ relations evaluated at $[\mathrm{Fe/H}]=-2.4$~dex and $[\mathrm{Fe/H}]=-0.4$~dex, respectively. The orange circles and cyan triangles are for the RR0 and RR1, respectively. The PWZ relations for the combined sample of RR0 and RR1 (after fundametalized the periods) are similar, hence they are not included in this Figure.}\label{fig_pwz}
\end{figure}

In Figure \ref{fig_pwz} we present the fitted PWZ relations evaluated at $[\mathrm{Fe/H}]=-2.4$~dex and $-0.4$~dex. These two values of metallicities were chosen to  bracket the metallicity listed in Table \ref{tab_gc}. As can be seen from left panel of Figure \ref{fig_pwz} and Table \ref{tab_plpw}, the $W^{gr}$ PWZ relations show a very weak, or even vanishing, dependence on the metallicity for either RR0 or RR1, or the combined sample. Interestingly, similar metallicity-independent PW(Z) relations were found in \citet{marconi2015} for Wesenheit indexes defined in the Johnson $BVR$ filters based on a series of theoretical models. We noted that the $gr$-band transmission curves covering a similar wavelength range as the Johnson $BVR$-band transmission curves. As discussed in \citet{marconi2015}, such a nearly metallicity-independent PW(Z) relation can be used as a robust distance indicator, however, the downside is our derived $W^{gr}$ PWZ relations exhibit a larger dispersion than other two Wesenheit indexes. Based on Table \ref{tab_plpw}, we also notice that the fitted PLZ and PWZ relations involving $i$-band have the smallest dispersion.

In the following subsections, we compared our derived PLZ relations to published PL(Z) relations in $gri$-band, including three empirical and two theoretical PL(Z) relations. For an ease of comparisons, we summarized slopes of these PL(Z) relations in Table \ref{tab_compare}.

\begin{deluxetable}{lcccc}
  \tabletypesize{\scriptsize}
  \tablecaption{Comparison of the slopes of PL(Z) relations from various studies.\label{tab_compare}}
  \tablewidth{0pt}
  \tablehead{
    \colhead{Ref.} & \colhead{$g$-band} & \colhead{$r$-band} & \colhead{$i$-band} & \colhead{G.C.}
  }
  \startdata
  \multicolumn{5}{c}{For RR0} \\ 
  TW & $-0.302\pm0.193$ & $-1.090\pm0.154$ & $-1.432\pm0.144$ & 46 G.C. \\
  S17                  & $-1.7\pm0.3$     & $-1.6\pm0.1$     & $-1.77\pm0.08$ & 5 G.C \\
  V17                  & $-0.57\pm0.17$   & $-1.28\pm0.11$   & $-1.59\pm0.09$ & M5 \\
  B21             & $-0.111\pm0.160$ & $\cdots$         & $-1.292\pm0.184$ & M15\\
  C08              & $\cdots$         & $\cdots$         & $1.035$ & $\cdots$ \\
  M06\tablenotemark{a}& $-0.311$      & $\cdots$         & $\cdots$ & $\cdots$ \\
  \multicolumn{5}{c}{For RR1} \\
  TW & $-0.342\pm0.289$ & $-1.393\pm0.238$ & $-1.706\pm0.246$ & 46 G.C.\\
  V17                  & $-0.72\pm0.32$   & $-1.35\pm0.21$   & $-1.61\pm0.16$ & M5  \\
  B21             & $-0.019\pm0.138$ & $\cdots$         & $-1.329\pm0.112$ & M15 \\
  M06\tablenotemark{a}& $-0.322$      & $\cdots$         & $\cdots$  & $\cdots$ \\
  \multicolumn{5}{c}{For RR0+RR1} \\
  TW & $-0.032\pm0.110$ & $-0.874\pm0.089$ & $-1.362\pm0.093$ & 46 G.C. \\
  B21             & $+0.185\pm0.066$ & $\cdots$         & $-1.222\pm0.060$ & M15 \\
  \enddata
  \tablenotetext{a}{Semi-theoretical relations based on equation (3) and (5), see Section \ref{sec4.2} for more details.}
  \tablecomments{Slopes without errors are for the (semi-)theoretical PL(Z) relations. In case of RR0+RR1, periods for RR1 have been fundametalized. The references (Ref.) are: TW = this work (Table \ref{tab_plpw}); S17 = \citet{ses2017}; V17 = \citet{viv2017}; B21 = \citet{bhardwaj2021}; C08 = \citet{caceres2008}; and M06 = \citet{marconi2006}.}
\end{deluxetable}


\subsection{Comparisons to Published Results: Empirical Relations} \label{sec4.1}

\begin{figure}
  \epsscale{1.1}
  \plotone{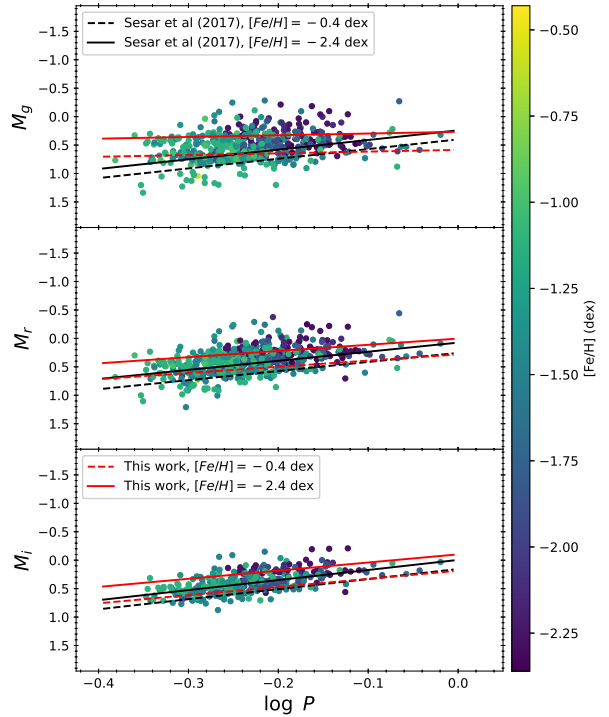}
  \caption{Comparison of the PLZ relations for RR0 from \citet[][in black colors]{ses2017} and this work (i.e. Table \ref{tab_plpw}, in red colors), evaluated at two metallicities: $[\mathrm{Fe/H}]=-2.4$~dex and $-0.4$~dex as solid and dashed lines, respectively.}\label{fig_sesar}
\end{figure}

\begin{figure}
  \epsscale{1.1}
  \plotone{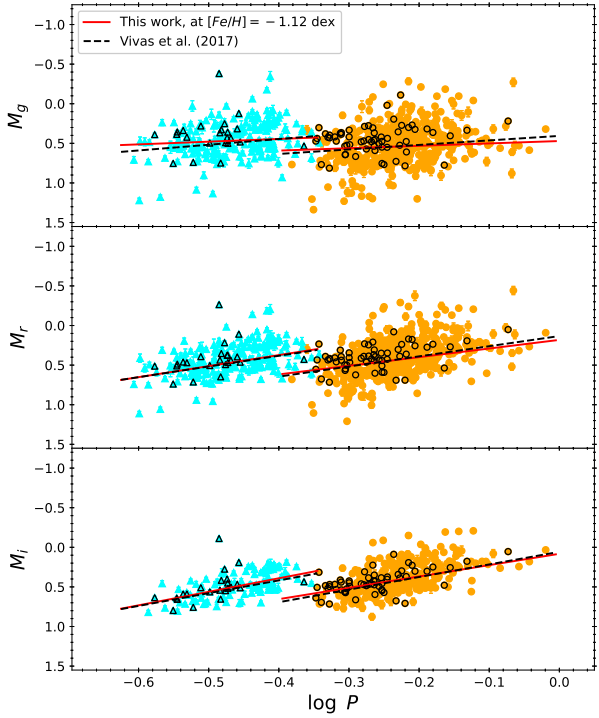}
  \caption{Comparison of the PL relations from \citet[][as black dashed lines]{viv2017} and this work (i.e. Table \ref{tab_plpw}, as red solid lines), evaluated at $[\mathrm{Fe/H}]=-1.12$~dex for M5 (see Table \ref{tab_gc}). The orange circles and cyan triangles are for the RR0 and RR1 in our sample, respectively. Data points with black circles or triangles represent the RR Lyrae in M5.}\label{fig_vivas}
\end{figure}

\begin{figure}
  \epsscale{1.1}
  \plotone{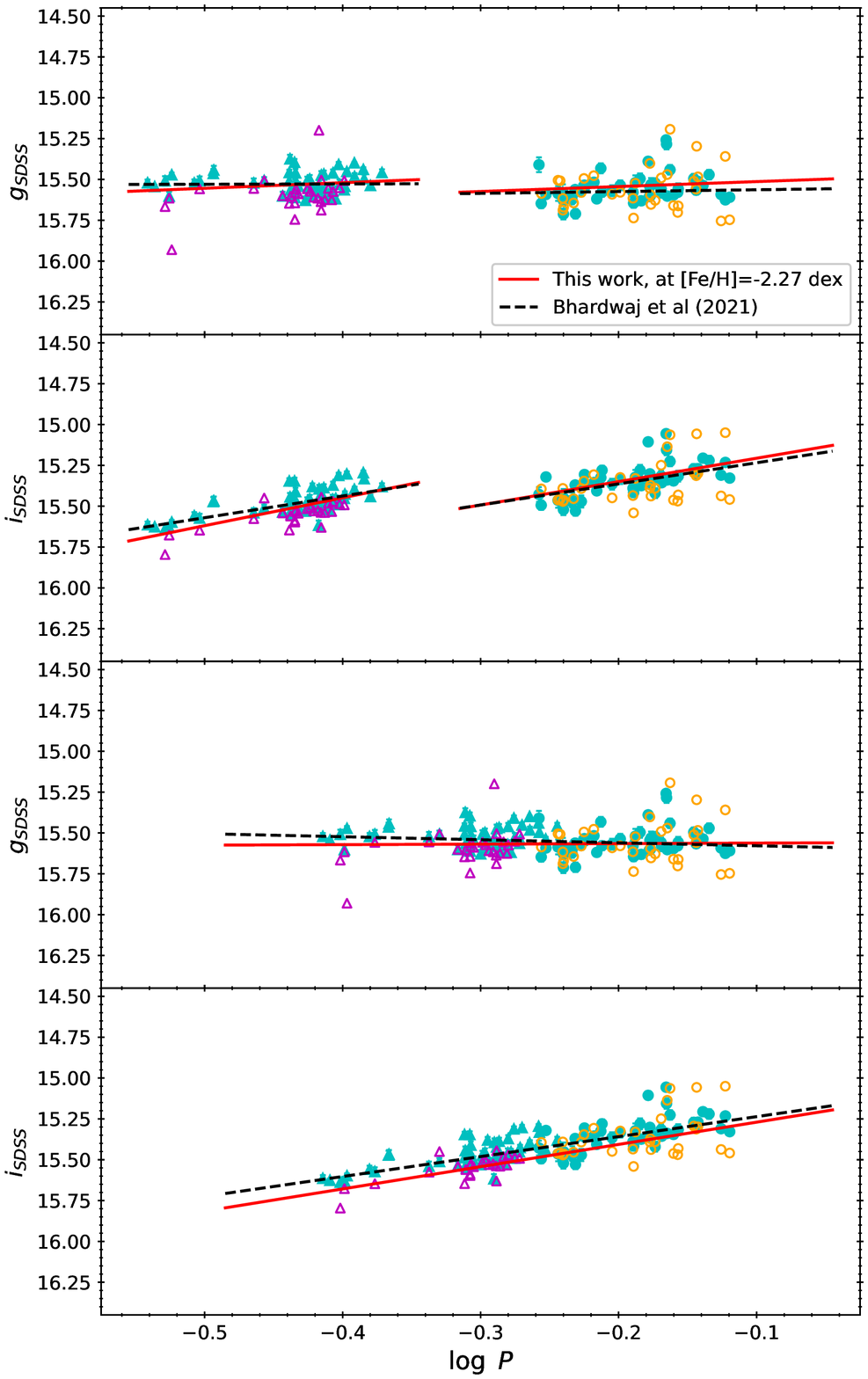}
  \caption{Comparison of the PL relations from \citet[][as black dashed lines]{bhardwaj2021} and this work (i.e. Table \ref{tab_plpw}, as red solid lines), evaluated at $[\mathrm{Fe/H}]=-2.27$~dex for M15 (see Table \ref{tab_gc}). The orange open circles and magenta open triangles are for the RR0 and RR1 in our sample, respectively, with mean magnitudes transformed to the SDSS system. Data points with filled light green circles or triangles represent the RR Lyrae taken from \citet{bhardwaj2021}. Consistent extinction of $A_g=0.379$ and $A_i=0.209$ \citep{bhardwaj2021} in the SDSS system were applied to all data points. Our PL relations were shifted vertically using the distance of M15 as listed in Table \ref{tab_gc}. The bottom two panels are for the combined sample of RR0 and RR1 after fundametalized the pulsation periods of RR1.}\label{fig_ab}
\end{figure}

\citet{ses2017} derived PLZ relations in the $gri(z)$-band using a sample of {\bf 55} RR0 located in 5 globular clusters. We compared the $gri$-band PLZ relations derived in their work with ours in Figure \ref{fig_sesar}, and found that these two sets of PLZ relations disagree with each other. For example, the metallicity terms for our PLZ relations are approximately twice the values reported in \citet{ses2017}. Slopes of these PLZ relations even show a larger disagreement, especially for the $g$-band PLZ relations. Our $i$-band PLZ relation has a slope closer to the one derived in \citet[][$-1.432$ vs. $-1.77$]{ses2017}, however these two slopes are different by $\sim 2\sigma$. Despite in common Pan-STARRS1 photometric system, the derivations of the PLZ relations between \citet{ses2017} and our work are significantly different, including data sources used, the adopted distances to the globular clusters, and the methodology of fitting the PLZ relations. Using a Bayesian inference approach, \citet{ses2017} adopted a tight Gaussian prior when fitting the slopes of the $ri$-band PLZ relations, but not in the $g$-band. This might explain why the slope of the $g$-band PLZ relation is much steeper in \citet{ses2017} than our derived value as presented in Table \ref{tab_plpw}.

A similar comparison was also done on the PL relations derived in \citet{viv2017}, using RR Lyrae in M5 observed with DECam. As evident in Figure \ref{fig_vivas}, these two sets of PL relations are in remarkably good agreement for both RR0 and RR1 PL relations. \citet{viv2017} calibrated their photometry to the native DECam system, which has been extensively employed by the Dark Energy Survey \citep[DES,][]{des2016}, therefore the PL relations presented in \citet{viv2017} strictly speaking are in the DES photometric system. To transform the PL relations given in \citet{viv2017} from DES system to Pan-STARRS1 system, we added a correction term $\delta_m$ to their PL relations, separately in $gri$-band for RR0 and RR1. These $\delta_m$ were determined using the averaged colors of RR0 and RR1 based on the $\sim60$ RR Lyrae listed in \citet{viv2017}, together with the transformation provided in \citet{des2021}, and found to be small (range from $-0.024$~mag to $+0.005$~mag). We have also checked the mean magnitudes between our work and \citet{viv2017}, by transforming the mean magnitudes of \citet{viv2017} from DES system to Pan-STARRS1 system. The averaged differences, $\langle VIVAS_{PS1}-ZTF_{PS1}\rangle$ (where $VIVAS_{PS1}$ are the transformed mean magnitudes, and $ZTF_{PS1}$ are the ZTF mean magnitudes given in Table \ref{tab_rrl}), were found to be small: $-0.028$~mag, $-0.002$~mag, and $+0.005$~mag (the corresponding standard deviations are $0.046$~mag, $0.052$~mag, and $0.060$~mag) in the $gri$-band, respectively.

Finally, we compared our PL relations with the $gi$-band PL relations derived in \citet{bhardwaj2021}, using RR Lyrae in a metal-poor globular cluster M15. \citet{bhardwaj2021} calibrated their data in the SDSS photometric system, and the transformation between SDSS and Pan-STARRS1 system can be found in \citet{tonry2012}. Such transformation, however, relies on the $(g-r)$ colors but the $r$-band data were absent in \citet{bhardwaj2021}. Therefore, we can only transform our data and PL relations to the SDSS photometric system. Furthermore, \citet{bhardwaj2021} did not apply an absolute calibration to their PL relations, hence we only compare the PL relations for RR Lyrae in M15. For $\sim 60$ common RR Lyrae after transforming the mean magnitudes to the SDSS photometric system, the averaged difference in the $gi$-band is $-0.023$~mag and $0.023$~mag (with corresponding standard deviation of $0.121$~mag and $0.086$~mag), respectively. The $gi$-band PL relation were compared in Figure \ref{fig_ab} after transforming our mean magnitudes, as well as adding the correction term $\delta_m$ (ranging from $0.046$~mag to $-0.003$~mag, applied separately to RR0, RR1, and the RR0+RR1 combined sample) to our PL relations, to the SDSS photometric system. Overall, these two sets of PL relations fairly agree, although there is evidence that the zero-points of our $i$-band PL relations are slight fainter for the RR1 and the RR0+RR1 combined sample.

\subsection{Comparisons to Published Results: Theoretical Relations} \label{sec4.2}

\begin{figure*}
  \epsscale{1.1}
  \plottwo{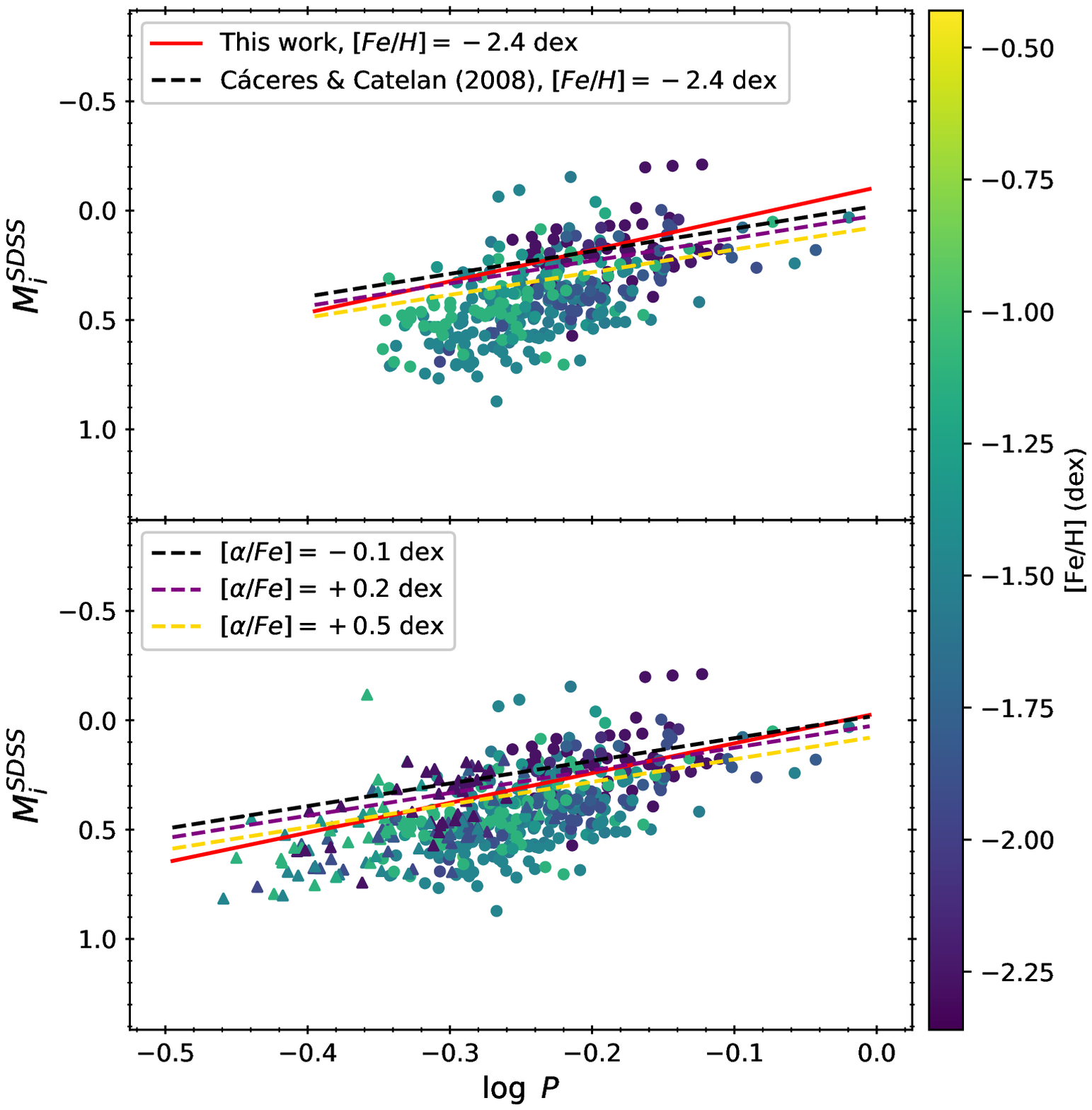}{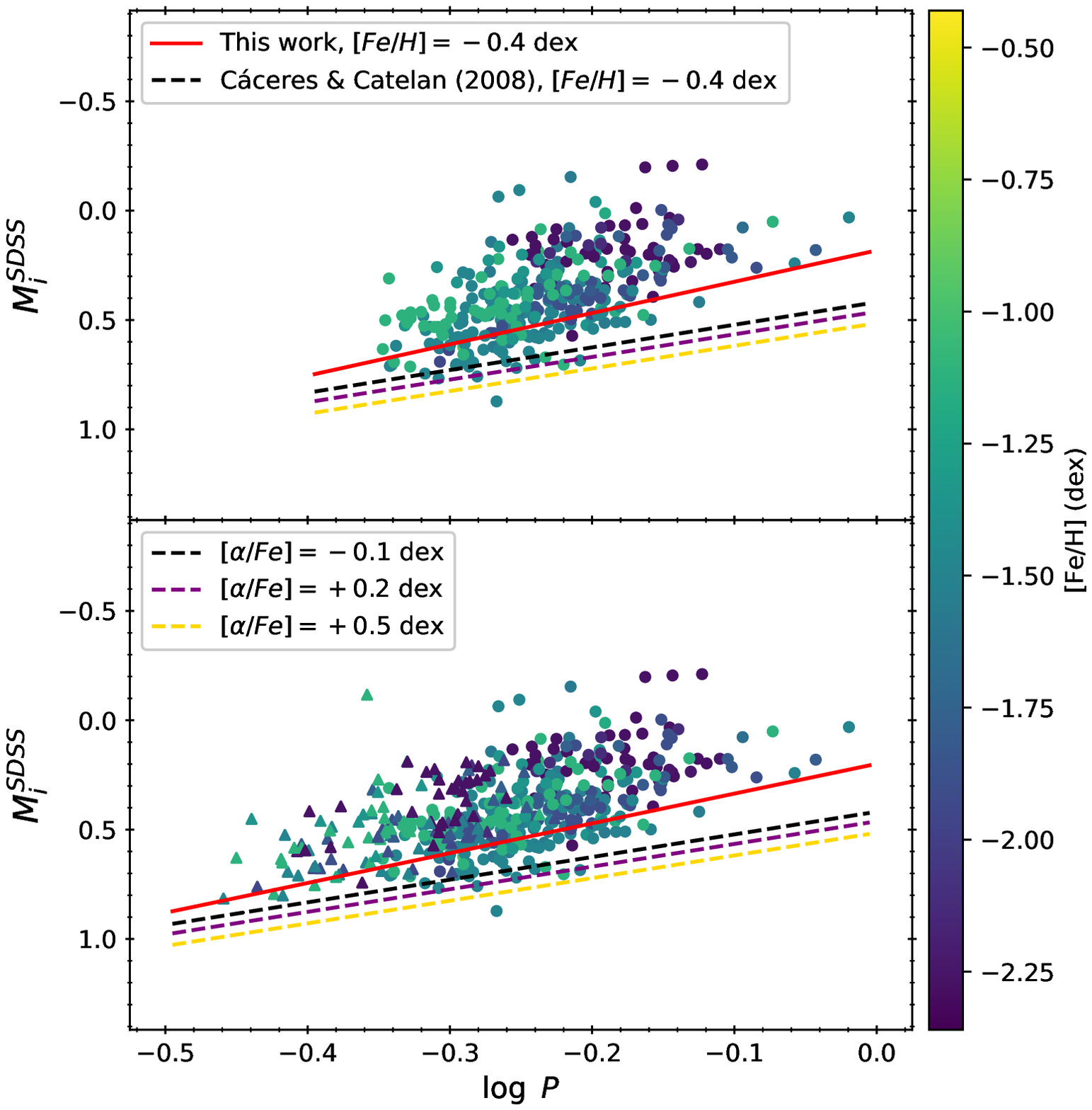}
  \caption{Comparison of the $i$-band PLZ relations given in \citet[][as dashed lines]{caceres2008} and our derived relations presented in Table \ref{tab_plpw} (in red solid lines), evaluated at $[\mathrm{Fe/H}]=-2.4$~dex (left panel) and $-0.4$~dex (right panel). For the theoretical PLZ relations, we also included the corresponding PLZ relations at three different adopted values of $[\alpha/\mathrm{Fe}]$ (see text for details). Note that the data points and our $i$-band PLZ relations have been transformed to the SDSS photometric system following the procedures described in Section \ref{sec4.1}. Top and bottom panels are for RR0 only, and RR0 and RR1 (after fundametalized the pulsation periods) combined samples, respectively.}\label{fig_cc08}
\end{figure*}

Two theoretical investigations of PLZ relations in the SDSS photometric system were presented in \citet{marconi2006} and \citet{caceres2008}. Metallicity in these theoretical investigations were expressed as $\log Z$ instead of $[\mathrm{Fe/H}]$. Therefore, a $\log Z$-$[\mathrm{Fe/H}]$ conversion was taken from \citet[][and reference therein]{caceres2008} in the form of:

\begin{eqnarray}
  \log Z & = & \log (0.638\times 10^{[\alpha/\mathrm{Fe}]} + 0.362)  \nonumber \\
         &   & + \mathrm{[Fe/H]} - 1.765.
\end{eqnarray}

\noindent  For globular clusters, $[\alpha/\mathrm{Fe}]$ varies between $\sim -0.1$~dex and $\sim +0.5$~dex \citep[based on the compilation presented in][]{pritzl2005}. Therefore, we adopted $[\alpha/\mathrm{Fe}]=-0.1$ and $+0.5$~dex, as well as the mid-point $+0.2$~dex, to construct the theoretical PLZ relation.

\begin{figure}
  \epsscale{1.1}
  \plotone{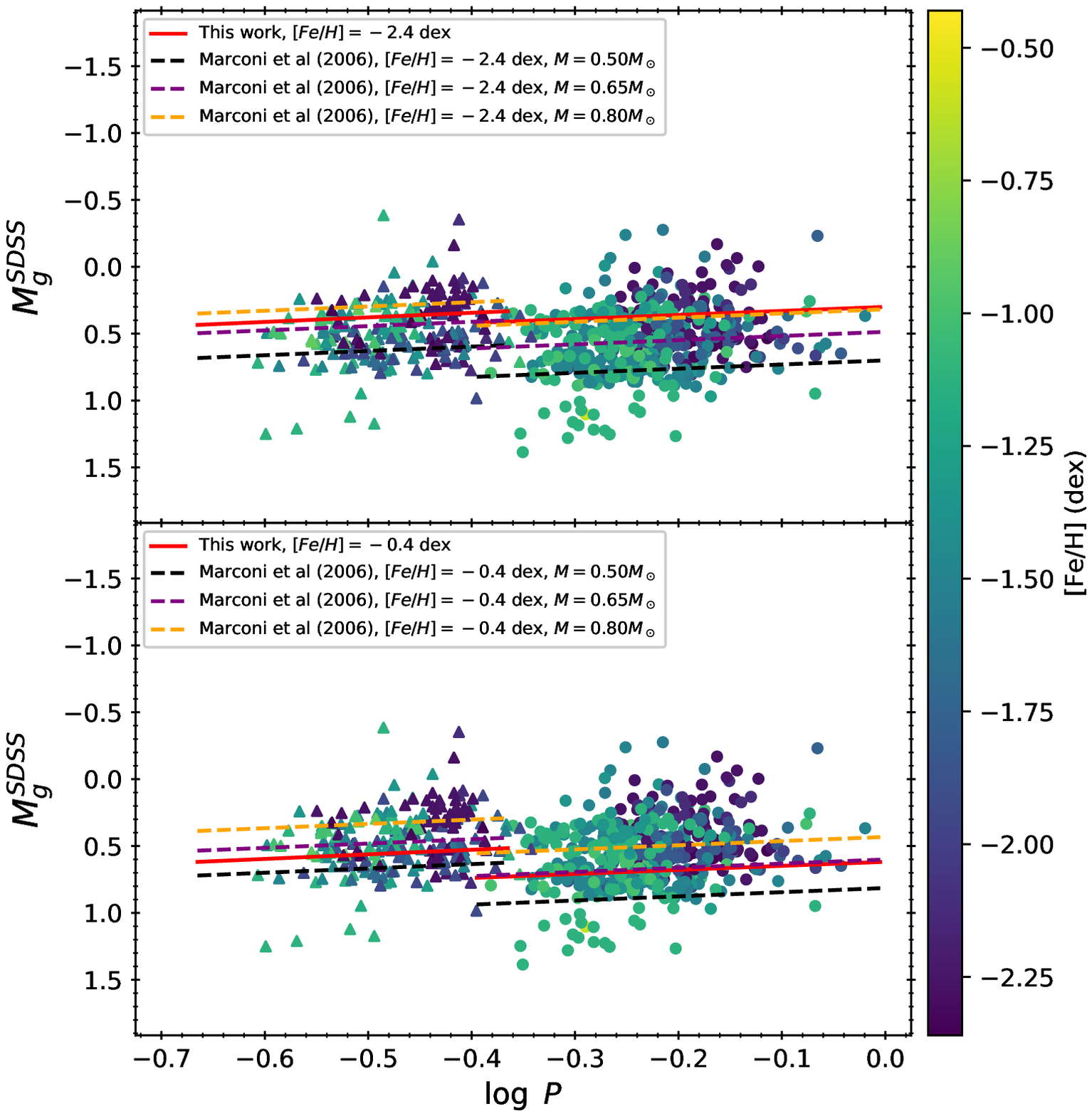}
  \caption{Comparison of the $g$-band PLZ relations given in \citet[][as dashed lines]{marconi2006} and our derived relations presented in Table \ref{tab_plpw} (in red solid lines), evaluated at $[\mathrm{Fe/H}]=-2.4$~dex (top panel) and $-0.4$~dex (bottom panel), with a fixed $[\alpha/\mathrm{Fe}]=+0.2$~dex. For the semi-theoretical PLZ relations (see text for details), we also included the corresponding PLZ relations at three different adopted values of the mass $M$, covering the expected mass range for RR Lyrae. Note that the data points and our $g$-band PLZ relations have been transformed to the SDSS photometric system following the procedures described in Section \ref{sec4.1}. Circles and triangles represent the RR0 and RR1, respectively.}\label{fig_marconi}
\end{figure}

\citet{caceres2008} provided a simple $i$-band theoretical PLZ relation in the form of $M_i = 0.908 - 1.035\log P + 0.220 \log Z$, this PLZ relation was compared to our derived PLZ relation in Figure \ref{fig_cc08} for the two adopted values of $[\mathrm{Fe/H}]$. Figure \ref{fig_cc08} revealed that slope of this theoretical $i$-band PLZ relation ($-1.035$) is shallower than the one we derived here ($-1.432\pm0.144$ for RR0, or $-1.362\pm0.093$ for RR0+RR1), or the one derived in \citet[][$-1.59\pm0.09$ for RR0]{viv2017}. Furthermore, the predicted $i$-band absolute magnitudes, using the theoretical and our derived PLZ relations, show a moderate agreement for RR Lyrae with periods between $\sim 0.5$ to $\sim 0.6$~days at low metallicity (i.e. $[\mathrm{Fe/H}]=-2.4$~dex). However, these two sets of PLZ relations are in disagreement when the metallicity is higher (as shown in right panel of Figure \ref{fig_cc08} with $[\mathrm{Fe/H}]=-0.4$~dex), due to the difference in the zero-point ($0.908$ vs. $-0.243\pm0.055$) and the metallicity term ($0.220$ vs. $0.144\pm0.018$) of the $i$-band PLZ relations in both studies. \citet{caceres2008} cautions the use of equation (2) for $Z>0.003$ \citep{vdb2000}, corresponding to $[\mathrm{Fe/H}]=-0.9$~dex (at $[\alpha/\mathrm{Fe}]=+0.2$~dex). This might contribute to the disagreement at high metallicity.

From equation (2) and Figure \ref{fig_cc08}, we note that the adopted value of $[\alpha/\mathrm{Fe}]$ only affects the zero-point of the PLZ relation. The difference on the PLZ zero-point between $[\alpha/\mathrm{Fe}]=-0.1$ and $+0.2$~dex is $c\times \Delta \log Z = -0.199c \sim \pm 0.044$~mag, and for $[\alpha/\mathrm{Fe}]=+0.2$ and $+0.5$~dex the difference is $c\times \Delta \log Z = -0.239c \sim \pm0.053$~mag, at fixed $[\mathrm{Fe/H}]$, where $c=0.220$ is the metallicity term in the theoretical $i$-band PLZ relation from \citet{caceres2008}. Hence, we adopted a single value of $[\alpha/\mathrm{Fe}]=+0.2$~dex throughout the rest of the paper.

The theoretical PLZ relations derived in \citet{marconi2006} are available in the $ug$-band, with additional terms on colors and masses. We can only compare our PLZ relations to their $g$-band PLZ relations together with the $(g-r)$ and $(r-i)$ colors. We first realized that slopes of the theoretical $g$-band PLZ relations \citep{marconi2006} are steep: $-2.87$ and $\sim -3.1$ for the RR0 and RR1, respectively. These slopes are much steeper than those presented in Table \ref{tab_plpw} or in \citet{viv2017}. However, the \citet{marconi2006} PLZ relations also included a color term, and the colors for RR Lyrae are expected to follow a PC relation \citep[for examples, see Section \ref{sec5} and][for colors in the $gri$-band]{caceres2008}. Combining the derived period-color-metallicity (PCZ) relations from Section \ref{sec5} (see Table \ref{tab_plpw}) with theoretical PLZ relations from \citet{marconi2006}, we obtained the following semi-theoretical PLZ relations\footnote{We intend to keep the two metallicity terms, $[\mathrm{Fe/H}]$ and $\log Z$, separately because $[\mathrm{Fe/H}]$ is taken from observations and $\log Z$ is based on theoretical modelings, which in principle could follow a different $\log Z$-$[\mathrm{Fe/H}]$ conversion as given in equation (2).} for the RR0:

\begin{eqnarray}
  M_g & = & 0.066 - 0.311\log P + 0.117\mathrm{[Fe/H]} -1.87\log \frac{M}{M_\odot} \nonumber \\
  &  &  -0.06\log Z + 3.35\delta_c\ \ \ \ [\mathrm{with\ (g-r)\ color}], \\
  & = & -0.097 - 0.350\log P - 0.003\mathrm{[Fe/H]} -1.94\log \frac{M}{M_\odot} \nonumber \\
  &  &  -0.024\log Z + 2.27\delta_c\ \ \ \ [\mathrm{with\ (g-i)\ color}].
\end{eqnarray}

\noindent Slopes of these semi-theoretical relations are now closer to the empirical value given in Table \ref{tab_plpw} ($-0.302\pm0.193$). Similarly, the semi-theoretical PLZ relations for the RR1 are:

\begin{eqnarray}
  M_g & = & -0.109 - 0.322\log P + 0.061\mathrm{[Fe/H]} -1.63\log \frac{M}{M_\odot} \nonumber \\
  &  &  -0.042\log Z + 3.57\delta_c\ \ \ \ [\mathrm{with\ (g-r)\ color}], \\
  & = & 0.285 + 0.536\log P + 0.033\mathrm{[Fe/H]} -1.72\log \frac{M}{M_\odot} \nonumber \\
  &  &  -0.017\log Z + 2.334\delta_c\ \ \ \ [\mathrm{with\ (g-i)\ color}].
\end{eqnarray}

\noindent Slope given in equation (5) is closer to the empirical value of $-0.342\pm0.289$. However, the slope found in equation (6) is drastically different, at which we do not have a formal explanation to account for this. Perhaps due to a much steeper slope for the RR1 $(g-i)$ PCZ relation.\footnote{To produce a slope of $\sim -0.34$ for equation (6), the slope for the RR1 $(g-i)$ PCZ relation would have to be $\sim 1.17$. The referee suggested using a relation of $(g-i)=(g-r)+(r-i)$ to account for the slope of $(g-i)$ PCZ relation. Using such a relation, the slope of equation (6) would be reduced to $0.081$, suggesting there might be some problems associated to the $i$-band RR1 data.} We will adopt equation (3) and (5) to be compared to our derived PLZ relations, because the slopes of these two semi-theoretical relations are closer to the empirical values. The $\delta_c$ in equation (3) to (6) is the correction term to covert the colors from Pan-STARRS1 photometric system to the SDSS photometric system, similar to the $\delta_m$ described in Section \ref{sec4.1}. The comparison is presented in Figure \ref{fig_marconi}.

An additional mass term, $\log M/M_\odot$, was included in equation (3) and (5). Figure \ref{fig_marconi} reveals that this mass term has a larger impact on the zero-point of the PLZ relation, in contrast to the change of $[\alpha/\mathrm{Fe}]$ that affect the zero-point of the PLZ relation by a small amount ($\lesssim 0.05$~mag). At a fixed metallicity, the $g$-band semi-theoretical PLZ relations could be in agreement or in disagreement with our derived $g$-band PLZ relations depending on the adopted mass. For example, Figure \ref{fig_marconi} shows that at $[\mathrm{Fe/H}]=-2.4$~dex, these two sets of PLZ relation for RR0 agree when the mass is $\sim 0.8M_\odot$, but not for other lower masses. Similarly at $[\mathrm{Fe/H}]=-0.4$~dex, the RR0 semi-theoretical PLZ relations with mass of $\sim 0.65M_\odot$ are closer to the empirical PLZ relations given in Table \ref{tab_plpw}.

\section{The Period-Color-Metallicity Relations and the Color-Color Diagram}\label{sec5}

\begin{figure}
  \epsscale{1.1}
  \plotone{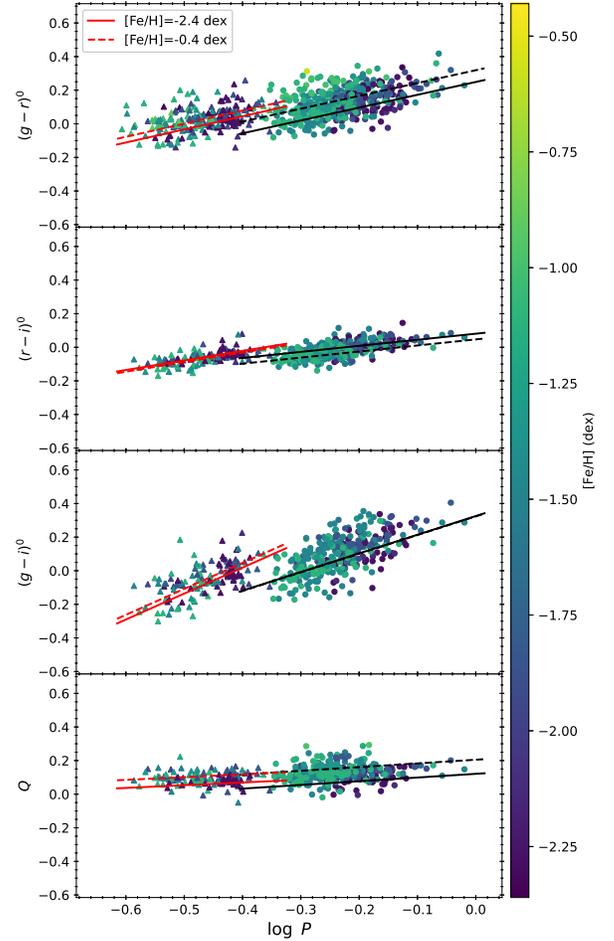}
  \caption{The extinction corrected PC relations (top three panels) and the extinction-free PQ relation (bottom panel) for our sample of RR Lyrae. The solid and dashed lines (in both red and black colors) represent the PCZ and PQZ relations evaluated at $[\mathrm{Fe/H}]=-2.4$~dex and $[\mathrm{Fe/H}]=-0.4$~dex, respectively. The relations for the combined sample of RR0 and RR1 (after fundametalized the periods) are similar, hence they are not included in this Figure.}\label{fig_pcq}
\end{figure}

\begin{figure*}
  \epsscale{1.1}
  \plotone{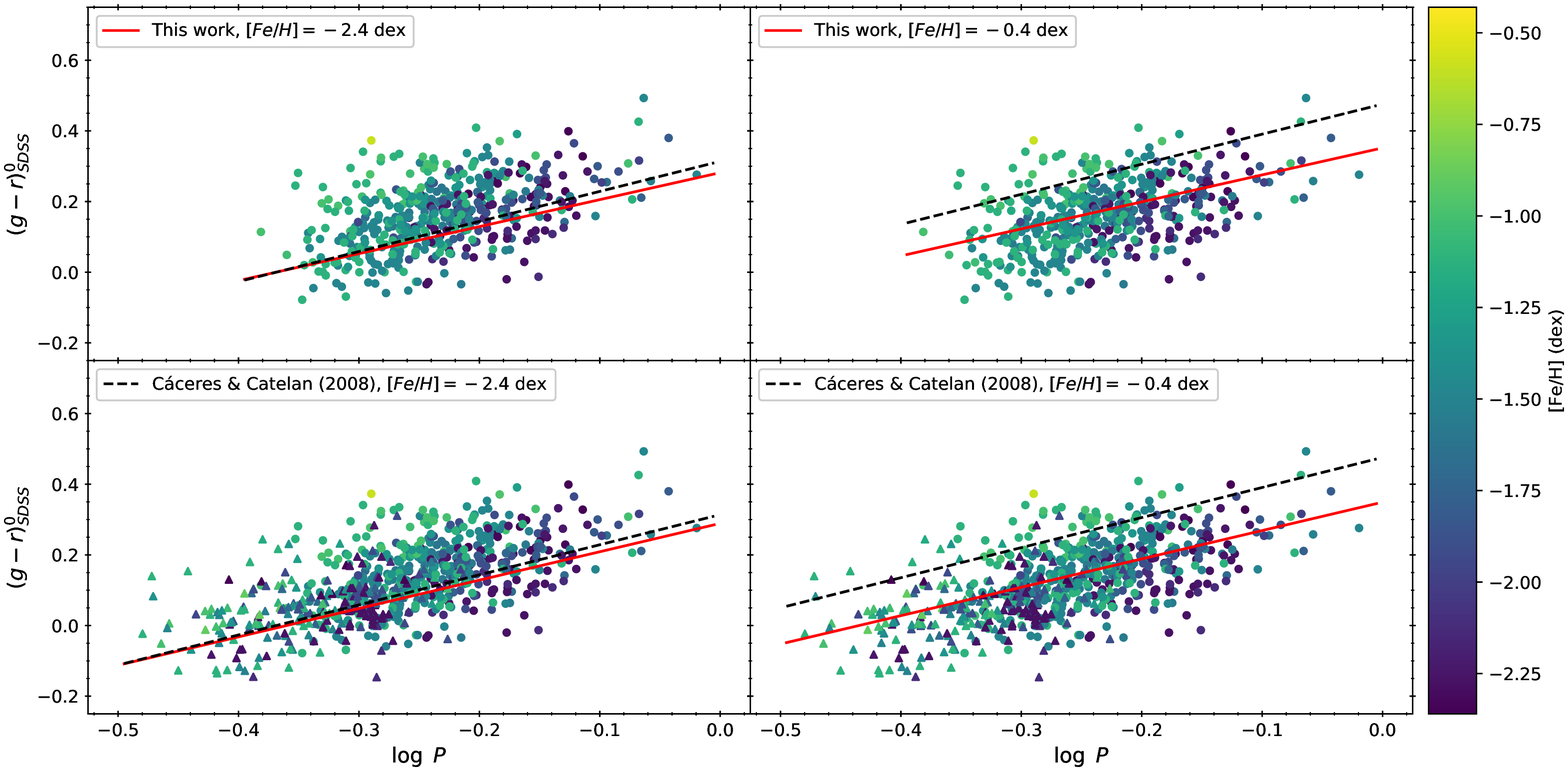}
  \plotone{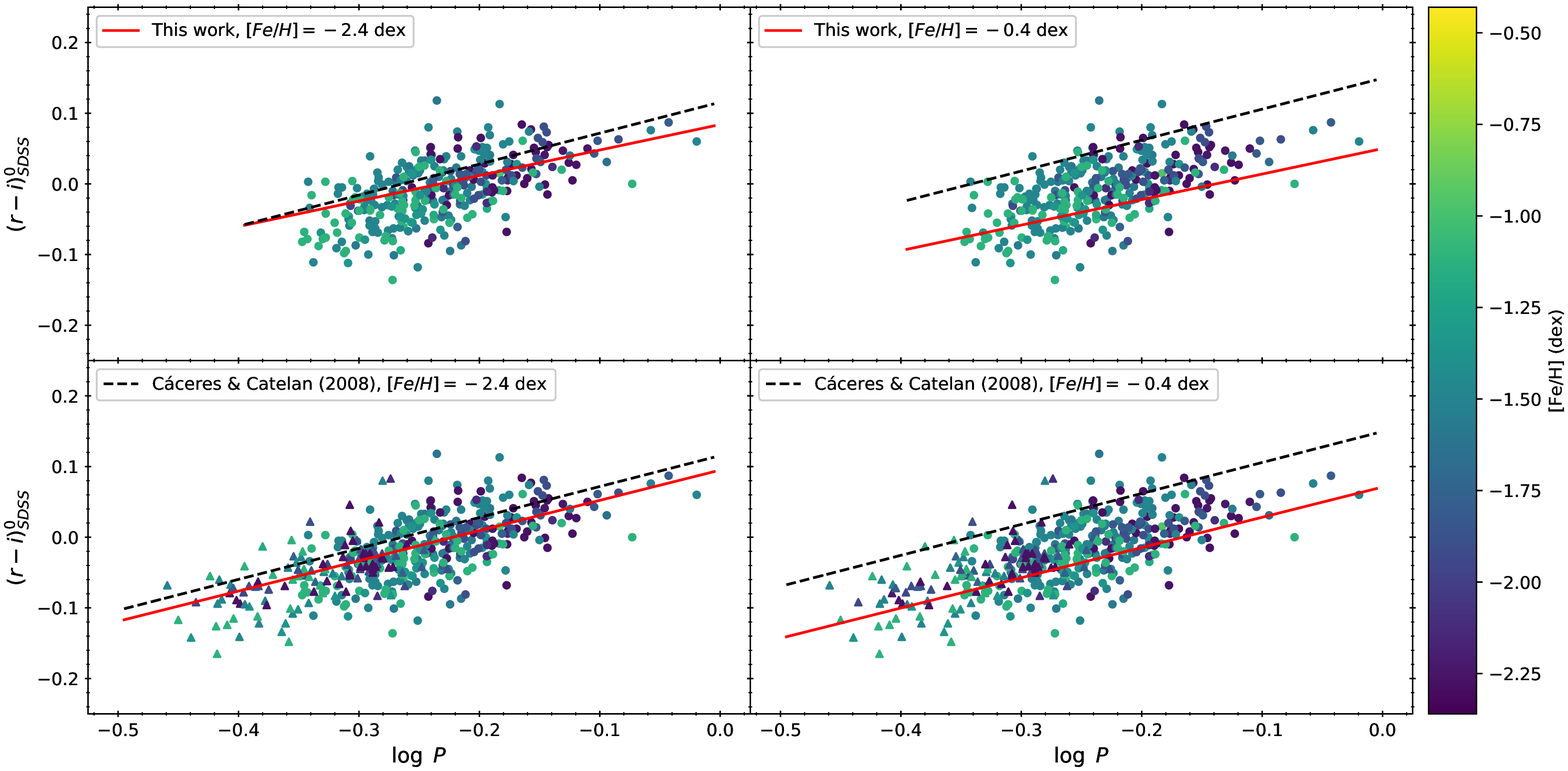}
  \caption{Comparison of the PCZ relations given in \citet[][as dashed lines]{caceres2008} and our derived relations presented in Table \ref{tab_plpw} (in red solid lines), evaluated at $[\mathrm{Fe/H}]=-2.4$~dex (left panels) and $-0.4$~dex (right panels), for the $(g-r)$ and the $(r-i)$ colors at the top and bottom panels, respectively. For the theoretical PLZ relations, we adopted a value of $[\alpha/\mathrm{Fe}]=+0.2$~dex (see text for details). Note that the data points and our PCZ relations have been transformed to the SDSS photometric system following the procedures described in Section \ref{sec4.1}. In each panels, the top and bottom sub-panels are for RR0 only, and RR0 and RR1 (after fundametalized the pulsation periods) combined samples, respectively.}\label{fig_pctheory}
\end{figure*}

\begin{figure*}
  \epsscale{1.1}
  \plotone{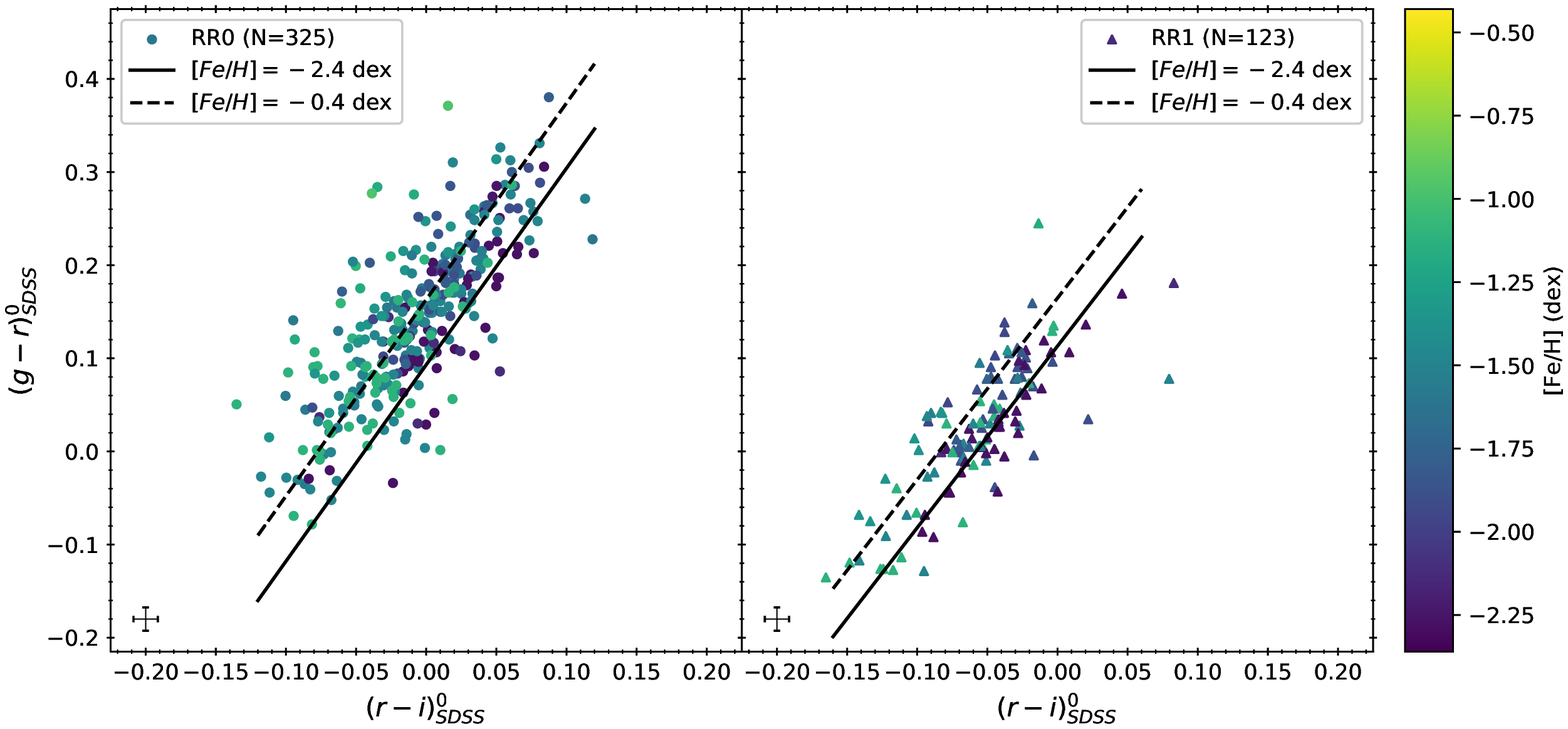}
  \caption{The extinction corrected color-color relations for RR0 (left panel) and RR1 (right panel) in our sample, where the mean magnitudes have been transformed to the SDSS system using the transformations provided in \citet{tonry2012}. Typical error bars of the colors are shown in the lower-left corners. Both the solid and dashed lines represent the theoretical color-color relations taken from \citet{marconi2006} at a given metallicity (see text for more details).}\label{fig_cc}
\end{figure*}

We have fitted the extinction-corrected PCZ relations and the extinction-free PQZ relation with our sample of RR Lyrae, using equation (1) and the same methodology as described in Section \ref{sec4}. The fitted results are summarized in Table \ref{tab_plpw} and shown in Figure \ref{fig_pcq}. We note that the metallicity terms for the PCZ and PQZ relations, the parameter $c$ in equation (1), in general are smaller than the PLZ and PWZ relations (except the $W^{gr}$ PWZ relations, see Section \ref{sec4}). For example, the metallicity term in $(r-i)$ PCZ relation for RR1 is consistent with zero. The $(g-i)$ PCZ relations have the steepest slopes and almost vanishing metallicity terms, for either RR0, RR1, or the combined sample of both. Dispersions of the PCZ and PQZ relations are also smaller, with $(r-i)$ PCZ relations have the smallest dispersion in all of the fitted relations.

Similar to the theoretical $i$-band PLZ relation, \citet{caceres2008} provided the theoretical PCZ relations in the $(g-r)$ and the $(r-i)$ colors. We compared their PCZ relations with those listed in Table \ref{tab_plpw}, evaluated at two metallicities, as shown in Figure \ref{fig_pctheory}. These theoretical PCZ relations behave similar to the $i$-band PLZ relation (shown in Figure \ref{fig_cc08}, see Section \ref{sec4.2}), such that they are in moderate agreement with our empirical relations at low metallicity, and in disagreement when the metallicity is high.

In addition to theoretical PLZ relations (see Section \ref{sec4.2}), \citet{marconi2006} has also derived the metallicity-dependent color-color relations based on a series of pulsation models. In Figure \ref{fig_cc}, these theoretical color-color relations were shown as solid and dashed lines for the two adopted metallicity (see Section \ref{sec4.2} for more details), together with the extinction-corrected colors for our sample of RR Lyrae. It can be seen from Figure \ref{fig_cc} that the theoretical color-color relations trace a relatively narrow ``stripe'' in the color-color diagram, while the observed RR Lyrae show a much larger scatter along this stripe. As discussed in \citet{marconi2006}, several sources (both intrinsic and observational), in addition to metallicity, may contribute to the scatter seen in the color-color relations. This imply that the $(g-r)$ and $(r-i)$ color-color diagram is not a good diagnostic to discriminate metallicity for RR Lyrae.

\section{An Example of Application: Distance to Dwarf Galaxy Crater II} \label{sec6}

Based on DECam observations, \citet{vivas2020} identified 83 and 5 RR0 and RR1, respectively, in a dwarf galaxy Crater II. There is one RR0 \citep[V24 identified in][]{joo2018}, however, missed in \citet{vivas2020} because it is located outside the footprint of DECam observations. This set of RR Lyrae provides an opportunity to compare and test various PL(Z) relations in a differential way. These include the empirical PL(Z) relations from \citet{ses2017}, \citet{viv2017}, \citet[][by adopting the distance to M15 as listed in Table \ref{tab_gc}]{bhardwaj2021}, and our results presented in Table \ref{tab_plpw}. We excluded the theoretical $g$-band PLZ relation from \citet{marconi2006} because of the mass-dependency, which has a significant impact on the zero-point of the PLZ relation (see discussion in Section \ref{sec4.2}). We also excluded the theoretical $i$-band PLZ relation from \citet{caceres2008}, because \citet{vivas2020} applied this relation to the RR Lyrae in Crater II, and found a distance modulus of $20.333\pm0.004$~mag to this dwarf galaxy, by adopting $[\mathrm{Fe/H}]=-2.0$~dex and $[\alpha/\mathrm{Fe}]=+0.3$~dex.

Since the mean magnitudes of the observed RR Lyrae are available in the SDSS $gi$-band \citep{vivas2020} and lack of $r$-band data, same as in the case of \citet{bhardwaj2021}, making the transformation between various photometric systems non-trivial. We first transformed our RR Lyrae in the globular clusters to the SDSS photometric system, and a subset of these RR Lyrae with the same mean $(g-i)$ colors as the RR Lyrae in Crater II were used to determine the correction term, $\delta_m$, for the PLZ relations based on Pan-STARRS1 photometric system. These correction terms were found to be $0.040$~mag and $0.029$~mag in the $g$-band, and $-0.003$~mag and $-0.004$~mag in the $i$-band, where the first and the second numbers are for RR0 and RR1, respectively. In Table \ref{tab_craterII_a}, we summarized the derived distance moduli ($\mu$) using the four mentioned empirical PL(Z) relations at $[\mathrm{Fe/H}]=-2.0$~dex, separately in the $g$- and $i$-band, while keeping the periods, mean magnitudes, and extinctions the same when fitting the PLZ relations to these RR Lyrae in Crater II. 

\begin{deluxetable*}{lccccccccc}
  \tabletypesize{\scriptsize}
  \tablecaption{Distance moduli to Crater II derived with various PL(Z) relations, using $[\mathrm{Fe/H}]=-2.0$~dex.\label{tab_craterII_a}}
  \tablewidth{0pt}
  \tablehead{
    \colhead{PLZ Relations} & \multicolumn{3}{c}{RR0} & \multicolumn{3}{c}{RR1} & \multicolumn{3}{c}{RR0+RR1}  \\
    \colhead{} & \colhead{$\mu_g$} & \colhead{$\mu_i$} & \colhead{$\Delta \mu_{gi}$} & \colhead{$\mu_g$} & \colhead{$\mu_i$} & \colhead{$\Delta \mu_{gi}$} & \colhead{$\mu_g$} & \colhead{$\mu_i$} & \colhead{$\Delta \mu_{gi}$} 
  }
  \startdata
  This work (Table \ref{tab_plpw}) & $20.488\pm0.006$ & $20.427\pm0.004$ & 0.061 & $20.465\pm0.016$ & $20.406\pm0.014$ & 0.059 & $20.450\pm0.006$ & $20.375\pm0.004$ & 0.075 \\
  \citet{viv2017}                  & $20.374\pm0.006$ & $20.277\pm0.005$ & 0.097 & $20.396\pm0.017$ & $20.287\pm0.014$ & 0.109 & $\cdots$ & $\cdots$  & $\cdots$ \\
  \citet{ses2017}                  & $20.262\pm0.008$ & $20.287\pm0.005$ & -0.025 & $\cdots$ & $\cdots$  & $\cdots$ & $\cdots$ & $\cdots$  & $\cdots$ \\
  \citet{bhardwaj2021}\tablenotemark{a} & $20.494\pm0.006$  & $20.452\pm0.004$  & 0.042 & $20.475\pm0.017$ & $20.433\pm0.015$   & 0.042 & $20.506\pm0.006$  & $20.455\pm0.004$ & 0.051  \\
  \enddata
  \tablenotetext{a}{Assuming the distance to M15 is $10.71$~kpc, as listed in Table \ref{tab_gc}.}
  \tablecomments{$\Delta \mu_{gi} =\mu_g-\mu_i$ is the difference of the distance moduli in the $g$- and $i$-band. Errors on each of the distance moduli are random errors only.}
\end{deluxetable*}

\begin{deluxetable*}{lccccccccc}
  \tabletypesize{\scriptsize}
  \tablecaption{Same as Table \ref{tab_craterII_a}, but with reddening values from the {\tt Bayerstar2019} 3D reddening map.\label{tab_craterII_b}}
  \tablewidth{0pt}
  \tablehead{
    \colhead{PLZ Relations} & \multicolumn{3}{c}{RR0} & \multicolumn{3}{c}{RR1} & \multicolumn{3}{c}{RR0+RR1}  \\
    \colhead{} & \colhead{$\mu_g$} & \colhead{$\mu_i$} & \colhead{$\Delta \mu_{gi}$} & \colhead{$\mu_g$} & \colhead{$\mu_i$} & \colhead{$\Delta \mu_{gi}$} & \colhead{$\mu_g$} & \colhead{$\mu_i$} & \colhead{$\Delta \mu_{gi}$} 
  }
  \startdata
  This work (Table \ref{tab_plpw}) & $20.426\pm0.007$ & $20.395\pm0.005$ & 0.031 & $20.393\pm0.016$ & $20.369\pm0.012$ & 0.024 & $20.387\pm0.007$ & $20.343\pm0.005$ & 0.044 \\
  \citet{viv2017}                  & $20.312\pm0.008$ & $20.245\pm0.005$ & 0.067 & $20.325\pm0.014$ & $20.250\pm0.013$ & 0.075 & $\cdots$ & $\cdots$  & $\cdots$ \\
  \citet{ses2017}                  & $20.200\pm0.009$ & $20.255\pm0.005$ &-0.055 & $\cdots$ & $\cdots$  & $\cdots$ & $\cdots$ & $\cdots$  & $\cdots$ \\
  \citet{bhardwaj2021}\tablenotemark{a} & $20.432\pm0.007$  & $20.420\pm0.005$  & 0.012 & $20.404\pm0.019$ & $20.396\pm0.015$ & 0.008 & $20.444\pm0.007$  & $20.423\pm0.005$ & 0.021 \\
  \enddata
  \tablenotetext{a}{Assuming the distance to M15 is $10.71$~kpc, as listed in Table \ref{tab_gc}.}
  \tablecomments{$\Delta \mu_{gi}=\mu_g-\mu_i$ is the difference of the distance moduli in the $g$- and $i$-band. Errors on each of the distance moduli are random errors only.}
\end{deluxetable*}

Table \ref{tab_craterII_a} reveals that the derived distance moduli are in general larger than $20.333$~mag as determined in \citet{vivas2020}, except the \citet{ses2017} PLZ relations and the $i$-band PL relations from \citet{viv2017}. To bring the distance moduli fitted from other PL(Z) relations closer to $20.333$~mag, the metallicity of Crater II has to increase. For the two sets of PL relations without the metallicity term, we note that \citet{bhardwaj2021} PL relations give a larger distance modulus than the PL relations derived in \citet{viv2017}. The \citet{bhardwaj2021} and \citet{viv2017} PL relations were derived using RR Lyrae in M15 and M5, respectively, at which is M15 is more metal-poor than M5, hence the \citet{bhardwaj2021} PL relations should be favored to be applied to RR Lyrae in Crater II (assuming its metallicity of $[\mathrm{Fe/H}]=-2.0$~dex is correct). Therefore, either the true distance modulus to Crater II is larger (than $20.333$~mag) or its metallicity is higher (than $-2.0$~dex), or both. 

Since the mean magnitudes have been corrected for extinction, we expect the $g$- and $i$-band distance moduli should be roughly the same, or $\Delta \mu_{gi}=\mu_g-\mu_i \sim 0$. Except for distance moduli fitted with \citet{ses2017} PLZ relations (presumably due to the problem in $g$-band, see Section \ref{sec4.1}), we found that the $g$-band distance moduli are larger than their counterparts in the $i$-band (Table \ref{tab_craterII_a}). One possibility is the slight incorrect estimation of extinction adopted in \citet{viv2017}. Instead of using the $E(B-V)$ values provided in \citet{viv2017}, we use the same {\tt Bayerstar2019} 3D reddening map as described in Section \ref{sec2.2} to correct the extinction for the $gi$-band mean magnitudes. The fitted distance moduli are given in Table \ref{tab_craterII_b}. Again with the exception of distance moduli based on \citet{ses2017} PLZ relations, now the $gi$-band distance moduli are closer to each others. \citet{viv2017} PL relations give the largest $\Delta \mu_{gi}$, indicating an extra metallicity term is needed to be applied to the derived distance moduli, because the \citet{viv2017} PL relations are based on RR Lyrae in M5, which has a higher metallicity than the assumed metallicity for Crater II. We note that $\Delta \mu_{gi}$ from using our PLZ relations (Table \ref{tab_plpw}) are larger than those using the PL relations from \citet{bhardwaj2021}, possibly due to the photometric transformation because without the photometric transformation, the $\Delta \mu_{gi}$ using our PLZ relations are even larger. Since by definition the difference of the distance moduli in two filters is also a measure of extinction \citep[for examples, see][]{kelson1996,turner1998,freedman2001}, and the foreground extinctions have been corrected using the {\tt Bayerstar2019} 3D reddening map, we suggest the excess of $\Delta \mu_{gi}$ is due to the internal extinction of Crater II, that is $\Delta_{gi} = (\mu_g + A_g) - (\mu_i + A_i) = E(g-i)\sim 0.03$ or $\sim 0.01$~mag depending on the adopted PL(Z) relations given in Table \ref{tab_plpw} or \citet{bhardwaj2021}.\footnote{For a comparison, the {\tt Bayerstar2019} interactive 3D reddening map returned a foreground extinction of $E(g-r)=0.06^{+0.02}_{-0.01}$~mag for Crater II.}

Since there should only have one distance modulus ($\mu_0$) to Crater II, it is possible to simultaneously fit four PLZ relations, the $gi$-band PLZ relations for both RR0 and RR1, to the data by defining the following merit function:

\begin{eqnarray}
  \chi^2 & = & \sum_{j=1}^{N_{\lambda,\psi}} \sum_{\lambda=\{g,i\}} \sum_{\psi=\{0,1\}} \frac{(m^j_{\lambda,\psi} - M^j_{\lambda,\psi} - \mu_0)^2}{\sigma^2_{\lambda,\psi}} \\
  \mathrm{where} & & \ M^j_{\lambda,\psi} = a_{\lambda,\psi} + b_{\lambda,\psi}\log P^j_{\lambda,\psi} + c_{\lambda,\psi} \mathrm{[Fe/H]}. \nonumber
\end{eqnarray}

\noindent In above equation, $\lambda$ represents the filters (either $g$ or $i$), and $\psi$ represents pulsation modes (either $0$ or $1$). Parameters $a$, $b$, and $c$ are same as the PLZ relation defined in equation (1), and $\sigma$ is the dispersion of the respected PLZ relation. Adopting $\mathrm{[Fe/H]}=-2.0$~dex and using the values given in Table \ref{tab_plpw} for our derived PLZ relations, we obtained $\mu_0=20.361\pm0.004$~mag (random error only) after including the additional correction of internal extinction of Crater II, $E(g-i)\sim 0.03$~mag, to the data. Allowing $\mathrm{[Fe/H]}$ as another free parameter to be fitted, we derived $\mu_0=20.327\pm0.100$~mag and $\mathrm{[Fe/H]}=-1.76\pm0.69$~dex using the ordinary least squares (OLS) regression. As a sanity check, we have also derived the distance modulus to Crater II using the $W^{gi}$ PWZ relations, evaluated at $\mathrm{[Fe/H]}=-2.0$~dex. After transforming our PWZ relations to SDSS photometric system, and using $W^{gi}_{SDSS} = g_{SDSS} - 2.058(g_{SDSS}-i_{SDSS})$, we obtained $\mu_0 = 20.354\pm0.004$~mag (random error only) by simultaneous fitting the RR0 and RR1 sample with equation (7).\footnote{If only using the RR0 or RR1 sample, then the resulted distance modulus is $20.353\pm0.004$~mag and $20.355\pm0.016$~mag (random error only), respectively, from the $W^{gi}$ PWZ relations.} Since multiple transformations were involved for the Wesenheit index $W^{gi}$, we adopted our derived distance modulus based on the PLZ relations instead of the PWZ relations.

\begin{figure}
  \epsscale{1.15}
  \plotone{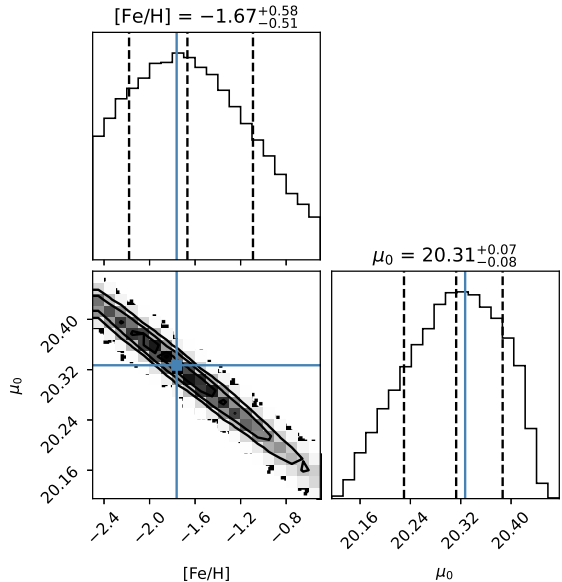}
  \caption{The corner plot based on the {\tt emcee} sampling on the posterior distributions for $\mu_0$ and $\mathrm{[Fe/H]}$, the sampling was run with 128 walkers for 10000 steps. The vertical dashed lines on the histograms represent the $16^{\mathrm{th}}$, $50^{\mathrm{th}}$, and $84^{\mathrm{th}}$ percentiles of the distributions. The blue lines marked the results obtained from the OLS regression.}\label{fig_bayes}
\end{figure}

We recall that in general a more metal-poor system would imply a larger distance modulus due to the positive metallicity term, the parameter $c$ in equation (1), in the PLZ relation. Our results are consistent with this expectation ($\mu_0=20.361$~mag at $\mathrm{[Fe/H]}=-2.0$~dex vs. $\mu_0=20.327$~mag at $\mathrm{[Fe/H]}=-1.76$~dex). Clearly, there is a degeneracy, or correlation, between $\mu_0$ and $\mathrm{[Fe/H]}$. Therefore, instead of using OLS regression, we fit equation (7) using technique of Bayesian linear regression to the data. We adopted a flat (uniform) prior on both $\mu_0$ (between 10 and 30) and $\mathrm{[Fe/H]}$ (between $-2.5$ and $-0.5$), and the affine-invariant Markov Chain Monte Carlo sampler package {\tt emcee}\footnote{\url{https://emcee.readthedocs.io/en/stable/}} \citep{fm2013} was used to sample the posterior distribution. The result is shown in Figure \ref{fig_bayes}, produced from the {\tt corner} package\footnote{\url{https://corner.readthedocs.io/en/latest/}} \citep{fm2016}, at which the (anti-)correlation of these two parameters can be clearly seen. By adopting median as the estimator of the parameters, we found the following best-fit values: $\mu_0= 20.313_{-0.083}^{+0.074}$~mag, and $\mathrm{[Fe/H]}=-1.67_{-0.51}^{+0.58}$~dex, where the error-bars represent the $16^{\mathrm{th}}$ and $84^{\mathrm{th}}$ percentiles of the distributions.

Nevertheless, metallicity of Crater II has been measured to be around $-2.0$~dex from multiple studies \citep{torrealba2016,caldwell2017,fu2019,walker2019}, and unlikely to have a higher metallicity around $-1.7$~dex. Therefore, the true distance modulus of Crater II should be larger than $20.333$~mag as reported in \citet{vivas2020}.

\section{Conclusion}\label{sec7}

In this work, we derived the $gri$-band PLZ and PWZ relations, in the Pan-STARRS1 photometric system, based on $\sim 750$ RR Lyrae located in 46 globular clusters. These PLZ and PWZ relations were derived as homogeneous as possible, such as only using the light curves data from ZTF, correcting the extinction using the same {\tt Bayerstar2019} 3D reddening map, and adopted the same sources for the metallicity and distances from the GOTHAM survey and \citet{baumgardt2021}, respectively. Table \ref{tab_plpw} presents the fitted results for these PLZ and PWZ relations. Other results obtained in this work are summarized as follow.

\begin{enumerate}
\item Based on our simulations, we found that the derived mean magnitudes may not be reliable or accurate with template light curve fitting approach when the light curve has a smaller number of data points. We recommend only fitting the observed light curves with 10 or more data points for deriving the mean magnitudes. We have also derived an empirical fitting equation to estimate the errors on the fitted mean magnitudes.

\item Given that blending is unavoidable for RR Lyrae located in globular clusters, we applied several selection criteria to exclude RR Lyrae that might affected by blending, as well as other RR Lyrae with problematic light curves. These selection criteria utilized information based on amplitudes, amplitudes ratios, colors, and/or residuals from the PL and PW relations. In case of the PL/PW residuals, we found there are indications that for RR Lyrae with brighter mean magnitudes (than expected) they also tend to have smaller amplitudes, strongly indicating the presence of blending.

\item We notice the $W^{gr}$ PWZ relations exhibit very weak metallicity dependence, similar to the theoretical PW relations constructed in $BVR$-band \citep{marconi2015}. We also notice the PLZ and PWZ relations involving $i$-band show the smallest dispersions, therefore increasing the $i$-band ZTF observations in the near future would be beneficial.

\item Compared to the literature PL(Z) relations, we found that our PLZ relations disagree with those presented in \citet[][especially in the $g$-band]{ses2017}, but in very good agreement with the PL relations derived in \citet{viv2017}. In comparison to \citet{bhardwaj2021} $gi$-band PL relations, we found a fairly good agreement, although our PL(Z) relations are fainter in the $i$-band for the RR1 and RR0+RR1 sample.

\item The theoretical $i$-band PLZ relation derived in \citet{caceres2008} was comparable to our derived PLZ relation at low metallicity ($\mathrm{[Fe/H]}=-2.4$~dex). However, disagreement was found for these two sets of $i$-band PLZ relation at high metallicity ($\mathrm{[Fe/H]}=-0.4$~dex), presumably due to invalid $\log Z$-$\mathrm{[Fe/H]}$ conversion at high metallicity.
  
\item The comparison to the theoretical PLZ relations presented in \citet{marconi2006} was more complicated due to the inclusion of colors and mass terms in the theoretical relations. To eliminate the color term, we substituted our derived PCZ relations to the theoretical relations, and obtained semi-theoretical period-luminosity-metallicity-mass relations. The mass term in these relations has a non-negligible impact on the zero-point of the corresponding PLZ relation.

\item In addition to PLZ and PWZ relations, we have also derived the PCZ and PQZ relations with our sample of RR Lyrae in globular clusters. Again, our empirical PCZ relations are in moderate agreement with the theoretical PCZ relations \citep{caceres2008} at low metallicity but not at the high metallicity, probably due to same reason(s) as in the case of $i$-band PLZ relation.

\item The empirical color-color diagram shows a larger scatter than the theoretical color-color relations at various metallicities. This implies the $(g-r)$ and $(r-i)$ color-color diagram is not a good diagnostic for determining the metallicity for RR Lyrae.

\item We applied the empirical PL(Z) relations to RR Lyrae found in a dwarf galaxy Crater II. Using published data, we found that the derived distance modulus to Crater II is in general larger than $20.333$~mag \citep{vivas2020}. Further analysis revealed that Crater II needs to be more metal-rich in order to bring the distance modulus derived using our PLZ relations closer to the published values, or Crater II is indeed located in a further distance. Based on our analysis, we suggested Crater II may have internal extinction of $E(g-i)\sim 0.01$ or $\sim 0.03$~mag.
  
\end{enumerate}

Our derived PLZ and PWZ relations in $gri$-band would be beneficial for various distance scale applications. For examples, precise distance can be derived to faint RR Lyrae located in outer Galactic halo or newly discovered (ultra-faint or diffuse) dwarf galaxies, such as those discovered from the HyperSuprime-Cam Subaru Strategic Program \citep[HSC-SSP,][]{aihara2018} in northern sky and the Vera Rubin Observatory Legacy Survey of Space and Time \citep[LSST,][]{lsst2019} in southern sky, as well as other similar optical surveys. This in turn is of great interest for the study of sub-structure in our Galactic halo \citep[for examples, see][]{lancester2019}. At present RR Lyrae observations are possible up to $\sim2$ Mpc distance \citep{dac2010}, just slightly beyond the Local Group, and we anticipated new dwarf galaxies will be routinely discovered in various optical surveys (such as HSC-SSP and/or LSST), an example is Eridanus IV recently reported in \citet{cerny2021}.

\acknowledgments

We thank the useful discussions and comments from an anonymous referee to improve the manuscript, and H. Baumgardt regarding the distance to the globular clusters. We are thankful for funding from the Ministry of Science and Technology (Taiwan) under the contract 107-2119-M-008-014-MY2, 107-2119-M-008-012, 108-2628-M-007-005-RSP and 109-2112-M-008-014-MY3.

Based on observations (prior to Dec 1, 2020) obtained with the Samuel Oschin Telescope 48-inch and the 60-inch Telescope at the Palomar Observatory as part of the Zwicky Transient Facility project. ZTF is supported by the National Science Foundation under Grant No. AST-1440341 and a collaboration including Caltech, IPAC, the Weizmann Institute for Science, the Oskar Klein Center at Stockholm University, the University of Maryland, the University of Washington, Deutsches Elektronen-Synchrotron and Humboldt University, Los Alamos National Laboratories, the TANGO Consortium of Taiwan, the University of Wisconsin at Milwaukee, and Lawrence Berkeley National Laboratories. Operations are conducted by COO, IPAC, and UW.

Based on observations (after Dec 1, 2020) obtained with the Samuel Oschin Telescope 48-inch and the 60-inch Telescope at the Palomar Observatory as part of the Zwicky Transient Facility project. ZTF is supported by the National Science Foundation under Grant No. AST-2034437 and a collaboration including Caltech, IPAC, the Weizmann Institute for Science, the Oskar Klein Center at Stockholm University, the University of Maryland, Deutsches Elektronen-Synchrotron and Humboldt University, the TANGO Consortium of Taiwan, the University of Wisconsin at Milwaukee, Trinity College Dublin, Lawrence Livermore National Laboratories, and IN2P3, France. Operations are conducted by COO, IPAC, and UW.

This research has made use of the SIMBAD database and the VizieR catalogue access tool, operated at CDS, Strasbourg, France. This research made use of Astropy,\footnote{\url{http://www.astropy.org}} a community-developed core Python package for Astronomy \citep{astropy2013, astropy2018}.

\facility{PO:1.2m}

\software{{\tt astropy} \citep{astropy2013,astropy2018},  {\tt dustmaps} \citep{green2018}, {\tt gatspy} \citep{vdp2015}, {\tt Matplotlib} \citep{hunter2007},  {\tt NumPy} \citep{harris2020}, {\tt SciPy} \citep{virtanen2020}, {\tt boundfit} \citep{cardiel2009}, {\tt corner} \citep{fm2016}, {\tt emcee} \citep{fm2013}.}

\newpage



\begin{thebibliography}{} 

\bibitem[Abbott et al.(2021)]{des2021} Abbott, T.~M.~C., Adam{\'o}w, M., Aguena, M., et al.\ 2021, \apjs, 255, 20

\bibitem[Aihara et al.(2018)]{aihara2018} Aihara, H., Arimoto, N., Armstrong, R., et al.\ 2018, \pasj, 70, S4
  
\bibitem[Astropy Collaboration et al.(2013)]{astropy2013} Astropy Collaboration, Robitaille, T.~P., Tollerud, E.~J., et al.\ 2013, \aap, 558, A33

\bibitem[Astropy Collaboration et al.(2018)]{astropy2018} Astropy Collaboration, Price-Whelan, A.~M., Sip{\H{o}}cz, B.~M., et al.\ 2018, \aj, 156, 123

\bibitem[Baumgardt \& Vasiliev(2021)]{baumgardt2021} Baumgardt, H. \& Vasiliev, E.\ 2021, \mnras, 505, 5957
  
\bibitem[Beaton et al.(2018)]{beaton2018} Beaton, R.~L., Bono, G., Braga, V.~F., et al.\ 2018, \ssr, 214, 113
  
\bibitem[Bellm \& Kulkarni(2017)]{bellm2017} Bellm, E. \& Kulkarni, S.\ 2017, Nature Astronomy, 1, 0071
  
\bibitem[Bellm et al.(2019)]{bel19} Bellm, E.~C., Kulkarni, S.~R., Graham, M.~J., et al.\ 2019, \pasp, 131, 018002

\bibitem[Benk{\H{o}} et al.(2006)]{ben2006} Benk{\H{o}}, J.~M., Bakos, G. {\'A}., \& Nuspl, J.\ 2006, \mnras, 372, 1657
  
\bibitem[Bhardwaj(2020)]{bhardwaj2020} Bhardwaj, A.\ 2020, Journal of Astrophysics and Astronomy, 41, 23

\bibitem[Bhardwaj(2022)]{bhardwaj2022} Bhardwaj, A.~B.\ 2022, Universe, 8, 122

\bibitem[Bhardwaj et al.(2016)]{bhardwaj2016} Bhardwaj, A., Kanbur, S.~M., Macri, L.~M., et al.\ 2016, \aj, 151, 88

\bibitem[Bhardwaj et al.(2020)]{bhardwaj2020a} Bhardwaj, A., Rejkuba, M., de Grijs, R., et al.\ 2020, \aj, 160, 220
  
\bibitem[Bhardwaj et al.(2021)]{bhardwaj2021} Bhardwaj, A., Rejkuba, M., Sloan, G.~C., et al.\ 2021, \apj, 922, 20

\bibitem[Bono(2003)]{bono2003} Bono, G.\ 2003, Stellar Candles for the Extragalactic Distance Scale, edited by D. Alloin \& W. Gieren, Lecture Notes in Physics, 635:85

\bibitem[Bono et al.(2001)]{bono2001} Bono, G., Caputo, F., Castellani, V., et al.\ 2001, \mnras, 326, 1183
  
\bibitem[Bono et al.(2003)]{bono2003a} Bono, G., Caputo, F., Castellani, V., et al.\ 2003, \mnras, 344, 1097

\bibitem[Bono et al.(2016)]{bono2016} Bono, G., Braga, V.~F., Pietrinferni, A., et al.\ 2016, \memsai, 87, 358

\bibitem[Borissova et al.(2007)]{bor2007} Borissova, J., Ivanov, V.~D., Stephens, A.~W., et al.\ 2007, \aap, 474, 121

\bibitem[Braga et al.(2016)]{braga2016} Braga, V.~F., Stetson, P.~B., Bono, G., et al.\ 2016, \aj, 152

\bibitem[Caldwell et al.(2017)]{caldwell2017} Caldwell, N., Walker, M.~G., Mateo, M., et al.\ 2017, \apj, 839, 20
  
\bibitem[Cardiel(2009)]{cardiel2009} Cardiel, N.\ 2009, \mnras, 396, 680

  
\bibitem[Catelan et al.(2004)]{catelan2004} Catelan, M., Pritzl, B.~J., \& Smith, H.~A.\ 2004, \apjs, 154, 633

\bibitem[C{\'a}ceres \& Catelan(2008)]{caceres2008} C{\'a}ceres, C. \& Catelan, M.\ 2008, \apjs, 179, 242

\bibitem[Cerny et al.(2021)]{cerny2021} Cerny, W., Pace, A.~B., Drlica-Wagner, A., et al.\ 2021, \apjl, 920, L44

\bibitem[Chambers et al.(2016)]{chambers2016} Chambers, K.~C., Magnier, E.~A., Metcalfe, N., et al.\ 2016, arXiv:1612.05560

\bibitem[Chicherov(1997)]{chi1997} Chicherov, A.~V.\ 1997, Astronomy Letters, 23, 600

\bibitem[Clement et al.(2001)]{clement2001} Clement, C.~M., Muzzin, A., Dufton, Q., et al.\ 2001, \aj, 122, 2587

\bibitem[Clement(2017)]{clement2017} Clement, C.~M.\ 2017, VizieR Online Data Catalog, V/150

\bibitem[Coppola et al.(2015)]{coppola2015} Coppola, G., Marconi, M., Stetson, P.~B., et al.\ 2015, \apj, 814, 71
  
\bibitem[Corwin \& Carney(2001)]{cor2001} Corwin, T.~M. \& Carney, B.~W.\ 2001, \aj, 122, 3183

\bibitem[Da Costa et al.(2010)]{dac2010} Da Costa, G.~S., Rejkuba, M., Jerjen, H., et al.\ 2010, \apjl, 708, L121
  
\bibitem[Dambis et al.(2014)]{dambis2014} Dambis, A.~K., Rastorguev, A.~S., \& Zabolotskikh, M.~V.\ 2014, \mnras, 439, 3765

\bibitem[Dark Energy Survey Collaboration et al.(2016)]{des2016} Dark Energy Survey Collaboration, Abbott, T., Abdalla, F.~B., et al.\ 2016, \mnras, 460, 1270.
  
\bibitem[Dekany et al.(2020)]{dec20} Dekany, R., Smith, R.~M., Riddle, R., et al.\ 2020, \pasp, 132, 038001

\bibitem[Dias et al.(2015)]{dias2015} Dias, B., Barbuy, B., Saviane, I., et al.\ 2015, \aap, 573, A13
  
\bibitem[Dias et al.(2016a)]{dias2016a} Dias, B., Barbuy, B., Saviane, I., et al.\ 2016a, \aap, 590, A9

\bibitem[Dias et al.(2016b)]{dias2016b} Dias, B., Saviane, I., Barbuy, B., et al.\ 2016b, The Messenger, 165, 19
  
\bibitem[Drake et al.(2013)]{drake2013} Drake, A.~J., Catelan, M., Djorgovski, S.~G., et al.\ 2013, \apj, 763, 32

\bibitem[Flaugher et al.(2015)]{flaugher2015} Flaugher, B., Diehl, H.~T., Honscheid, K., et al.\ 2015, \aj, 150, 150

\bibitem[Foreman-Mackey(2016)]{fm2016} Foreman-Mackey, D.\ 2016, The Journal of Open Source Software, 1, 24

\bibitem[Foreman-Mackey et al.(2013)]{fm2013} Foreman-Mackey, D., Hogg, D.~W., Lang, D., et al.\ 2013, \pasp, 125, 306

\bibitem[Freedman et al.(2001)]{freedman2001} Freedman, W.~L., Madore, B.~F., Gibson, B.~K., et al.\ 2001, \apj, 553, 47
  
\bibitem[Fu et al.(2019)]{fu2019} Fu, S.~W., Simon, J.~D., \& Alarc{\'o}n Jara, A.~G.\ 2019, \apj, 883, 11
  
\bibitem[Graham et al.(2019)]{gra19} Graham, M.~J., Kulkarni, S.~R., Bellm, E.~C., et al.\ 2019, \pasp, 131, 078001

\bibitem[Green(2018)]{green2018} Green, G.~M.\ 2018, The Journal of Open Source Software, 3, 695
  
\bibitem[Green et al.(2019)]{green2019} Green, G.~M., Schlafly, E., Zucker, C., et al.\ 2019, \apj, 887, 93

\bibitem[Guhathakurta et al.(1994)]{guh1994} Guhathakurta, P., Yanny, B., Bahcall, J.~N., et al.\ 1994, \aj, 108, 1786

\bibitem[Harris et al.(2020)]{harris2020} Harris, C.~R., Millman, K.~J., van der Walt, S.~J., et al.\ 2020, \nat, 585, 357

\bibitem[Heinze et al.(2018)]{henize2018} Heinze, A.~N., Tonry, J.~L., Denneau, L., et al.\ 2018, \aj, 156, 241
  
\bibitem[Hunter(2007)]{hunter2007} Hunter, J.~D.\ 2007, Computing in Science and Engineering, 9, 90
  
\bibitem[Iben(1974)]{iben1974} Iben, I.\ 1974, \araa, 12, 215
  
\bibitem[Ivezi{\'c} et al.(2019)]{lsst2019} Ivezi{\'c}, {\v{Z}}., Kahn, S.~M., Tyson, J.~A., et al.\ 2019, \apj, 873, 111

\bibitem[Joo et al.(2018)]{joo2018} Joo, S.-J., Kyeong, J., Yang, S.-C., et al.\ 2018, \apj, 861, 23
  
\bibitem[Kelson et al.(1996)]{kelson1996} Kelson, D.~D., Illingworth, G.~D., Freedman, W.~F., et al.\ 1996, \apj, 463, 26
  
\bibitem[Kochanek et al.(2017)]{kochanek2017} Kochanek, C.~S., Shappee, B.~J., Stanek, K.~Z., et al.\ 2017, \pasp, 129, 104502

\bibitem[Kopacki(2013)]{kopacki2013} Kopacki, G.\ 2013, \actaa, 63, 91

\bibitem[Kunder et al.(2013)]{kunder2013} Kunder, A., Stetson, P.~B., Cassisi, S., et al.\ 2013, \aj, 146, 119

\bibitem[Lancaster et al.(2019)]{lancester2019} Lancaster, L., Belokurov, V., \& Evans, N.~W.\ 2019, \mnras, 484, 2556
  
\bibitem[Madore(1982)]{madore1982} Madore, B.~F.\ 1982, \apj, 253, 575
  
\bibitem[Madore \& Freedman(1991)]{madore1991} Madore, B.~F. \& Freedman, W.~L.\ 1991, \pasp, 103, 933

\bibitem[Magnier et al.(2020)]{magnier2020} Magnier, E.~A., Sweeney, W.~E., Chambers, K.~C., et al.\ 2020, \apjs, 251, 5
  
\bibitem[Marconi et al.(2003)]{marconi2003} Marconi, M., Caputo, F., Di Criscienzo, M., et al.\ 2003, \apj, 596, 299

\bibitem[Marconi et al.(2006)]{marconi2006} Marconi, M., Cignoni, M., Di Criscienzo, M., et al.\ 2006, \mnras, 371, 1503

\bibitem[Marconi et al.(2015)]{marconi2015} Marconi, M., Coppola, G., Bono, G., et al.\ 2015, \apj, 808, 50
  
\bibitem[Masci et al.(2019)]{mas19} Masci, F.~J., Laher, R.~R., Rusholme, B., et al.\ 2019, \pasp, 131, 018003

\bibitem[Muraveva et al.(2018)]{muraveva2018} Muraveva, T., Delgado, H.~E., Clementini, G., et al.\ 2018, \mnras, 481, 1195
  
\bibitem[Neeley et al.(2017)]{neeley2017} Neeley, J.~R., Marengo, M., Bono, G., et al.\ 2017, \apj, 841, 84

\bibitem[Neeley et al.(2019)]{neeley2019} Neeley, J.~R., Marengo, M., Freedman, W.~L., et al.\ 2019, \mnras, 490, 4254

\bibitem[Nemec et al.(1994)]{nemec1994} Nemec, J.~M., Nemec, A.~F.~L., \& Lutz, T.~E.\ 1994, \aj, 108, 222
  
\bibitem[Ngeow et al.(2021)]{ngeow2021} Ngeow, C.-C., Liao, S.-H., Bellm, E.~C., et al.\ 2021, \aj, 162, 63
  
\bibitem[Pollacco et al.(2006)]{pollaco2006} Pollacco, D.~L., Skillen, I., Collier Cameron, A., et al.\ 2006, \pasp, 118, 1407

\bibitem[Pritzl et al.(2005)]{pritzl2005} Pritzl, B.~J., Venn, K.~A., \& Irwin, M.\ 2005, \aj, 130, 2140
  
\bibitem[Riess et al.(2020)]{riess2020} Riess, A.~G., Yuan, W., Casertano, S., et al.\ 2020, \apjl, 896, L43
  
\bibitem[Sandage \& Tammann(2006)]{sandage2006} Sandage, A. \& Tammann, G.~A.\ 2006, \araa, 44, 93

\bibitem[Sesar et al.(2010)]{ses2010} Sesar, B., Ivezi{\'c}, {\v{Z}}., Grammer, S.~H., et al.\ 2010, \apj, 708, 717
    
\bibitem[Sesar et al.(2011)]{sesar2011} Sesar, B., Stuart, J.~S., Ivezi{\'c}, {\v{Z}}., et al.\ 2011, \aj, 142, 190
  
\bibitem[Sesar et al.(2017)]{ses2017} Sesar, B., Hernitschek, N., Mitrovi{\'c}, S., et al.\ 2017, \aj, 153, 204

\bibitem[Smith(2004)]{smith2004} Smith, H.~A.\ 2004, RR Lyrae Stars, by Horace A. Smith, pp. 166. ISBN 0521548179. Cambridge, UK: Cambridge University Press, September 2004

\bibitem[Sollima et al.(2006)]{sollima2006} Sollima, A., Cacciari, C., \& Valenti, E.\ 2006, \mnras, 372, 1675

\bibitem[Tonry et al.(2012)]{tonry2012} Tonry, J.~L., Stubbs, C.~W., Lykke, K.~R., et al.\ 2012, \apj, 750, 99
  
\bibitem[Tonry et al.(2018)]{tonry2018} Tonry, J.~L., Denneau, L., Heinze, A.~N., et al.\ 2018, \pasp, 130, 064505

\bibitem[Torrealba et al.(2016)]{torrealba2016} Torrealba, G., Koposov, S.~E., Belokurov, V., et al.\ 2016, \mnras, 459, 2370

\bibitem[Turner et al.(1998)]{turner1998} Turner, A., Ferrarese, L., Saha, A., et al.\ 1998, \apj, 505, 207
  
\bibitem[VandenBerg et al.(2000)]{vdb2000} VandenBerg, D.~A., Swenson, F.~J., Rogers, F.~J., et al.\ 2000, \apj, 532, 430
  
\bibitem[VanderPlas \& Ivezi{\'c}(2015)]{vdp2015} VanderPlas, J.~T., \& Ivezi{\'c}, {\v{Z}}.\ 2015, \apj, 812, 18

\bibitem[VanderPlas(2016)]{vdp2016} VanderPlas, J.\ 2016, gatspy: General tools for Astronomical Time Series in Python, ascl:1610.007

\bibitem[V{\'a}squez et al.(2018)]{vasquez2018} V{\'a}squez, S., Saviane, I., Held, E.~V., et al.\ 2018, \aap, 619, A13
  
\bibitem[Virtanen et al.(2020)]{virtanen2020} Virtanen, P., Gommers, R., Oliphant, T.~E., et al.\ 2020, Nature Methods, 17, 261

\bibitem[Vivas et al.(2017)]{viv2017} Vivas, A.~K., Saha, A., Olsen, K., et al.\ 2017, \aj, 154, 85
  
\bibitem[Vivas et al.(2020)]{vivas2020} Vivas, A.~K., Walker, A.~R., Mart{\'\i}nez-V{\'a}zquez, C.~E., et al.\ 2020, \mnras, 492, 1061

\bibitem[Walker et al.(2019)]{walker2019} Walker, A.~R., Mart{\'\i}nez-V{\'a}zquez, C.~E., Monelli, M., et al.\ 2019, \mnras, 490, 4121
  
\end{thebibliography}
\end{document}